\documentclass[
 reprint,
 amsmath,amssymb,
 aps,superscriptaddress
]{revtex4-2}

\usepackage{times}

\usepackage{graphicx}
\usepackage{dcolumn}
\usepackage{bm}
\usepackage{mathdots}
\usepackage{hyperref}
\hypersetup{
    colorlinks=true,   
    allcolors=blue,    
}

\begin{document}

\title{Defeating Barren Plateaus with Task-Aligned Symmetry}

\author{Ruipeng Xing}
\affiliation{%
Faculty of Information Science and Engineering, Ocean University of China, Qingdao 266100, China
}%

\author{Yanan Li}
\affiliation{%
Faculty of Information Science and Engineering, Ocean University of China, Qingdao 266100, China
}%

\author{Zhen Shang}
\affiliation{%
Faculty of Information Science and Engineering, Ocean University of China, Qingdao 266100, China
}%

\author{Shengbin Wang}
\affiliation{%
China Telecom Quantum Information Technology Group Co., LTD., Hefei 230088, China
}%

\author{Yongjian Gu}
\affiliation{%
Faculty of Information Science and Engineering, Ocean University of China, Qingdao 266100, China
}%

\author{Zhimin Wang}
\email{email: wangzhimin@ouc.edu.cn}
\affiliation{%
Faculty of Information Science and Engineering, Ocean University of China, Qingdao 266100, China
}%

\begin{abstract}
\vspace{-0.2em}
\begin{center}
\textbf{ABSTRACT}
\end{center}
Barren plateaus---the exponential vanishing of gradients---are a fundamental obstacle to training scalable quantum neural networks. Whether they arise in quantum recurrent neural networks (QRNNs), a natural architecture for sequential data, remains a pressing question. Here we show that the decisive ingredient for trainability in QRNNs is not the recurrent circuit topology \emph{per se}, but enforcing time-translation symmetry through parameter sharing across time steps. We prove that, without parameter sharing, QRNNs suffer from barren plateaus, with gradient variance decaying exponentially with sequence length. Imposing parameter sharing across time steps fundamentally alters this scaling, transforming it into a polynomial dependence and thereby suppressing the barren plateau. Numerical simulations corroborate these analytical predictions. By rigorously showing how time-translation symmetry suppresses barren plateaus and enhances learning capability in QRNNs, our work establishes task-aligned symmetry as a constructive resolution to the expressivity-trainability tension in quantum neural networks. 
\end{abstract}

\maketitle

\section*{\label{A} Introduction}
Variational quantum algorithms represent the leading paradigm for harnessing near-term quantum devices, yet their promise is fundamentally threatened by the problem of barren plateaus (BPs) \cite{bharti2022noisy,cerezo2021variational,mcclean2016theory,tilly2022variational,blekos2024review}. This threat is especially pernicious for quantum neural networks (QNNs) \cite{wan2017quantum,farhi2018classification,mitarai2018quantum,benedetti2019parameterized,schuld2020circuit}, for the very expressivity that underpins their potential quantum advantage over classical models also provokes a trainability crisis \cite{schuld2019quantum,havlivcek2019supervised,du2021learnability,du2023problem,abbas2021power}. BPs can be viewed as a quantum manifestation of the curse of dimensionality, where a highly expressive yet weakly structured ansatz explores vast regions of Hilbert space, resulting in exponentially suppressed gradients \cite{larocca2025barren,liu2022presence,mcclean2018barren,cerezo2021cost,holmes2022connecting,pesah2021absence,sharma2022trainability,wang2021noise,patti2021entanglement,letcher2024tight}. The central challenge for QNNs is therefore to construct a structured ansatz that is both sufficiently powerful for a given learning task and inherently trainable. 

Classical deep learning offers a blueprint: its success is built not on brute-force parameter scaling, but on the deliberate injection of architectural inductive biases \cite{lecun2015deep,goodfellow2016deep}. Convolutional networks exploit spatial translation invariance \cite{khan2020survey,he2016deep}; recurrent networks leverage temporal structure \cite{hochreiter1997long,lipton2015critical}; attention mechanisms encode permutation equivariance \cite{vaswani2017attention}. These symmetries, carefully aligned with the structure of the data, constrain expressivity precisely where it is harmful for optimization while preserving the capacity to model meaningful patterns. This motivates the incorporation of architectural biases from classical models into QNNs. A notable example is the quantum convolutional neural network \cite{cong2019quantum}, whose logarithmic circuit depth and restricted expressivity can mitigate BPs \cite{pesah2021absence}.

For sequential data---a modality as fundamental as images---quantum recurrent neural networks (QRNNs) have been proposed as the natural counterpart to classical recurrent models \cite{bausch2020recurrent,takaki2021learning,chen2022quantum,di2022dawn,li2023quantum}. A QRNN processes a sequence of length $T$ by iteratively applying a recurrent block that couples a persistent quantum memory with new input at each time step. Our recent work established that QRNNs can, in principle, overcome the long-term dependency bottlenecks that limit classical recurrent architectures, owing to the completely positive, trace-preserving nature of quantum channels \cite{li2025quantum}. However, the ability to maintain long-range memory does not guarantee trainability. This naturally raises a pressing question: do QRNNs exhibit BPs and, if so, what architectural choices determine their trainability?

In this work, we answer this question by disentangling the roles of the two key inductive biases present in QRNNs, namely the recurrent architecture and the time-translation symmetry enforced through parameter sharing across time steps, and identifying the latter as the decisive mechanism governing their trainability. Specifically, we first prove that, without parameter sharing, QRNNs inevitably suffer from BPs, with gradient variance decaying exponentially with sequence length, indicating that the recurrent architecture alone offers no inherent protection. We then prove that imposing time-translation symmetry through parameter sharing changes this scaling from exponential decay to polynomial dependence on sequence length, thereby suppressing BPs. Numerical simulations corroborate these analytical predictions. 

Our results establish task-aligned symmetry as a practical and theoretically grounded design principle for building trainable quantum sequential models. More broadly, they illuminate a path through the central tension in QNNs: task-aligned symmetry offers a principled means of constraining the immense expressivity of Hilbert space, without succumbing to the exponential optimization curse that unstructured expressivity provokes.

\vspace{-0.2em}
\section*{\label{B} Results}
\vspace{-0.2em}
\subsection{\label{B} Architecture of QRNNs}
\vspace{-0.2em}

\begin{figure}[t]
    \includegraphics[width=\columnwidth]{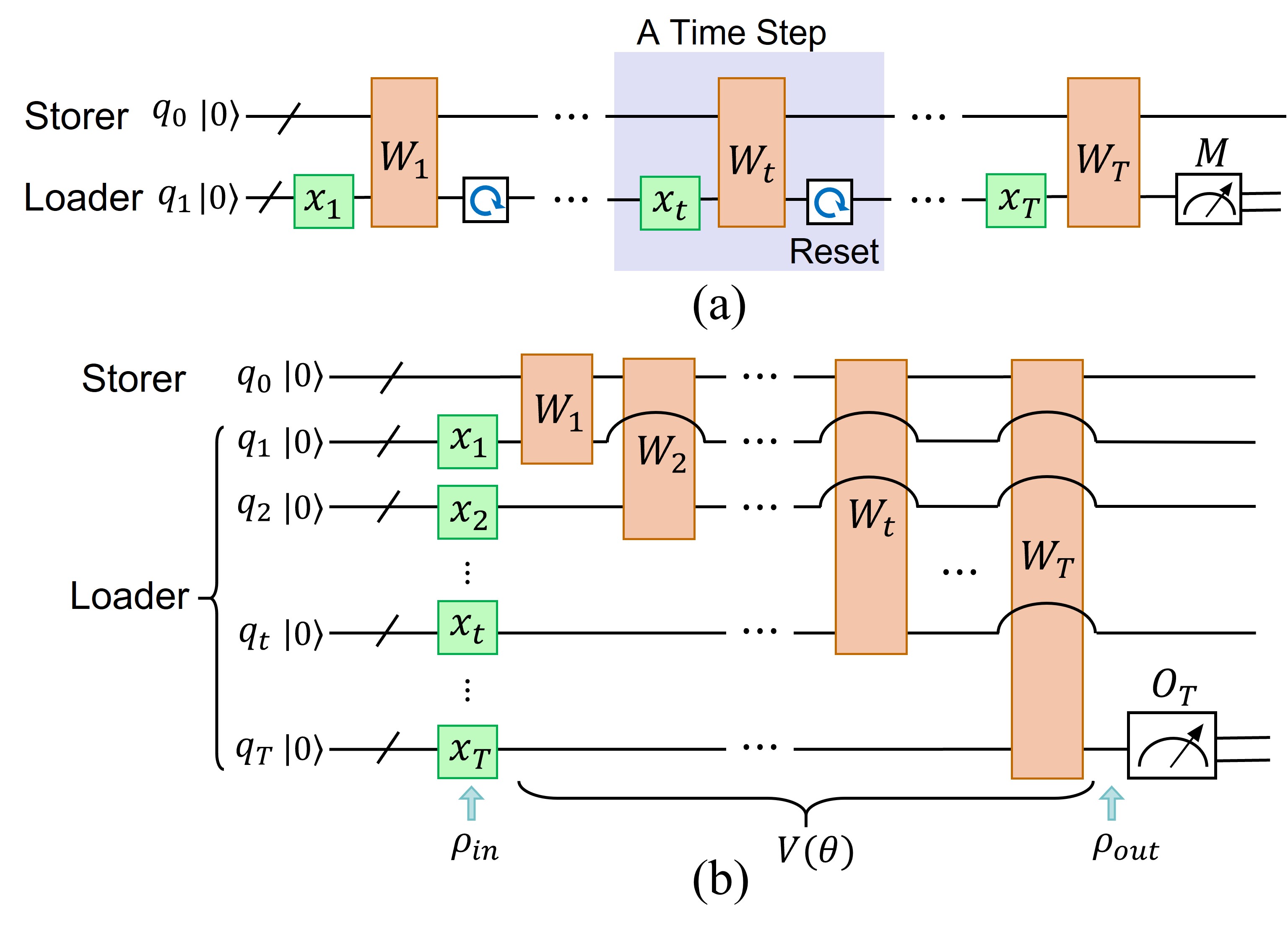}
    \caption{QRNN architectures. (a) Standard model $\mathcal Q_1$: a recurrent block $W_t$ couples the Storer register ($q_0$) and the Loader register ($q_1$); the Loader can be reset and reused at each step. (b) Equivalent model $\mathcal Q_2$: each time step uses a distinct Loader register, and all intermediate partial traces are postponed until the end. Our theoretical analysis is based on $\mathcal Q_2$.}
    \label{fig:Fg1}
\end{figure}

A QRNN processes a length-$T$ sequence by iterating a recurrent quantum block, as illustrated in Fig.~\ref{fig:Fg1}(a). Each block operates on two registers: a \emph{Storer} that carries the hidden state across time steps (encoding history information), and a \emph{Loader} that encodes the current input \cite{li2023quantum}. This separation naturally makes QRNNs compatible with variable-length sequences while also enabling qubit-efficient implementations through resetting and reusing the Loader register after each step. 

At time step $t$, the input $x_t$ is embedded into the Loader as a quantum state
\begin{equation}\label{eq0}
    \rho_{x_t} = U_{in}(f(x_t))|\mathbf{0}\rangle\langle \mathbf{0}| U_{in}^\dagger(f(x_t)),
\end{equation}
which is then combined with the previous hidden state $\rho_{h_{t-1}}$ through a variational block $W_t$ acting on the Storer$\otimes$Loader. The updated hidden state is obtained by tracing out the Loader,
\begin{equation}\label{eq1}
    \rho_{h_t} = \text{Tr}_L \left[ W_t (\rho_{h_{t-1}} \otimes \rho_{x_t}) W_t^\dagger \right].
\end{equation}
A task-dependent readout can be obtained by measuring the Loader at intermediate steps. Operationally, the Loader qubits can be reset and reused after each step, reducing the total qubit count. Additionally, a staggered connection pattern further shortens the effective depth, thereby relaxing coherence-time requirements \cite{li2023quantum}. Therefore, mid-circuit measurements and qubit reuse naturally appear in this model, making it compatible with current quantum hardware execution capabilities \cite{decross2023qubit, ivashkov2024high, zhang2025generalized}.

For theoretical analysis, we introduce an equivalent QRNN model, denoted by $\mathcal Q_2$ (Fig.~\ref{fig:Fg1}(b)), in which each time step uses a fresh Loader register and all intermediate partial traces are postponed until the end. Let $V=\prod_{t=1}^{T} W_t$ denote the global unitary, where $W_t$ couples the Storer to the $t$-th Loader register at step $t$. Then the final readout on the last Loader can be written compactly as    
\begin{equation}\label{eq3}
    \langle O \rangle = \text{Tr} \left[V \left( \rho_{h_0} \otimes \bigotimes_{t=1}^{T} \rho_{x_t} \right)  V^\dagger O_T \right],
\end{equation}
where $\rho_{h_0} = |0\rangle\langle 0|^{\otimes m}$ denotes the initial state of the Storer, and $O_T$ acts only on the final Loader register. The equivalence between $\mathcal Q_1$ and $\mathcal Q_2$ follows from the commutativity of the partial trace with unitaries acting on disjoint subsystems, as detailed in the Supplemental Material (Sec. B) \cite{SupplementalMaterial}. Eq.~(\ref{eq3}) serves as the basis for our gradient variance analysis.

\subsection{\label{C} Existence of BPs in QRNNs}

We first analyze the existence of BPs in QRNNs using the $\mathcal{Q}_{2}$ model, under the assumption that no parameter sharing is imposed among the recurrent blocks $W_t$. The QRNN ansatz is given by:
\begin{equation}\label{eq6}
    V( \bm{\theta}) = \prod_{t=1}^T W_t = \prod_{t=1}^T W_{t0}(\bm{\theta_t})\otimes I_{\overline{t0}},
\end{equation}
where  $W_{t0}$  represents the parameterized quantum circuit (PQC) acting on subsystems $q_0$ and $q_t$, with parameter vector  $\bm{\theta_t}$. Each $W_{t0}$ can be decomposed into elementary gates of the form $ G_\nu(\theta^\nu_t) = R_\nu(\theta^\nu_t)Q_\nu$, where $Q_\nu$ is a parameter-free quantum gate, and $R_\nu(\theta^\nu_t) = \exp(-i\theta^\nu_t\sigma_\nu/2)$, with $\sigma_\nu$ being a Pauli operator.

To analyze BPs, we compute the partial derivatives of the loss function with respect to the individual parameters $ \{\theta_t^{\nu} \}$ \cite{mcclean2018barren,cerezo2021cost}. The loss function, defined over $N$ data points $\{x^{(1)}, ..., x^{(N)}\}$, is $\mathcal{L}_N = \frac{1}{N} \sum_{i=1}^{N} L(\langle O \rangle^{(i)})$, and the gradient of $L$ can be split into two factors: $\frac{\partial L}{\partial \langle O \rangle}$ and $\frac{\partial \langle O \rangle}{\partial \theta_t^{\nu}}$. The first factor depends on the specific form of the loss function and can be bounded with an appropriate choice. Accordingly, our primary focus is on the second factor, which depends on the measurement outcome $C \equiv \langle O \rangle = \text{Tr}[OV(\bm{\theta})\rho_{in}V(\bm{\theta})^\dagger]$, with $\rho_{in}$ denoting the input state given in Eq.~(\ref{eq3}). 

For a given parameter $\theta_t^{\nu}$, the ansatz circuit $V(\bm{\theta})$ is split at the gate $G_\nu(\theta_t^{\nu})$ within the block $W_{t0}(\bm{\theta_t})$, and the corresponding partial derivative can be written as:
\begin{equation}\label{eq10}
    \begin{split}
        \partial_\nu C \equiv \frac{\partial C}{\partial \theta_t^{\nu}} =- \text{Tr}\left\{( W_B^\dagger \otimes \mathbb{I}_{\overline{t0}}) V_\mathcal{R}^\dagger O V_\mathcal{R} ( W_B \otimes \mathbb{I}_{\overline{t0}} ) \times \right .\\
        \left. \left[ \frac{i}{2}\sigma_\nu \otimes \mathbb{I}_{\overline{t0}}, (W_A \otimes \mathbb{I}_{\overline{t0}}) V_\mathcal{L} \rho_{in} V_\mathcal{L}^\dagger (W_A^\dagger \otimes \mathbb{I}_{\overline{t0}}) \right] \right\}.
    \end{split}
\end{equation}
Note that $W_{t0}(\bm{\theta_t})=W_BW_A$ and $V(\boldsymbol{\theta}) =  V_{\mathcal{R}} \left( \left( W_{B}W_{A} \right ) \otimes \mathbb{I}_{\overline{t0}} \right)V_{\mathcal{L}}$. Assuming the ansatz circuit $V(\bm{\theta})$ is fully random (i.e., the family $\{ W_{t0}(\bm{\theta_t}) \}$ forms a unitary $1$-design), the expectation value of the partial derivative under the Haar measure vanishes,
\begin{equation}\label{eq12}
    \left\langle \partial_{\nu}C \right\rangle_{V(\bm{\theta})} = 0.
\end{equation}
Thus, the loss function gradient is unbiased. A detailed derivation is provided in the Supplemental Material (Sec. C) \cite{SupplementalMaterial}.

Next, we compute the variance of the partial derivative, $\text{Var}[\partial_{\nu}C] = \langle (\partial_\nu C)^2 \rangle_{V(\bm{\theta})}$ \cite{mcclean2018barren}. When this variance decreases exponentially with system size or sequence length, Chebyshev's inequality implies that the gradient concentrates around zero with high probability, which is the defining feature of a barren plateau. Under the condition that $W_A$ and $W_B$ form independent 2-designs, the variance can be expressed as \cite{cerezo2021cost}:
\begin{equation}\label{eq15}
    \langle (\partial_{\nu} C)^2 \rangle = \frac{2^{2m-1} \text{Tr}[\sigma_{\nu}^2]}{(2^{4m}-1)^2} \sum_{\bm{pq}, \bm{p'q'}} \langle \Delta \Omega_{\bm{pq}}^{\bm{p'q'}} \rangle_{V_\mathcal{R}} \langle \Delta \Psi_{\bm{pq}}^{\bm{p'q'}} \rangle_{V_\mathcal{L}},
\end{equation}
where $m$ is the number of qubits in each of the Storer and Loader registers. 

By exploiting the recursive structure of the QRNN ansatz circuit, we can efficiently evaluate the first factor $\langle \Delta \Omega_{\bm{pq}}^{\bm{p'q'}} \rangle_{V_\mathcal{R}}$. The main idea is to formulate the operator $\Omega_{\bm{pq}}$ in a recursive manner, and perform Haar integration over the sequential modules $W_{t}$ iteratively (see Supplementary Information, Sec. D \cite{SupplementalMaterial}). The result for this factor is given by 
\begin{equation}\label{eq16}
    \begin{split}
        \left\langle\Delta \Omega_{\bm{pq}}^{\bm{p^{\prime} q^{\prime}}}\right\rangle_{V_{\mathcal{R}}}=\delta_{(\bm{p q})_{<t}} \delta_{\left(\bm{p^{\prime} q^{\prime}}\right)_{<t}} \delta_{\left(\bm{p q^{\prime}}\right)_{>t}} \delta_{\left(\bm{p^{\prime} q}\right)_{>t}} \times\\
        \left(\frac{2^{m}}{2^{2 m}+1}\right)^{T-t} D_{H S}\left(O_{T}, \frac{\operatorname{Tr}\left[O_{T}\right] \mathbb{I}_{T}}{2^{m}}\right),
    \end{split}
\end{equation}
where $D_{HS}(\rho, \sigma) = \text{Tr}[(\rho - \sigma)^2]$ denotes the Hilbert-Schmidt distance. Combining this result with $\Delta \Psi_{\bm{pq}}^{\bm{p'q'}}$, we obtain the final result for the variance,
\begin{equation}\label{eq16-2}
    \begin{split}
    \left\langle \left( \partial_{\nu}C \right)^{2} \right\rangle_{V} = \frac{2^{2m - 1}\text{Tr}\left[ \sigma_{\nu}^{2} \right]}{{(2^{4m} - 1)}^{2}} & {\left(\frac{2^{m}}{2^{2m} + 1}\right)}^{T - t} \text{Tr}\left[ {\widetilde{\rho}}_{> t}^{2} \right] \times\\
   D_{HS}\left( O_{T}, \frac{\text{Tr}[ O_{T}]\mathbb{I}_{T}}{2^{m}} \right) & \left\langle D_{HS}\left( {\widetilde{\rho}}_{t0}, \frac{\mathbb{I}_{t0}}{2^{2m}} \right) \right\rangle_{V_{\mathcal{L}}}.
    \end{split}
\end{equation}
Here, $\text{Tr}[\sigma_\nu^2] = 2^{2m}$ and $ D_{HS} \left( \hat{\rho}_{t0},{\mathbb{I}_{t0}}/{2^{2m}} \right) \in [0,2]$. The scaling of $D_{HS} \left( O_T, {\text{Tr}[O_T] \mathbb{I}_T}/{2^m} \right)$ is $\mathcal{O}(1)$ for global measurement and $\mathcal{O}(2^m/m)$ for local measurement. Based on Eq. (\ref{eq16-2}), we obtain the following theorem:

\textit{Theorem 1.} For a QRNN with independent parameters and $W_A$, $W_B$, and $W_t$ forming 2-designs, the gradient variance is bounded by: 
\begin{equation}\label{eq17}
    \text{Var}[\partial_\nu C] \leq g(T), \quad \text{with} \quad g(T) \in \mathcal{O}\left(\frac{1}{2^m}\right)^T,
\end{equation}
where $T$ is the sequence length and $m$ is the register size of Storer and Loader in QRNNs. 

Theorem 1 shows that the gradient variance decays exponentially with both the number of time steps and system size, leading to BPs. Therefore, recurrent architecture alone does not prevent BPs in QRNNs, which is consistent with related observations for dissipative QNNs \cite{sharma2022trainability,beer2020training}.

\subsection{\label{D} Mitigating BPs in QRNNs}

We next analyze BP behavior in QRNNs with parameter sharing across recurrent blocks in $\mathcal{Q}_{2}$. Although the Haar measure approach can in principle be used to compute the gradient variance, parameter sharing introduces significant computational challenges. Specifically, in a QRNN with parameter sharing, the partial derivative of the loss function contains $2T$ terms, compared to just two in the non-sharing case (see Eq. (\ref{eq10})). Each term is a degree-$T$ polynomial, requiring $W_A$ or $W_B$ to form unitary $T$-designs. The corresponding Haar integral involves summing over $T!$ permutations, leading to a combinatorial explosion that makes closed-form evaluation intractable. 

To address this, we exploit the trigonometric-moment representation of the loss function to compute the gradient variance, rather than using the Haar measure approach. Before turning to the general case, we present three examples that illustrate, from different perspectives, how parameter sharing mitigates BPs. 

In the first example, where the ansatz circuit is extended along the longitudinal direction (i.e., circuit depth), parameter sharing results in a gradient variance of $\text{Var}(\partial_\theta C_{\text{share}}^{(1)}) = T^2/8$, while non-sharing gives $\text{Var}(\partial_\theta C_{\text{non}}^{(1)}) = 1/8$. The second example, extending the ansatz circuit along the transverse direction (i.e., qubit number), shows that non-sharing yields a gradient variance of $\text{Var}(\partial_\theta C_{\text{non}}^{(2)}) = \frac{1}{8}\left(\frac{3}{8}\right)^{T-1}$, exhibiting exponential decay with $T$, whereas parameter sharing results in $ \text{Var}(\partial_\theta C_{\text{share}}^{(2)}) = \frac{T}{2^{4T}} \binom{4T - 2}{2T - 1}$, which scales as $\mathcal{O}(\sqrt{T})$ and therefore exhibits polynomial growth with $T$. These two examples show that parameter sharing can effectively amplify gradient variance along the two fundamental structural dimensions of variational circuits. 

In the third example, for a QRNN model with a specific ansatz structure, non-sharing results in a gradient variance of $\text{Var}(\partial_\theta C_{\text{non}}^{(3)}) = \frac{1}{2^{T+2}}$, exhibiting BPs. In contrast, parameter sharing yields $\text{Var}(\partial_\theta C_{\text{share}}^{(3)}) \approx \frac{\sqrt{T}}{8\sqrt{\pi}}$, illustrating a transition from exponential decay $\mathcal{O}(2^{-T})$ to polynomial growth $\mathcal{O}(\sqrt{T})$, thereby mitigating BPs. Additional details of these examples are presented in the Supplementary Information (Sec. E) \cite{SupplementalMaterial}.  

For a general QRNN with $T$ time steps, the loss function can be expressed as
\begin{equation}\label{s11}
    C = \sum_k \beta_k \prod_{j=1}^{\xi} \cos^{a_{kj}}(\theta_j) \sin^{b_{kj}}(\theta_j),
\end{equation}
where $\{\theta_j\}_{j=1}^\xi$ are the parameters of each recurrent block $W_t$, shared across all $T$ blocks. Here, $a_{kj}, b_{kj} \ge 0$ and $a_{kj}+b_{kj} \le T$, where the upper bound $T$ arises from the $T$-fold repetition of each gate, and $\beta_k$ is a constant coefficient that depends primarily on the observable and the quantum gates. The gradient variance of the loss function is then given by
\begin{equation}\label{s12}
\mathrm{Var}[\partial_{\nu}C_{\mathrm{share}}] \approx
     \sum_k \beta_k^2 \,\gamma_\nu(T)\,
           \prod_{j=1}^{\xi}
           \frac{\Gamma\!\bigl(a_{kj}+\tfrac12\bigr)\Gamma\!\bigl(b_{kj}+\tfrac12\bigr)}{\Gamma(a_{kj}+b_{kj}+1)},
\end{equation}
where $\Gamma(\cdot)$ denotes the Gamma function and $\gamma_\nu(T)\propto T^2$. If there exists an index $k$ such that $a_{k i} = \mathcal{O}(1)$ (or $a_{k i}\ll T$) for all $i$ in Eq. (\ref{s11}), then the Gamma-function factor in Eq. (\ref{s12}) scales as $\Omega\!\left(T^{-c}\right)$ for some constant $c$ independent of $T$. Moreover, as shown in the Supplementary Information (Sec. E) \cite{SupplementalMaterial}, QRNNs with parameter sharing admit at least one index $k$ for which $a_{k i} = \mathcal{O}(1)$ holds for all $i$. This is because parameter sharing induces correlations that yield a trigonometric expansion containing terms of the form $\cos^c(\theta) \sin^{T-c}(\theta)$. Therefore, we obtain the following theorem:

\textit{Theorem 2.} For a QRNN with parameter sharing across $T$ time steps, the gradient variance satisfies \begin{equation}\label{s13} 
    \mathrm{Var}[\partial_{\nu}C_{\mathrm{share}}] \approx h(T), \quad \text{with} \quad h(T) \in \Omega\!\left(\frac{T^2}{\mathrm{poly}(T)}\right).
\end{equation}
This theorem demonstrates that parameter sharing changes the scaling of the gradient variance in QRNNs from exponential decay to a polynomial dependence on the sequence length $T$, thereby mitigating BPs.

\subsection{\label{E} Numerical experiments}
To validate the theoretical predictions, we conduct two sets of numerical experiments. The first set examines the scaling of gradient variance in the three examples discussed above, with the results provided in the Supplementary Information (Sec. F) \cite{SupplementalMaterial}.    

\begin{figure}[h]
    \includegraphics[width=\columnwidth]{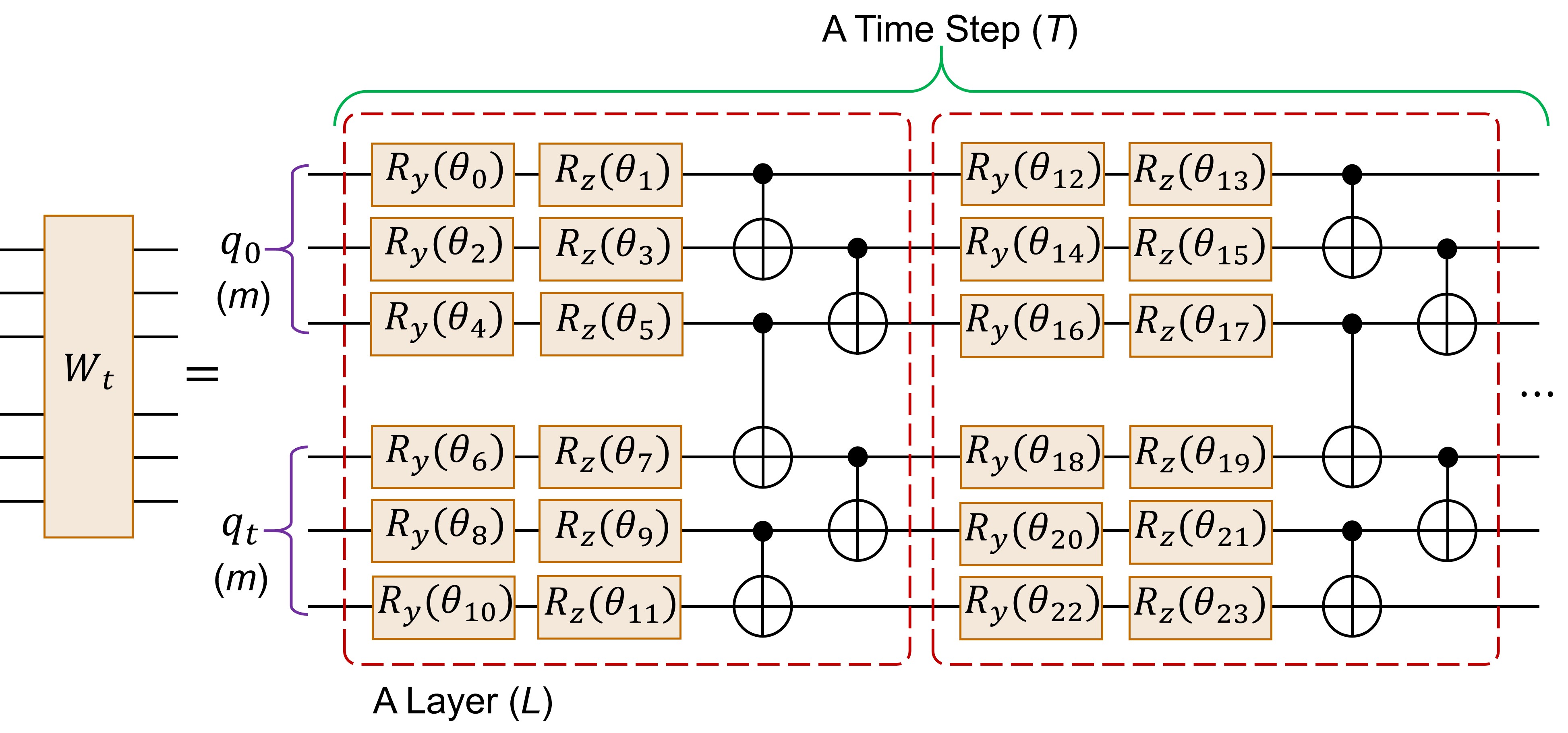}
    \caption{\label{fig:2} A multi-layer ($L$) ansatz circuit used as the module $W_t$ to construct a general QRNN model, as shown in Fig. \ref{fig:Fg1}. The circuit acts on the $q_0$ and $q_t$ registers, and implements a single time step in the QRNN architecture.}
\end{figure}

\begin{figure}[h]
    \includegraphics[width=\columnwidth]{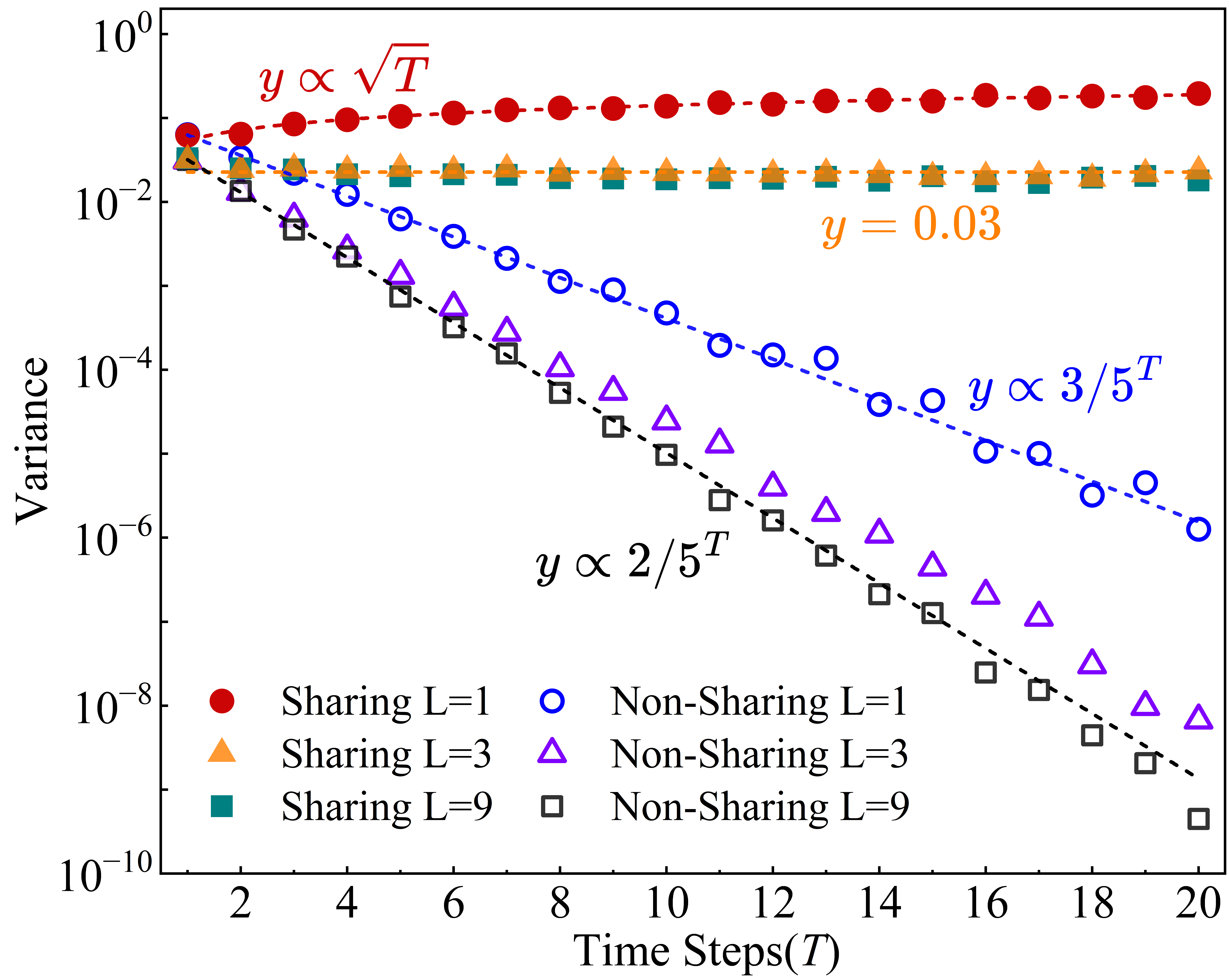}
    \caption{\label{fig:Exp4} Gradient variance of the loss function for a general QRNN model with parameter sharing and non-sharing, using register size $m = 1$ and layers $L=1,3,9$. Fitted curves for $L=1, 9$ are included to illustrate the scaling trends.}
\end{figure}

The second set focuses on a general QRNN model, where a strongly entangled, multi-layer ansatz circuit, as illustrated in Fig. \ref{fig:2}, is used to implement the recurrent block $W_t$ in Fig. \ref{fig:Fg1}. Increasing the number of layers ($L$) enhances the expressiveness of the ansatz, bringing its output state closer to the Haar distribution, thereby enabling the QRNN model to exhibit general learning capabilities. The numerical results for the gradient variance are presented in Fig. \ref{fig:Exp4}. 

In the non-sharing case, even when $L = 1$ (below the 2-design threshold), the variance decays exponentially with respect to the sequence length $T$. This extends the prediction of Theorem 1: even when the recurrent blocks $W_t$ do not approach the Haar random distribution, QRNNs with non-shared parameters still exhibit BPs. Therefore, the recurrent architecture alone is insufficient to mitigate BPs in QRNNs. 

In contrast, with parameter sharing, the variance increases as $\mathcal{O}(\sqrt{T})$ when $L =1$. As $L$ increases from 1 to 9 (bringing the circuit closer to the Haar distribution), the magnitude of the parameter-sharing variance decreases. However, it stabilizes within a certain range and no longer decays with sequence length $T$. These results support Theorem 2, confirming that parameter sharing significantly alters the gradient-scaling behavior in QRNNs and effectively prevents gradient vanishing during training. 

\section*{ Discussion}
The challenge of BPs is far more than a technical inconvenience for variational quantum algorithms; it embodies a fundamental tension at the heart of QNNs. On one hand, we seek to exploit the immense expressivity of Hilbert space to encode and process information in ways that have no classical analogue. On the other, this very expressivity, when left unstructured, flattens optimization landscapes exponentially, rendering learning impossible. How to resolve this tension is arguably the central design problem of the field---the very problem we set out to address for quantum sequential models.

In this work, we provide a rigorous answer by analytically decoupling the two inductive biases intertwined in QRNNs, i.e., the recurrent circuit topology and time-translation symmetry enforced by parameter sharing. We prove that the recurrent architecture alone, despite its structured appearance, offers no protection whatsoever against BPs. Without parameter sharing, the gradient variance decays exponentially with sequence length, inevitably leading to untrainable circuits. In stark contrast, the injection of symmetry through parameter sharing fundamentally changes this scaling, transforming an exponential decay into a polynomial dependence. The architectural element that defeats the curse is not recurrence, but symmetry.

The mechanism underlying this mitigation can be understood in terms of the \textit{effective} independent degrees of freedom that contribute to the gradient. Without parameter sharing, the gradient typically involves $N=\mathcal{O}(T)$ effectively independent averaging factors. Since each factor contributes a constant suppression $c\in(0,1)$, the overall scaling behaves as $c^{\mathcal{O}(T)}$, leading to the exponential decay characteristic of BPs. Parameter sharing, by contrast, imposes rigid correlations across time steps, and the gradient receives contributions from all steps simultaneously and in a coherent manner, collapsing the effective number of independent random variables from $\mathcal{O}(T)$ to $\mathcal{O}(1)$. Consequently, the variance is governed by higher-order moments of a small, fixed set of shared parameters, resulting in at most polynomial suppression. 

Beyond constraining effective independence and suppressing BPs, parameter sharing serves an equally important function for generalization. Imposing time-translation symmetry ensures that the same type of information can be recognized regardless of its temporal position in a sequence \cite{goodfellow2016deep}, thereby directly enhancing the learning capability of QRNNs. Parameter sharing thus acts as a dual-purpose design principle: it simultaneously protects against the exponential curse by limiting effective degrees of freedom, and promotes generalization by encoding the inherent time-translation symmetry of sequential tasks.
 
This dual role offers a constructive resolution to the tension identified at the outset. It demonstrates that the curse of dimensionality in Hilbert space can be addressed with the same principle that has long served to tame complexity in classical machine learning: the deliberate imposition of structured inductive biases. Task-aligned symmetry acts as a powerful regulator; it prunes the vast wilderness of Hilbert space by removing those directions irrelevant to the task while preserving the expressive capacity to represent meaningful solutions. In this view, the exponential scale of quantum state space is channeled, disciplined by the constraints that the problem itself dictates.

In conclusion, our findings resolve the pressing question of BPs in QRNNs and demonstrate that their trainability is born not from any particular circuit motif, but from the principled embrace of symmetry. By providing a rigorous understanding of how task-aligned parameter sharing collapses the effective degrees of freedom that cause BPs, this work offers a solid foundation for building trainable quantum sequential models and suggests a guiding principle that may extend to broader classes of variational quantum models. 

\section*{Methods}
\emph{Equivalent QRNN model.} For the convenience of theoretical analysis, we introduce the equivalent QRNN architecture $\mathcal{Q}_{2}$ as shown in Fig.~\ref{fig:Fg1}(b). The equivalence between the two models is established in the Supplementary Information (Section B) \cite{SupplementalMaterial}. To visualize the sequential action of the recurrent blocks in QRNNs, we adopt a \emph{vertical tensor-product tableau notation}. In this notation, each row corresponds to a subsystem (either the Storer or Loader register), while the columns record the ordered sequence of operators acting over time. For instance, the QRNN evolution $U = \left(\prod_{t=1}^{T}W_t \left(\bigotimes_{t=0}^{T}\rho_t \right) \prod_{t=1}^{T}W_t^\dagger\right) O_T$ can be represented as:
\begin{equation}\label{eq5} 
    \begin{aligned}
        &\begin{pmatrix}
        I & W_{T0} & \cdots & W_{20} & W_{10} & \rho_0 & W_{10}^\dagger & W_{20}^\dagger & \cdots & W_{T0}^\dagger \\
        I & I & \cdots & I & W_{10} & \rho_1 & W_{10}^\dagger & I & \cdots & I \\
        I & I & \cdots & W_{20} & I & \rho_2 & I & W_{20}^\dagger & \cdots & I \\
        \vdots & \vdots & \iddots & \vdots & \vdots & \vdots & \vdots & \vdots & \ddots & \vdots \\
        O_T & W_{T0} & \cdots & I & I & \rho_T & I & I & \cdots & W_{T0}^\dagger
        \end{pmatrix}.
    \end{aligned}
\end{equation}
In this representation, whenever two operators $W_{t0}$ appear in the same column, they act jointly on the subsystems $q_{0}$ and $q_{t}$, generating an entangled state that cannot be factorized into a product state. This notation makes the multi-register correlations induced by the recurrent unitaries explicit, thereby providing an intuitive tool for the subsequent theoretical analysis.

\emph{Haar measure approach.} For QRNNs without parameter sharing, we employ Haar measure integration to evaluate the variance of the loss function gradient and thereby characterize the BP behavior. The general expression for the variance of a partial derivative is given in Eq. (\ref{eq15}), where the integrands corresponding to the right and left parts are
\begin{equation}\label{eq15-1}
    \begin{split}
        {\Delta\Omega}_{\boldsymbol{pq}}^{\boldsymbol{p'q'}} 
        &= \text{Tr}\left[ \Omega_{\boldsymbol{qp}}\Omega_{\boldsymbol{q'p'}} \right] - \frac{\text{Tr}\left[ \Omega_{\boldsymbol{qp}} \right] \text{Tr}\left[ \Omega_{\boldsymbol{q'p'}} \right]}{2^{2m}},\\
        {\Delta\Psi}_{\boldsymbol{pq}}^{\boldsymbol{p'q'}} 
        &= \text{Tr}\left[ \Psi_{\boldsymbol{pq}}\Psi_{\boldsymbol{p'q'}} \right] - \frac{\text{Tr}\left[ \Psi_{\boldsymbol{pq}} \right] \text{Tr}\left[ \Psi_{\boldsymbol{p'q'}} \right]}{2^{2m}}.
    \end{split}
\end{equation}
The operators $\Omega_{\bm{qp}}$ and $\Psi_{\bm{pq}}$ constitute the fundamental building blocks of these integrands; their explicit forms are provided in the Supplementary Information (Section D) \cite{SupplementalMaterial}. 

\begin{figure*}[htbp]
    \includegraphics[width=0.99\textwidth]{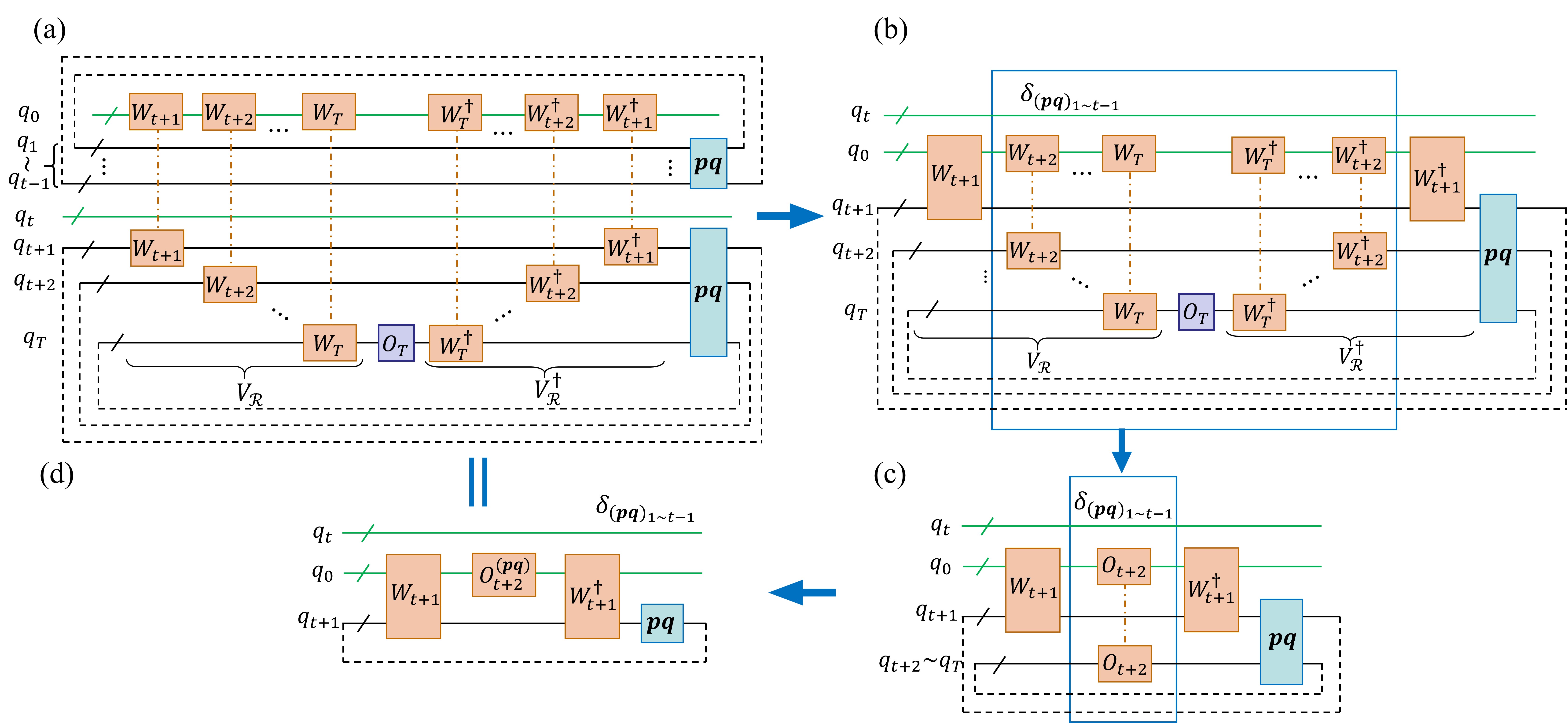}
    \caption{\label{fig:4} Reduction of the operator $\Omega_{\bm{qp}}$ in Eq. (\ref{eq15-1}) to $O_{t + 2}^{\boldsymbol{(pq)}}$, which enables the integral in Eq. (\ref{eq15-2}) to be evaluated recursively. (a) Tensor network representation of $\Omega_{\boldsymbol{q} \boldsymbol{p}}=\operatorname{Tr}_{\bar{t0}}\left[\left(|\boldsymbol{p}\rangle\langle\boldsymbol{q}| \otimes \mathbb{I}_{t 0}\right) V_{\mathcal{R}}^{\dagger} O_T V_{\mathcal{R}}\right]$, where $|\boldsymbol{p}\rangle$ and $|\boldsymbol{q}\rangle$ denote computational basis states on all qubits except those in registers $q_0$ and $q_t$. Paired squares connected by dash-dot lines represent a full circuit module (e.g., $W_T$ denotes the circuit module acting on registers $q_0$ and $q_T$, analogous to the construction in Fig. \ref{fig:Fg1}). Dashed lines indicate partial trace operations. (b) The partial trace of the operator $\left | \boldsymbol{p} \right \rangle \left \langle \boldsymbol{q} \right |$ over the subsystem $q_{1} \sim q_{t - 1}$ simplifies to $\delta_{\boldsymbol{(pq)}_{1\sim t - 1}}$. (c) All operations outside time step $t+1$ are grouped into a single module $O_{t+2}$, which acts on registers $q_0$, $q_t$, and $q_{t+2}\sim q_{T}$. (d) After tracing out the subsystem $q_{t+2}\sim q_T$, the combination of $O_{t+2}$ and $|p\rangle\langle q|$ reduces to an equivalent operator $O_{t + 2}^{\boldsymbol{(pq)}}$ acting solely on register $q_{0}$. As detailed in the Supplemental Material (Section D) \cite{SupplementalMaterial}, $O_{t+2}^{\boldsymbol{(pq)}}$ can be further reduced to $O_{t+3}^{\boldsymbol{(pq)}}$ involving $W_{t+2}$, and this reduction proceeds recursively until reaching the final operators $O_T$ and $W_T$.}       
\end{figure*}

The factor $\langle \Delta \Omega_{\bm{pq}}^{\bm{p'q'}} \rangle_{V_\mathcal{R}}$ in Eq. (\ref{eq15}) can be evaluated by exploiting the recursive structure of the QRNN ansatz circuit. Specifically, because the PQC modules $W_{t0}(\bm{\theta_t})$ shown in Fig.~\ref{fig:Fg1} are independent, the expectation value can be computed sequentially over time, i.e., $\langle {\Delta\Omega}_{\boldsymbol{pq}}^{\boldsymbol{p'q'}} \rangle_{V_{\mathcal{R}}} = \langle \cdots\langle \langle {\Delta\Omega}_{\boldsymbol{pq}}^{\boldsymbol{p'q'}} \rangle_{W_{t + 1}} \rangle_{W_{t + 2}}\cdots \rangle_{W_{T}}$. We begin by averaging over the module $W_{t+1}$. The key step is to express $\Omega_{\bm{pq}}$ in Eq. (\ref{eq15-1}) in a recursive form, as illustrated in Fig. \ref{fig:4}. Under this recursive decomposition, the expectation over $W_{t+1}$ can be expressed as
\begin{widetext}
\begin{equation}\label{eq15-2}
    \begin{split}
        \left\langle {\Delta\Omega}_{\boldsymbol{pq}}^{\boldsymbol{p'q'}} \right\rangle_{W_{t + 1}} 
        &= \int_{}^{}{d\mu(W_{t + 1})\left \{\text{Tr}\left[ \Omega_{\boldsymbol{qp}}\Omega_{\boldsymbol{q'p'}} \right] - \frac{\text{Tr}\left[ \Omega_{\boldsymbol{qp}} \right] \text{Tr}\left[ \Omega_{\boldsymbol{q'p'}} \right]}{2^{2m}} \right \}}\\
        &= \delta_{\boldsymbol{(pq)}_{< t}}\delta_{\boldsymbol{(p'q')}_{< t}}\delta_{\boldsymbol{(pq')}_{t + 1}}\delta_{\boldsymbol{(p'q)}_{t + 1}}\frac{2^{m}(2^{2m} - 1)}{2^{4m} - 1}\left \{ \text{Tr}[ O_{t + 2}^{\boldsymbol{(pq)}} O_{t + 2}^{\boldsymbol{(p'q')}}] - \frac{\text{Tr}[ O_{t + 2}^{\boldsymbol{(pq)}}]\text{Tr}[ O_{t + 2}^{\boldsymbol{(p'q')}}]}{{2^{m}}} \right \}.
    \end{split}
\end{equation}
\end{widetext}
Since $O_{t+2}^{(\bm{pq})}$ possesses the same circuit structure as $\Omega_{\bm{pq}}$, the Haar average over the next module $W_{t+2}$ can be evaluated in the same manner, yielding an expression of the same form in terms of $O_{t+3}^{(\bm{pq})}$. Repeating this procedure over all subsequent time steps yields the expectation $\langle \Delta \Omega_{\bm{pq}}^{\bm{p'q'}} \rangle_{V_\mathcal{R}}$ shown in Eq. (\ref{eq16}). 

Next, by combining the $\delta_{(\cdot)}$ factors from Eq. (\ref{eq16}) with $\Delta \Psi_{\bm{pq}}^{\bm{p'q'}}$ in Eq. (\ref{eq15}), we obtain
\begin{equation}\label{eq16-1}
    \begin{split}
        &\sum_{\boldsymbol{pq}}^{}{\delta_{\boldsymbol{(pq)}_{< t}}\delta_{\boldsymbol{(p'q')}_{< t}}\delta_{\boldsymbol{(pq')}_{> t}}\delta_{\boldsymbol{(p'q)}_{> t}}{\Delta\Psi}_{\boldsymbol{pq}}^{\boldsymbol{p'q'}}}\\
        &= \text{Tr}\left[ {\widetilde{\rho}}_{\overline{< t}}^{2} \right] - \frac{\text{Tr}\left[ {\widetilde{\rho}}_{> t}^{2} \right]}{2^{2m}} 
        = D_{HS}\left( {\widetilde{\rho}}_{t0}, \frac{\mathbb{I}_{t0}}{2^{2m}} \right) \text{Tr}\left[ {\widetilde{\rho}}_{> t}^{2} \right],
    \end{split}
\end{equation}
where ${\widetilde{\rho}}_{\overline{< t}} = {\text{Tr}}_{< t}\left[ V_{\mathcal{L}}\rho_{in}V_{\mathcal{L}}^{\dagger} \right]$ and ${\widetilde{\rho}}_{> t} =\text{Tr}_{t0}\left[ \widetilde{\rho}_{<t}\right]$. Finally, combining Eq. (\ref{eq16}) with Eq. (\ref{eq16-1}) leads to Eq. (\ref{eq16-2}), from which Theorem 1, stated in Eq. (\ref{eq17}), follows directly.

\emph{Trigonometric-moment approach.} For QRNNs with parameter sharing, we evaluate the gradient variance by exploiting the trigonometric-moment form of the loss function. The measurement outcome of the QRNN circuit is given by Eqs. (\ref{eq3}) and (\ref{eq6}), and each recurrent block takes the form
\begin{equation}\label{eq-t1}
    W_t
    = \prod_{i=1}^{\xi}
    \left[
    \cos\left(\frac{\theta_i^{(t)}}{2}\right)\mathbb{I}
    - i \sin\left(\frac{\theta_i^{(t)}}{2}\right)\sigma_i^{(t)}
    \right] Q_i^{(t)},
\end{equation}
where $\sigma_i^{(t)}$ denotes a Pauli string acting on a single qubit $j$, i.e., $\sigma_i^{(t)} = (\sigma_i^{(t)})_{j} \otimes \mathbb{I}_{\overline{j}}$ with $(\sigma_i^{(t)})_{j} \in \{\sigma_x,\sigma_y,\sigma_z\}$, and $Q_i^{(t)}$ denotes a parameter-free quantum gate. In the parameter sharing setting, the parameter set $\{\theta_i^{(t)}\}_{i=1}^\xi$ is shared across all $T$ time steps. In this setting, the loss function admits the general form
\begin{equation}\label{E38}
    C = \sum_k \beta_k \prod_{j=1}^{\xi} \cos^{a_{kj}}(\theta_j) \sin^{b_{kj}}(\theta_j),
\end{equation}
where $a_{kj}, b_{kj} \ge 0$ and $a_{kj}+b_{kj} \le T$, with the upper bound $T$ arising from the $T$-fold repetition of each gate. The coefficients $\beta_k$ depend on the observable $O$, the input state $\rho$, and the circuit components $\sigma_i^{(t)}$ and $Q_i^{(t)}$ in Eq. (\ref{eq-t1}). 

The variance of the partial derivative, $ \mathrm{Var}\left[\partial_{\nu}C_{\mathrm{share}}\right]$, can then be approximated by 
\begin{equation}\label{E38_Var}
    \sum_k \beta_k^2 \,\gamma_\nu(T)\,
           \prod_{j=1}^{\xi}
           \frac{1}{2\pi}\int_0^{2\pi} \cos^{2a_{kj}}(\theta_j) \sin^{2b_{kj}}(\theta_j) \,d\theta_j,
\end{equation}
where cross terms between different $k$ are neglected, as they do not affect the leading-order scaling. The factor $\gamma_\nu(T)$ collects all prefactors arising from differentiation of the trigonometric functions; for the leading contributions, the chain rule yields $\gamma_\nu(T) \propto T^2$.

The trigonometric-moment integrals can be evaluated using Gamma functions, yielding a closed-form expression for the gradient variance, as shown in Eq. (\ref{s12}). In the Supplementary Information (Section E) \cite{SupplementalMaterial}, we analyze the behavior of the ratio of Gamma functions, $f(a,T)=\frac{\Gamma(a+\tfrac12)\Gamma(T-a+\tfrac12)}{\Gamma(T+1)}$ with $a\in(0,T)$, and demonstrate that when $a$ is constant, $f(a,T)$ exhibits polynomial decay in $T$; whereas when $a$ scales proportionally with $T$, $f(a,T)$ decays exponentially. 

Importantly, according to Eq. (\ref{E38}), the trigonometric expansion of the gradient variance for QRNNs with parameter sharing necessarily includes terms with $a\ll T$ (or $T-a\ll T$). Based on this observation, we obtain Theorem 2, as stated in Eq. (\ref{s13}). In addition, the Supplementary Information provides three concrete examples of variational circuits with specific architectures to illustrate the application of this method.   

\section*{Data Availability}
The data that support the findings of this study are available within the paper and its Supplementary Information.

\section*{Code availability}
The code used for the numerical experiments carried out for this work is available
upon request from the authors

\par\vspace{-1em}
\section*{Acknowledgements}
This work is supported by the Natural Science Foundation of Shandong Province of China (ZR2021ZD19, ZR2024LLZ003) and the Fundamental Research Funds for the Central Universities (202461012).

\par\vspace{2em}
\section*{Author contributions}
Z.W. and Y.G. conceived and supervised the project. R.X. and Z.W. developed the theoretical framework. R.X. and Y.L. performed the numerical experiments. Z.S. and S.W. verified the theoretical derivations and numerical results. Z.W., R.X., and Y.G. wrote the manuscript. All authors contributed to the revision of the manuscript and approved the final version.

\par\vspace{-1em}
\section*{Competing interests}
The authors declare no competing interests.


\section*{References}
\begin{thebibliography}{44}

\bibitem{bharti2022noisy}
Bharti, K. et al. Noisy intermediate-scale quantum algorithms. \emph{Rev. Mod. Phys.} \textbf{94}, 015004 (2022).

\bibitem{cerezo2021variational}
Cerezo, M. et al. Variational quantum algorithms. \emph{Nat. Rev. Phys.} \textbf{3}, 625--644 (2021).

\bibitem{mcclean2016theory}
McClean, J. R., Romero, J., Babbush, R. \& Aspuru-Guzik, A. The theory of variational hybrid quantum-classical algorithms. \emph{New J. Phys.} \textbf{18}, 023023 (2016).

\bibitem{tilly2022variational}
Tilly, J. et al. The variational quantum eigensolver: a review of methods and best practices. \emph{Phys. Rep.} \textbf{986}, 1--128 (2022).

\bibitem{blekos2024review}
Blekos, K. et al. A review on quantum approximate optimization algorithm and its variants. \emph{Phys. Rep.} \textbf{1068}, 1--66 (2024).

\bibitem{wan2017quantum}
Wan, K. H., Dahlsten, O., Kristj\'ansson, H., Gardner, R. \& Kim, M. Quantum generalisation of feedforward neural networks. \emph{npj Quantum Inf.} \textbf{3}, 36 (2017).

\bibitem{farhi2018classification}
Farhi, E. \& Neven, H. Classification with quantum neural networks on near term processors. Preprint at arXiv:1802.06002 (2018).

\bibitem{mitarai2018quantum}
Mitarai, K., Negoro, M., Kitagawa, M. \& Fujii, K. Quantum circuit learning. \emph{Phys. Rev. A} \textbf{98}, 032309 (2018).

\bibitem{benedetti2019parameterized}
Benedetti, M., Lloyd, E., Sack, S. \& Fiorentini, M. Parameterized quantum circuits as machine learning models. \emph{Quantum Sci. Technol.} \textbf{4}, 043001 (2019).

\bibitem{schuld2020circuit}
Schuld, M., Bocharov, A., Svore, K. M. \& Wiebe, N. Circuit-centric quantum classifiers. \emph{Phys. Rev. A} \textbf{101}, 032308 (2020).

\bibitem{schuld2019quantum}
Schuld, M. \& Killoran, N. Quantum machine learning in feature Hilbert spaces. \emph{Phys. Rev. Lett.} \textbf{122}, 040504 (2019).

\bibitem{havlivcek2019supervised}
Havl\'{\i}\v{c}ek, V. et al. Supervised learning with quantum-enhanced feature spaces. \emph{Nature} \textbf{567}, 209--212 (2019).

\bibitem{abbas2021power}
Abbas, A. et al. The power of quantum neural networks. \emph{Nat. Comput. Sci.} \textbf{1}, 403--409 (2021).

\bibitem{du2021learnability}
Du, Y., Hsieh, M.-H., Liu, T., You, S. \& Tao, D. Learnability of quantum neural networks. \emph{PRX Quantum} \textbf{2}, 040337 (2021).

\bibitem{du2023problem}
Du, Y., Yang, Y., Tao, D. \& Hsieh, M.-H. Problem-dependent power of quantum neural networks on multiclass classification. \emph{Phys. Rev. Lett.} \textbf{131}, 140601 (2023).

\bibitem{larocca2025barren}
Larocca, M. et al. Barren plateaus in variational quantum computing. \emph{Nat. Rev. Phys.} \textbf{7}, 174--189 (2025).

\bibitem{liu2022presence}
Liu, Z., Yu, L.-W., Duan, L.-M. \& Deng, D.-L. Presence and absence of barren plateaus in tensor-network based machine learning. \emph{Phys. Rev. Lett.} \textbf{129}, 270501 (2022).

\bibitem{mcclean2018barren}
McClean, J. R., Boixo, S., Smelyanskiy, V. N., Babbush, R. \& Neven, H. Barren plateaus in quantum neural network training landscapes. \emph{Nat. Commun.} \textbf{9}, 4812 (2018).

\bibitem{cerezo2021cost}
Cerezo, M., Sone, A., Volkoff, T., Cincio, L. \& Coles, P. J. Cost function dependent barren plateaus in shallow parametrized quantum circuits. \emph{Nat. Commun.} \textbf{12}, 1791 (2021).

\bibitem{holmes2022connecting}
Holmes, Z., Sharma, K., Cerezo, M. \& Coles, P. J. Connecting ansatz expressibility to gradient magnitudes and barren plateaus. \emph{PRX Quantum} \textbf{3}, 010313 (2022).

\bibitem{pesah2021absence}
Pesah, A. et al. Absence of barren plateaus in quantum convolutional neural networks. \emph{Phys. Rev. X} \textbf{11}, 041011 (2021).

\bibitem{sharma2022trainability}
Sharma, K., Cerezo, M., Cincio, L. \& Coles, P. J. Trainability of dissipative perceptron-based quantum neural networks. \emph{Phys. Rev. Lett.} \textbf{128}, 180505 (2022).

\bibitem{wang2021noise}
Wang, S. et al. Noise-induced barren plateaus in variational quantum algorithms. \emph{Nat. Commun.} \textbf{12}, 6961 (2021).

\bibitem{patti2021entanglement}
Patti, T. L., Najafi, K., Gao, X. \& Yelin, S. F. Entanglement devised barren plateau mitigation. \emph{Phys. Rev. Res.} \textbf{3}, 033090 (2021).

\bibitem{letcher2024tight}
Letcher, A., Woerner, S. \& Zoufal, C. Tight and efficient gradient bounds for parameterized quantum circuits. \emph{Quantum} \textbf{8}, 1484 (2024).

\bibitem{lecun2015deep}
LeCun, Y., Bengio, Y. \& Hinton, G. Deep learning. \emph{Nature} \textbf{521}, 436--444 (2015).

\bibitem{goodfellow2016deep}
Goodfellow, I., Bengio, Y. \& Courville, A. \emph{Deep Learning} (MIT Press, Cambridge, MA, 2016).

\bibitem{khan2020survey}
Khan, A., Sohail, A., Zahoora, U. \& Qureshi, A. S. A survey of the recent architectures of deep convolutional neural networks. \emph{Artif. Intell. Rev.} \textbf{53}, 5455--5516 (2020).

\bibitem{he2016deep}
He, K., Zhang, X., Ren, S. \& Sun, J. Deep residual learning for image recognition. In \emph{Proceedings of the IEEE Conference on Computer Vision and Pattern Recognition} 770--778 (2016).

\bibitem{hochreiter1997long}
Hochreiter, S. \& Schmidhuber, J. Long short-term memory. \emph{Neural Comput.} \textbf{9}, 1735--1780 (1997).

\bibitem{lipton2015critical}
Lipton, Z. C., Berkowitz, J. \& Elkan, C. A critical review of recurrent neural networks for sequence learning. Preprint at arXiv:1506.00019 (2015).

\bibitem{vaswani2017attention}
Vaswani, A. et al. Attention is all you need. In \emph{Advances in Neural Information Processing Systems 30} 5998--6008 (2017).

\bibitem{cong2019quantum}
Cong, I., Choi, S. \& Lukin, M. D. Quantum convolutional neural networks. \emph{Nat. Phys.} \textbf{15}, 1273--1278 (2019).

\bibitem{bausch2020recurrent}
Bausch, J. Recurrent quantum neural networks. In \emph{Advances in Neural Information Processing Systems 33} 1368--1379 (2020).

\bibitem{takaki2021learning}
Takaki, Y., Mitarai, K., Negoro, M., Fujii, K. \& Kitagawa, M. Learning temporal data with a variational quantum recurrent neural network. \emph{Phys. Rev. A} \textbf{103}, 052414 (2021).

\bibitem{chen2022quantum}
Chen, S. Y.-C., Yoo, S. \& Fang, Y.-L. L. Quantum long short-term memory. In \emph{ICASSP 2022--2022 IEEE International Conference on Acoustics, Speech and Signal Processing (ICASSP)} 8622--8626 (IEEE, 2022).

\bibitem{di2022dawn}
Di Sipio, R., Huang, J.-H., Chen, S. Y.-C., Mangini, S. \& Worring, M. The dawn of quantum natural language processing. In \emph{ICASSP 2022--2022 IEEE International Conference on Acoustics, Speech and Signal Processing (ICASSP)} 8612--8616 (IEEE, 2022).

\bibitem{li2023quantum}
Li, Y. et al. Quantum recurrent neural networks for sequential learning. \emph{Neural Netw.} \textbf{166}, 148--161 (2023).

\bibitem{li2025quantum}
Li, Y. et al. Quantum gated recurrent neural networks. \emph{IEEE Trans. Pattern Anal. Mach. Intell.} \textbf{47}, 2493--2504 (2025).

\bibitem{decross2023qubit}
DeCross, M., Chertkov, E., Kohagen, M. \& Foss-Feig, M. Qubit-reuse compilation with mid-circuit measurement and reset. \emph{Phys. Rev. X} \textbf{13}, 041057 (2023).

\bibitem{ivashkov2024high}
Ivashkov, P., Uchehara, G., Jiang, L., Wang, D. S. \& Seif, A. High-fidelity, multiqubit generalized measurements with dynamic circuits. \emph{PRX Quantum} \textbf{5}, 030315 (2024).

\bibitem{zhang2025generalized}
Zhang, Z., Chen, S., Liu, Y. \& Jiang, L. Generalized cycle benchmarking algorithm for characterizing midcircuit measurements. \emph{PRX Quantum} \textbf{6}, 010310 (2025).

\bibitem{SupplementalMaterial}
The Supplementary Information of this paper,

\bibitem{beer2020training}
Beer, K. et al. Training deep quantum neural networks. \emph{Nat. Commun.} \textbf{11}, 808 (2020).

\end{thebibliography}
\end{document}


\preprint{APS/123-QED}

\title{Supplementary Information for: ``Defeating Barren Plateaus with Task-Aligned Symmetry"}

\author{Ruipeng Xing}
\affiliation{%
Faculty of Information Science and Engineering, Ocean University of China, Qingdao 266100, China
}%

\author{Yanan Li}
\affiliation{%
Faculty of Information Science and Engineering, Ocean University of China, Qingdao 266100, China
}%

\author{Zhen Shang}
\affiliation{%
Faculty of Information Science and Engineering, Ocean University of China, Qingdao 266100, China
}%

\author{Shengbin Wang}
\affiliation{%
China Telecom Quantum Information Technology Group Co., LTD., Hefei 230088, China
}%

\author{Yongjian Gu}
\affiliation{%
Faculty of Information Science and Engineering, Ocean University of China, Qingdao 266100, China
}%

\author{Zhimin Wang}
\email{wangzhimin@ouc.edu.cn}
\affiliation{%
Faculty of Information Science and Engineering, Ocean University of China, Qingdao 266100, China
}%

\begin{abstract}
In this supplement, we provide additional details that support the main results and theorems presented in the main text. Section A introduces preliminaries on the Haar measure and unitary $t$-designs, as well as key lemmas for integrals over the Haar measure. Section B presents the equivalent model of QRNNs used in our theoretical analysis. Section C demonstrates that, in the absence of parameter sharing, the expectation of the partial derivatives of the QRNN loss function with respect to individual parameters vanishes. Section D analyzes the gradient variance of the loss function for QRNNs without parameter sharing, deriving an upper bound and thereby establishing Theorem 1 in the main text. Section E analyzes the gradient variance for QRNNs with parameter sharing using the trigonometric-function representation of the loss function, establishing Theorem 2 in the main text. Finally, Section F presents two sets of numerical experiments that validate our theoretical predictions. 
\end{abstract}

\maketitle 

\section{Section A. Preliminaries}\label{sec:A}
In this section, we summarize key results on Haar integrals of polynomial functions over the unitary group, where the integration is taken with respect to the Haar measure—a normalized measure invariant under unitary translations. These results will be used extensively in the subsequent theoretical derivations.

\subsection{\label{sec:level2}1. Haar Measure}
The Haar measure $\mu$ assigns to each measurable subset $\mathcal{U}_s \subset \mathcal{U}(d)$ a non-negative real number $\mu(\mathcal{U}_s)$:
\begin{equation}\label{A1}
    \mu(\mathcal{U}_s) = \int_{\mathcal{U}_s} d\mu_H(U),
\end{equation}
where \(U \in \mathcal{U}_s\) denotes the unitary matrix in the subset, and \(d\mu_H(U) \equiv d\mu(U)\) represents the volume element of the matrix in the Haar measure. Unlike other measures defined on \(\mathcal{U}(d)\), the Haar measure is both left- and right-invariant. For any integrable function \(f(U)\) of the unitary matrix \(U\), we have
\begin{equation}\label{A2}
    \int_{\mathcal{U}(d)} d\mu(U)f(U) = \int_{\mathcal{U}(d)} d\mu(U)f(WU) = \int_{\mathcal{U}(d)} d\mu(U)f(UW), \quad \forall U, W \in \mathcal{U}(d),
\end{equation}
which implies \(d\mu(U) = d\mu(WU) = d\mu(UW)\), indicating that the Haar measure treats all unitary matrix uniformly. This suggests that the unitary matrix group is uniformly distributed throughout the space. To give concrete meaning to these integrals, normalization is typically applied:
\begin{equation}\label{A3}
    \mu(\mathcal{U}(d)) = \int_{\mathcal{U}(d)} d\mu(U) = 1.
\end{equation}
This assigns a probabilistic interpretation to the Haar measure: \(d\mu(U)\) can be understood as the probability of randomly selecting a unitary matrix in the neighborhood of \(U\), and \(\mu(\mathcal{U}_s)\) represents the probability that the unitary matrix lies within the subset \(\mathcal{U}_s\). Thus, the integral of a function under the Haar measure is defined as
\begin{equation}\label{A4}
    \langle f \rangle_{\mathcal{U}(d)} = \int_{\mathcal{U}(d)} d\mu(U)f(U),
\end{equation}
which denotes the expectation of the integrable function \(f(U)\) under the condition that \(U\) follows a continuous and uniform distribution. For further details on the Haar measure, see Refs.~\cite{renes2004symmetric,dankert2009exact,harrow2009random}.

\subsection{\label{sec:level2}2. Definition of $t$-design}
In practical scenarios, the distribution of unitary matrices is often represented by a discrete, finite set. Consider a finite set \(\{W_y\}_{y \in Y}\) of unitary matrices \(W_y\) in a \(d\)-dimensional Hilbert space, where the set size is \(|Y|\). Let \(P_{(t,t)}(W)\) be any polynomial function of the matrix \(W\) and its conjugate transpose \(W^\dagger\), where each term contains at most \(t\) products of \(W\) and \(t\) products of \(W^\dagger\). For such a finite set, according to Ref.~\cite{renes2004symmetric}, a \(t\)-design is defined as
\begin{equation}\label{A5}
    \frac{1}{|Y|} \sum_{y \in Y} P_{(t,t)}(W_y) = \int_{\mathcal{U}(d)} d\mu(W)P_{(t,t)}(W).
\end{equation}
This equation indicates that the expectation of the function \(P_{(t,t)}(W)\) over the finite set \(\{W_y\}_{y \in Y}\) that satisfies the \(t\)-design is equivalent to the integral with respect to the Haar measure over the continuous unitary matrix group \(\mathcal{U}(d)\) of the same dimension.

\subsection{\label{sec:level2}3. Integrals over Haar Measure}
Here, we present two commonly used integral formulas for unitary matrices under the Haar measure:
\begin{equation}\label{A6}
    \int_{\mathcal{U}(d)} d \mu(W) w_{\boldsymbol{i}_1, \boldsymbol{j}_1} w_{\boldsymbol{i}_1^\prime , \boldsymbol{j}_1^\prime}^* = \frac{\delta_{\boldsymbol{i}_1, \boldsymbol{i}_1^\prime} \delta_{\boldsymbol{j}_1, \boldsymbol{j}_1^\prime}}{d},
\end{equation}
\begin{equation} \label{A7}
    \begin{split}
    \int_{\mathcal{U}(d)} d \mu(W) w_{\boldsymbol{i}_1, \boldsymbol{j}_1} w_{\boldsymbol{i}_2, \boldsymbol{j}_2} w_{\boldsymbol{i}_1^{\prime}, \boldsymbol{j}_1^{\prime}}^* w_{\boldsymbol{i}_2^{\prime}, \boldsymbol{j}_2^{\prime}}^*
    = & \frac{1}{d^2-1}\left(\delta_{\boldsymbol{i}_1, \boldsymbol{i}_1^{\prime}} \delta_{\boldsymbol{i}_2, \boldsymbol{i}_2^{\prime}} \delta_{\boldsymbol{j}_1, \boldsymbol{j}_1^{\prime}} \delta_{\boldsymbol{j}_2, \boldsymbol{j}_2^{\prime}}+\delta_{\boldsymbol{i}_1, \boldsymbol{i}_2^{\prime}} \delta_{\boldsymbol{i}_2, \boldsymbol{i}_1^{\prime}} \delta_{\boldsymbol{j}_1, \boldsymbol{j}_2^{\prime}} \delta_{\boldsymbol{j}_2, \boldsymbol{j}_1^{\prime}}\right) \\
    & -\frac{1}{d\left(d^2-1\right)}\left(\delta_{\boldsymbol{i}_1, \boldsymbol{i}_1^{\prime}} \delta_{\boldsymbol{i}_2, \boldsymbol{i}_2^{\prime}} \delta_{\boldsymbol{j}_1, \boldsymbol{j}_2^{\prime}} \delta_{\boldsymbol{j}_2, \boldsymbol{j}_1^{\prime}}+\delta_{\boldsymbol{i}_1, \boldsymbol{i}_2^{\prime}} \delta_{\boldsymbol{i}_2, \boldsymbol{i}_1^{\prime}} \delta_{\boldsymbol{j}_1, \boldsymbol{j}_1^{\prime}} \delta_{\boldsymbol{j}_2, \boldsymbol{j}_2^{\prime}}\right),
    \end{split}
\end{equation}
where \(w_{\boldsymbol{i}\boldsymbol{j}}\) denotes the unitary matrix of \(W\), and the integration domain for the unitary matrix \(W\) is \(\mathcal{U}(d)\). Unless otherwise specified, \(W\) is assumed to be an operator acting on the \(d\)-dimensional Hilbert space \(\mathcal{H}_w\). When the dimension \(d = 2^m\), this corresponds to the case of \(m\) qubits. The bold symbol \(\boldsymbol{i} = (i_1, \ldots, i_m)\) represents bitstrings (computational basis) of length \(m\), where \(i_1, \ldots, i_m \in \{0,1\}\) denote the state of each qubit. The notation \(\boldsymbol{i}\) (\(\boldsymbol{j}\)) refers to the row (column) of the matrix \(W\) that corresponds to the quantum state \(|i_1, \ldots, i_m\rangle\) (\(|j_1, \ldots, j_m\rangle\)). If \(W\) acts on two subsystems \(\mathcal{H}_1\) and \(\mathcal{H}_2\), each with \(n\) qubits, we define the concatenation of two bitstrings \(\boldsymbol{i}\) and \(\boldsymbol{k}\) as \(\boldsymbol{i}\cdot\boldsymbol{k} = (i_1, \ldots, i_n, k_1, \ldots, k_n)\), which serves as the computational basis for the composite system \(\mathcal{H} = \mathcal{H}_1 \otimes \mathcal{H}_2\) (corresponding to the rows or columns of \(W\)).

Based on these expressions, we can derive the following lemmas. \textbf{Lemmas 1–4} are similar to those presented in Ref.~\cite{cerezo2021cost}, while \textbf{Lemmas 5–6} are newly introduced in this work.

\noindent{\textbf{Lemma 1.}} Let \(\{W_y\}_{y \in Y} \subset \mathcal{U}(d)\) form a \(t\)-design with \(t \geq 1\), and let \(A, B: \mathcal{H}_w \rightarrow \mathcal{H}_w\) be arbitrary linear operators. Then we have
\begin{equation}\label{A8}
    \frac{1}{|Y|} \sum_{y \in Y} \text{Tr}[W_y A W_y^\dagger B] = \int_{\mathcal{U}(d)} d\mu(W) \text{Tr}[W A W^\dagger B] = \frac{\text{Tr}[A] \text{Tr}[B]}{d}.
\end{equation}

\noindent{\textit{Proof.}} The first equality follows from the fact that \( \text{Tr}[W_y A W_y^\dagger B] \in P_{(1,1)}(W_y)\). For the second equality, note that \(\text{Tr}[W A W^\dagger B]\) can be expanded as
\begin{equation}\label{A9}
    \operatorname{Tr}\left[W A W^{\dagger} B\right] = \sum_{\boldsymbol{i}_1, \boldsymbol{j}_1, \boldsymbol{i}_1^{\prime}, \boldsymbol{j}_1^{\prime}} w_{\boldsymbol{i}_1, \boldsymbol{j}_1} a_{\boldsymbol{j}_1, \boldsymbol{j}_1^{\prime}} w_{\boldsymbol{i}_1^{\prime}, \boldsymbol{j}_1^{\prime}}^* b_{\boldsymbol{i}_1^{\prime}, \boldsymbol{i}_1}.
\end{equation}
where \(w_{\boldsymbol{i}_1^{\prime}, \boldsymbol{j}_1^{\prime}}^*\) is the matrix element of \(W^\dagger\) at row \(\boldsymbol{j}_1^{\prime}\) and column \(\boldsymbol{i}_1^{\prime}\), satisfying \(w_{\boldsymbol{i}_1^{\prime}, \boldsymbol{j}_1^{\prime}}^* = (W^\dagger)_{\boldsymbol{j}_1^{\prime}, \boldsymbol{i}_1^{\prime}}\). Applying Eq. (\ref{A6}) to degenerate several indices, we have
\begin{equation}\label{A10}
    \int_{\mathcal{U}(d)} d\mu(W) \text{Tr}[W A W^\dagger B] = \frac{1}{d} \sum_{i_1,j_1} a_{j_1,j_1} b_{i_1,i_1} = \frac{\text{Tr}[A] \text{Tr}[B]}{d}.
\end{equation}
\hfill $\square$

\noindent{\textbf{Lemma 2.}} Let \(\{W_y\}_{y \in Y} \subset \mathcal{U}(d)\) form a \(t\)-design with \(t \geq 2\), and let \(A, B, C, D: \mathcal{H}_w \rightarrow \mathcal{H}_w\) be arbitrary linear operators. Then we have
\begin{equation}\label{A11}
    \begin{split}
        \frac{1}{|Y|} \sum_{y \in Y} \text{Tr}[W_y A W_y^\dagger B W_y C W_y^\dagger D] 
        = & \int_{\mathcal{U}(d)} d\mu(W) \text{Tr}[W A W^\dagger B W C W^\dagger D] \\
        = & \frac{1}{d^2 - 1} \left( \text{Tr}[A] \text{Tr}[C] \text{Tr}[B D] + \text{Tr}[A C] \text{Tr}[B] \text{Tr}[D] \right) \\
        &- \frac{1}{d(d^2 - 1)} \left( \text{Tr}[A C] \text{Tr}[B D] + \text{Tr}[A] \text{Tr}[B] \text{Tr}[C] \text{Tr}[D] \right).
    \end{split}
\end{equation}
\noindent{\textit{Proof.}} The first equality follows from the definition of \(2\)-design, and \(\text{Tr}[W A W^\dagger B W C W^\dagger D] \in P_{(2,2)}(W)\) can be written as
\begin{equation}\label{A12}
    \operatorname{Tr}\left[W A W^{\dagger} B W C W^{\dagger} D\right] = \sum_{\substack{\boldsymbol{i}_1, \boldsymbol{j}_1, \boldsymbol{i}_1^{\prime}, \boldsymbol{j}_1^{\prime} \\ \boldsymbol{i}_2, \boldsymbol{j}_2^{\prime}, \boldsymbol{i}_2^{\prime}, \boldsymbol{j}_2^{\prime}}} w_{\boldsymbol{i}_1, \boldsymbol{j}_1} a_{\boldsymbol{j}_1, \boldsymbol{j}_1^{\prime}} w_{\boldsymbol{i}_1^{\prime}, \boldsymbol{j}_1^{\prime}}^* b_{\boldsymbol{i}_1^{\prime}, \boldsymbol{i}_2} w_{\boldsymbol{i}_2, \boldsymbol{j}_2} c_{\boldsymbol{j}_2, \boldsymbol{j}_2^{\prime}} w_{\boldsymbol{i}_2^{\prime}, \boldsymbol{j}_2^{\prime}}^* d_{\boldsymbol{i}_2^{\prime}, \boldsymbol{i}_1}.
\end{equation}
Applying Eq. (\ref{A7}) to degenerate half of the indices, we obtain 
\begin{equation}\label{A13}
\begin{aligned}
\int d \mu(W) \operatorname{Tr}\left[W A W^{\dagger} B W C W^{\dagger} D\right] = & \frac{1}{d^2-1} \sum_{\boldsymbol{i}_1, \boldsymbol{j}_1, \boldsymbol{i}_2, \boldsymbol{j}_2}\left(a_{\boldsymbol{j}_1, \boldsymbol{j}_1} b_{\boldsymbol{i}_1, \boldsymbol{i}_2} c_{\boldsymbol{j}_2, \boldsymbol{j}_2} d_{\boldsymbol{i}_2, \boldsymbol{i}_1} + a_{\boldsymbol{j}_1, \boldsymbol{j}_2} b_{\boldsymbol{i}_2, \boldsymbol{i}_2} c_{\boldsymbol{j}_2, \boldsymbol{j}_1} d_{\boldsymbol{i}_1, \boldsymbol{i}_1}\right) \\
& -\frac{1}{d\left(d^2-1\right)} \sum_{\boldsymbol{i}_1, \boldsymbol{j}_1, \boldsymbol{i}_2, \boldsymbol{j}_2}\left(a_{\boldsymbol{j}_1, \boldsymbol{j}_2} b_{\boldsymbol{i}_1, \boldsymbol{i}_2} c_{\boldsymbol{j}_2, \boldsymbol{j}_1} d_{\boldsymbol{i}_2, \boldsymbol{i}_1} + a_{\boldsymbol{j}_1, \boldsymbol{j}_1} b_{\boldsymbol{i}_2, \boldsymbol{i}_2} c_{\boldsymbol{j}_2, \boldsymbol{j}_2} d_{\boldsymbol{i}_1, \boldsymbol{i}_1}\right) \\
= & \frac{1}{d^2-1}(\operatorname{Tr}[A] \operatorname{Tr}[C] \operatorname{Tr}[B D] + \operatorname{Tr}[A C] \operatorname{Tr}[B] \operatorname{Tr}[D]) \\
& -\frac{1}{d\left(d^2-1\right)}(\operatorname{Tr}[A C] \operatorname{Tr}[B D] + \operatorname{Tr}[A] \operatorname{Tr}[B] \operatorname{Tr}[C] \operatorname{Tr}[D]) .
\end{aligned}
\end{equation}
\hfill $\square$

\noindent{\textbf{Lemma 3.}} Let \(\{W_y\}_{y \in Y} \subset \mathcal{U}(d)\) form a $t$-design with \(t \geq 2\), and let \(A, B, C, D: \mathcal{H}_w \rightarrow \mathcal{H}_w\) be arbitrary linear operators. Then we have
\begin{equation}\label{A14}
    \begin{split}
        \frac{1}{|Y|} \sum_{y \in Y} \text{Tr}[W_y A W_y^\dagger B] \text{Tr}[W_y C W_y^\dagger D] 
        = & \int_{\mathcal{U}(d)} d\mu(W) \text{Tr}[W A W^\dagger B] \text{Tr}[W C W^\dagger D] \\
        = &\frac{1}{d^2-1}(\operatorname{Tr}[A] \operatorname{Tr}[B] \operatorname{Tr}[C] \operatorname{Tr}[D]+\operatorname{Tr}[A C] \operatorname{Tr}[B D]) \\
        &-\frac{1}{d\left(d^2-1\right)}(\operatorname{Tr}[A C] \operatorname{Tr}[B] \operatorname{Tr}[D]+\operatorname{Tr}[A] \operatorname{Tr}[C] \operatorname{Tr}[B D]). 
    \end{split}
\end{equation}
\noindent{\textit{Proof.}} The first equality follows from the definition of $t$-design, and \(\text{Tr}[W A W^\dagger B] \text{Tr}[W C W^\dagger D] \in P_{(2,2)}(W)\) can be expanded as
\begin{equation}\label{A15}
    \text{Tr}[W A W^\dagger B] \text{Tr}[W C W^\dagger D] =\sum_{\substack{\boldsymbol{i}_1, \boldsymbol{j}_1, \boldsymbol{i}_1^{\prime}, \boldsymbol{j}_1^{\prime} \\ \boldsymbol{i}_2, \boldsymbol{j}_2^{\prime}, \boldsymbol{i}_2^{\prime}, \boldsymbol{j}_2^{\prime}}} w_{\boldsymbol{i}_1, \boldsymbol{j}_1} a_{\boldsymbol{j}_1, \boldsymbol{j}_1^{\prime}} w_{\boldsymbol{i}_1^{\prime}, \boldsymbol{j}_1^{\prime}}^* b_{\boldsymbol{i}_1^{\prime}, \boldsymbol{i}_1} w_{\boldsymbol{i}_2, \boldsymbol{j}_2} c_{\boldsymbol{j}_2, \boldsymbol{j}_2^{\prime}} w_{\boldsymbol{i}_2^{\prime}, \boldsymbol{j}_2^{\prime}}^* d_{\boldsymbol{i}_2^{\prime}, \boldsymbol{i}_2}.
\end{equation}
Applying Eq. (\ref{A7}) to degenerate half of the indices, we obtain
\begin{equation}\label{A16}
    \begin{split}
        \int d \mu(W) \operatorname{Tr}\left[W A W^{\dagger} B\right] \operatorname{Tr}\left[W C W^{\dagger} D\right]
        =&\frac{1}{d^2-1} \sum_{\boldsymbol{i}_1, \boldsymbol{j}_1, \boldsymbol{i}_2, \boldsymbol{j}_2}\left(a_{\boldsymbol{j}_1, \boldsymbol{j}_1} b_{\boldsymbol{i}_1, \boldsymbol{i}_1} c_{\boldsymbol{j}_2, \boldsymbol{j}_2} d_{\boldsymbol{i}_2, \boldsymbol{i}_2}+a_{\boldsymbol{j}_1, \boldsymbol{j}_2} b_{\boldsymbol{i}_2, \boldsymbol{i}_1} c_{\boldsymbol{j}_2, \boldsymbol{j}_1} d_{\boldsymbol{i}_1, \boldsymbol{i}_2}\right) \\
        &-\frac{1}{d\left(d^2-1\right)} \sum_{i_1, \boldsymbol{j}_1, \boldsymbol{i}_2, \boldsymbol{j}_2}\left(a_{\boldsymbol{j}_1, \boldsymbol{j}_2} b_{\boldsymbol{i}_1, \boldsymbol{i}_1} c_{\boldsymbol{j}_2, \boldsymbol{j}_1} d_{\boldsymbol{i}_2, \boldsymbol{i}_2}+a_{\boldsymbol{j}_1, \boldsymbol{j}_1} b_{\boldsymbol{i}_2, \boldsymbol{i}_1} c_{\boldsymbol{j}_2, \boldsymbol{j}_2} d_{\boldsymbol{i}_1, \boldsymbol{i}_2}\right) \\
        =&\frac{1}{d^2-1}(\operatorname{Tr}[A] \operatorname{Tr}[B] \operatorname{Tr}[C] \operatorname{Tr}[D]+\operatorname{Tr}[A C] \operatorname{Tr}[B D]) \\
        &-\frac{1}{d\left(d^2-1\right)}(\operatorname{Tr}[A C] \operatorname{Tr}[B] \operatorname{Tr}[D]+\operatorname{Tr}[A] \operatorname{Tr}[C] \operatorname{Tr}[B D]).
    \end{split}
\end{equation}
\hfill $\square$  

\noindent{\textbf{Lemma 4.}} Let \(\mathcal{H} = \mathcal{H}_{w} \otimes \mathcal{H}_{\overline{w}}\) be a bipartite Hilbert space of dimension \(d = d_{w} d_{\overline{w}}\), and let \(\{ W_{y} \}_{y \in Y} \subset \mathcal{U}(d_{w})\) form a unitary \(t\)-design with \(t \geq 1\). For arbitrary linear operators \(A, B: \mathcal{H} \rightarrow \mathcal{H}\), we have
\begin{equation}\label{A16}
    \int d\mu(W) (\mathbb{I}_{\overline{w}} \otimes W) A (\mathbb{I}_{\overline{w}} \otimes W^{\dagger}) B = \frac{\text{Tr}_{w}[A] \otimes \mathbb{I}_{w}}{d_{w}} B,
\end{equation} 
and
\begin{equation}\label{A17}
   \int d\mu(W) \text{Tr}[(\mathbb{I}_{\overline{w}} \otimes W) A (\mathbb{I}_{\overline{w}} \otimes W^{\dagger}) B] = \frac{1}{d_{w}} \text{Tr}[\text{Tr}_{w}[A] \text{Tr}_{w}[B]],
\end{equation}
where the symbol \(\mathbb{I}_{w}\) denotes the identity matrix in the subspace \(\mathcal{H}_{w}\), and \(\text{Tr}_{w}\) represents the partial trace over the subspace \(\mathcal{H}_{w}\). Additionally, we omit the conversion of the function expectation under the set \(\{ W_{y} \}_{y \in Y}\) to the Haar integral, i.e., the definition of the \(t\)-design. This conversion will also be omitted in subsequent lemmas, and we will directly present the results of the Haar integral.

\noindent{\textit{Proof.}} First, we expand \(W\) and \(W^{\dagger}\) using their matrix spectral decomposition
\begin{equation}
(\mathbb{I}_{\overline{w}} \otimes W) A (\mathbb{I}_{\overline{w}} \otimes W^{\dagger}) B = \sum_{\boldsymbol{i}_{1}, \boldsymbol{j}_{1}, \boldsymbol{i}_{1}^{\prime}, \boldsymbol{j}_{1}^{\prime}} (\mathbb{I}_{\overline{w}} \otimes w_{\boldsymbol{i}_{1}, \boldsymbol{j}_{1}} | \boldsymbol{i}_{1} \rangle \langle \boldsymbol{j}_{1} |) A (\mathbb{I}_{\overline{w}} \otimes w_{\boldsymbol{i}_{1}^{\prime}, \boldsymbol{j}_{1}^{\prime}}^{*} | \boldsymbol{j}_{1}^{\prime} \rangle \langle \boldsymbol{i}_{1}^{\prime} |) B.
\end{equation}  
Next, we use Eq. (\ref{A6}) to simplify several indices and obtain
\begin{equation}
    \begin{split}
        \int d\mu(W) (\mathbb{I}_{\overline{w}} \otimes W) A (\mathbb{I}_{\overline{w}} \otimes W^{\dagger}) B 
        &= \sum_{\boldsymbol{i}_{1}, \boldsymbol{j}_{1}, \boldsymbol{i}_{1}^{\prime}, \boldsymbol{j}_{1}^{\prime}} \int d\mu(W) w_{\boldsymbol{i}_{1}, \boldsymbol{j}_{1}} w_{\boldsymbol{i}_{1}^{\prime}, \boldsymbol{j}_{1}^{\prime}}^{*} (\mathbb{I}_{\overline{w}} \otimes | \boldsymbol{i}_{1} \rangle \langle \boldsymbol{j}_{1} |) A (\mathbb{I}_{\overline{w}} \otimes | \boldsymbol{j}_{1}^{\prime} \rangle \langle \boldsymbol{i}_{1}^{\prime} |) B\\
        &= \frac{1}{d_{w}} \sum_{\boldsymbol{i}_{1}, \boldsymbol{j}_{1}} (\mathbb{I}_{\overline{w}} \otimes | \boldsymbol{i}_{1} \rangle \langle \boldsymbol{j}_{1} |) A (\mathbb{I}_{\overline{w}} \otimes | \boldsymbol{j}_{1} \rangle \langle \boldsymbol{i}_{1} |) B\\
        &= \frac{1}{d_{w}} \sum_{\boldsymbol{i}_{1}, \boldsymbol{j}_{1}} \left[(\mathbb{I}_{\overline{w}} \otimes \langle \boldsymbol{j}_{1} |) A (\mathbb{I}_{\overline{w}} \otimes | \boldsymbol{j}_{1} \rangle)\right] \otimes | \boldsymbol{i}_{1} \rangle \langle \boldsymbol{i}_{1} | B\\
        &= \frac{1}{d_{w}} \text{Tr}_{w}[A] \otimes \mathbb{I}_{w} B.
    \end{split}
\end{equation}  
For any unitary matrix \(C\), we have \(\text{Tr}[C] = \text{Tr}[\text{Tr}_{w}[C]]\). Using the definition of the partial trace, \(\text{Tr}_{w}[C] = \sum_{\boldsymbol{i}} (\mathbb{I}_{\overline{w}} \otimes \langle \boldsymbol{i} |) C (\mathbb{I}_{\overline{w}} \otimes | \boldsymbol{i} \rangle)\), we obtain
\begin{equation}
    \begin{split}
        \int d\mu(W) \text{Tr}[(\mathbb{I}_{\overline{w}} \otimes W) A (\mathbb{I}_{\overline{w}} \otimes W^{\dagger}) B] 
        &= \text{Tr}\left[\frac{1}{d_{w}} \text{Tr}_{w}[A] \otimes \mathbb{I}_{w} B\right]\\
        &= \frac{1}{d_{w}} \text{Tr}[\text{Tr}_{w}[\text{Tr}_{w}[A] \otimes \mathbb{I}_{w} B]]\\
        &= \frac{1}{d_{w}} \text{Tr}\left[\sum_{\boldsymbol{i}} (\mathbb{I}_{\overline{w}} \otimes \langle \boldsymbol{i} |) [\text{Tr}_{w}[A] \otimes \mathbb{I}_{w} B] (\mathbb{I}_{\overline{w}} \otimes | \boldsymbol{i} \rangle)\right]\\
        &= \frac{1}{d_{w}} \text{Tr}\left[\text{Tr}_{w}[A] \sum_{\boldsymbol{i}} (\mathbb{I}_{\overline{w}} \otimes \langle \boldsymbol{i} |) B (\mathbb{I}_{\overline{w}} \otimes | \boldsymbol{i} \rangle)\right]\\
        &= \frac{1}{d_{w}} \text{Tr}[\text{Tr}_{w}[A] \text{Tr}_{w}[B]].
    \end{split}
\end{equation} 
\hfill $\square$

\noindent{\textbf{Lemma 5.}} Let the composite Hilbert space $\mathcal{H} = \mathcal{H}_{1} \otimes \mathcal{H}_{2}$ have dimension $d \times d = d^{2}$ (both subspaces have dimension $d$), and let $\{ W_{y} \}_{y \in Y} \subset \mathcal{U}(d^{2})$ form a unitary $t$-design with $t \geq 2$, acting on the composite space $\mathcal{H}$. For arbitrary linear operators $A, C: \mathcal{H}_{2} \rightarrow \mathcal{H}_{2}$ and $B, D: \mathcal{H}_{1} \rightarrow \mathcal{H}_{1}$, define the operator  
\begin{equation}
\Omega = \text{Tr}_{2}\left[\begin{matrix}  \mathbb{I} \\  \otimes \\  A \end{matrix} W^{\dagger} \begin{matrix}  B \\  \otimes \\  \mathbb{I}  \end{matrix} W \right] \text{Tr}_{2}\left[ \begin{matrix}  \mathbb{I} \\  \otimes \\  C  \end{matrix} W^{\dagger} \begin{matrix}  D \\ \otimes \\  \mathbb{I}  \end{matrix} W \right].
\end{equation}
Here we introduce a \textit{tensor-product tableau notation} where $\mathbb{I}_{1} \otimes A$ is substituted by $\begin{matrix}  \mathbb{I} \\ \otimes \\ A \end{matrix}$, representing the operator $A$ acting on subspace $\mathcal{H}_{2}$ while the identity $\mathbb{I}$ acts on $\mathcal{H}_{1}$. This stacked arrangement visually distinguishes operations that occur in independent subsystems. For the operator $\Omega$, the following equality holds
\begin{equation}
    \begin{split}
        \int d\mu(W) \text{Tr}[\Omega] 
        =& \frac{1}{d^{4} - 1} \left( \text{Tr}[A] \text{Tr}[B] \text{Tr}[C] \text{Tr}[D] \cdot d^{3} + \text{Tr}[AC] \text{Tr}[BD] \cdot d^{3} \right)\\
        &- \frac{1}{d^{2}(d^{4} - 1)} \left( \text{Tr}[AC] \text{Tr}[B] \text{Tr}[D] \cdot d^{4} + \text{Tr}[A] \text{Tr}[C] \text{Tr}[BD] \cdot d^{2} \right).
    \end{split}
\end{equation} 

\noindent{\textit{Proof.}} We begin by expanding the spectral decompositions of $W$ and $W^{\dagger}$ in the composite space
\begin{equation}\label{A24}
    \begin{split}
        W = \sum_{\boldsymbol{i}, \boldsymbol{j}} w_{\boldsymbol{i}, \boldsymbol{j}} | \boldsymbol{i} \rangle \langle \boldsymbol{j} | 
        &= \sum_{\boldsymbol{i} \cdot \boldsymbol{k}, \boldsymbol{j} \cdot \boldsymbol{l}} w_{\boldsymbol{i} \cdot \boldsymbol{k}, \boldsymbol{j} \cdot \boldsymbol{l}} \begin{matrix}  \left | \boldsymbol{i}  \right \rangle  \left \langle \boldsymbol{j} \right| \\  \left | \boldsymbol{k}  \right \rangle  \left \langle \boldsymbol{l} \right| \end{matrix},
        \\
        W^{\dagger} = \sum_{\boldsymbol{i}^{\prime}, \boldsymbol{j}^{\prime}} w_{\boldsymbol{i}^{\prime}, \boldsymbol{j}^{\prime}}^{*} | \boldsymbol{j}^{\prime} \rangle \langle \boldsymbol{i}^{\prime} | 
        &= \sum_{\boldsymbol{i}^{\prime} \cdot \boldsymbol{k}^{\prime}, \boldsymbol{j}^{\prime} \cdot \boldsymbol{l}^{\prime}} w_{\boldsymbol{i}^{\prime} \cdot \boldsymbol{k}^{\prime}, \boldsymbol{j}^{\prime} \cdot \boldsymbol{l}^{\prime}}^{*} 
        \begin{matrix} | \boldsymbol{j}_{1}^{\prime} \rangle \langle \boldsymbol{i}_{1}^{\prime} | \\ | \boldsymbol{l}_{1}^{\prime} \rangle \langle \boldsymbol{k}_{1}^{\prime} |\end{matrix} .
    \end{split}
\end{equation}  
Substituting these expansions into the expression for $\Omega$ and evaluating the Haar integral yields
\begin{equation}\label{A25}
    \begin{split}
        \int d\mu(W) \text{Tr}[\Omega] 
        &= \int d\mu(W) \text{Tr}\left\{ \text{Tr}_{2}\left[ \begin{matrix}  \mathbb{I} \\  \otimes \\  A  \end{matrix} W^{\dagger} \begin{matrix}  B \\  \otimes \\  \mathbb{I}  \end{matrix} W \right] \text{Tr}_{2}\left[ \begin{matrix}  \mathbb{I} \\  \otimes \\  C  \end{matrix} W^{\dagger} \begin{matrix} D \\  \otimes \\  \mathbb{I}  \end{matrix} W \right] \right\}
        \\
        &= \int d\mu(W) \sum_{\mathbf{I}\cdot \mathbf{J},\mathbf{K}\cdot \mathbf{L}} w_{\mathbf{I}\cdot \mathbf{J},\mathbf{K}\cdot \mathbf{L}}
        \text{Tr}\left\{ \text{Tr}_{2}\left[ \begin{matrix}  
        \mathbb{I} \\  \otimes \\  A  \end{matrix} \begin{matrix}  
        | \boldsymbol{j}_{1}^{\prime} \rangle \langle \boldsymbol{i}_{1}^{\prime} | \\  \otimes \\  | \boldsymbol{l}_{1}^{\prime} \rangle \langle \boldsymbol{k}_{1}^{\prime} |  \end{matrix} \begin{matrix}  B \\  \otimes \\  \mathbb{I}  \end{matrix} \begin{matrix}  | \boldsymbol{i}_{1} \rangle \langle \boldsymbol{j}_{1} | \\  \otimes \\  | \boldsymbol{k}_{1} \rangle \langle \boldsymbol{l}_{1} |  \end{matrix} \right] \text{Tr}_{2}\left[ \begin{matrix}  \mathbb{I} \\  \otimes \\  C  \end{matrix} \begin{matrix}  | \boldsymbol{j}_{2}^{\prime} \rangle \langle \boldsymbol{i}_{2}^{\prime} | \\  \otimes \\  | \boldsymbol{l}_{2}^{\prime} \rangle \langle \boldsymbol{k}_{2}^{\prime} |  \end{matrix} \begin{matrix}  D \\  \otimes \\  \mathbb{I}  \end{matrix} \begin{matrix}  | \boldsymbol{i}_{2} \rangle \langle \boldsymbol{j}_{2} | \\  \otimes \\  | \boldsymbol{k}_{2} \rangle \langle \boldsymbol{l}_{2} |  \end{matrix} \right] \right\}
        \\
        &= \sum_{\mathbf{I}\cdot \mathbf{J},\mathbf{K}\cdot \mathbf{L}} \int d\mu(W) w_{\mathbf{I}\cdot \mathbf{J},\mathbf{K}\cdot \mathbf{L}}
        \langle \boldsymbol{l}_{1} | A | \boldsymbol{l}_{1}^{\prime} \rangle \langle \boldsymbol{i}_{1}^{\prime} | B | \boldsymbol{i}_{1} \rangle \langle \boldsymbol{l}_{2} | C | \boldsymbol{l}_{2}^{\prime} \rangle \langle \boldsymbol{i}_{2}^{\prime} | D | \boldsymbol{i}_{2} \rangle \delta_{\boldsymbol{j}_{1}, \boldsymbol{j}_{2}^{\prime}} \delta_{\boldsymbol{j}_{2}, \boldsymbol{j}_{1}^{\prime}} \delta_{\boldsymbol{k}_{1}^{\prime}, \boldsymbol{k}_{1}} \delta_{\boldsymbol{k}_{2}^{\prime}, \boldsymbol{k}_{2}},
    \end{split}
\end{equation} 
where we have introduced the composite index notation $\mathbf{I}\cdot \mathbf{J},\mathbf{K}\cdot \mathbf{L}\equiv \{\boldsymbol{i}_{1} \cdot \boldsymbol{k}_{1}, \boldsymbol{j}_{1} \cdot \boldsymbol{l}_{1} ; \boldsymbol{i}_{2} \cdot \boldsymbol{k}_{2}, \boldsymbol{j}_{2} \cdot \boldsymbol{l}_{2} ; \boldsymbol{i}_{1}^{\prime} \cdot \boldsymbol{k}_{1}^{\prime}, \boldsymbol{j}_{1}^{\prime} \cdot \boldsymbol{l}_{1}^{\prime};  \boldsymbol{i}_{2}^{\prime} \cdot \boldsymbol{k}_{2}^{\prime}, \boldsymbol{j}_{2}^{\prime} \cdot \boldsymbol{l}_{2}^{\prime}   \}$ and $w_{\mathbf{I}\cdot \mathbf{J},\mathbf{K}\cdot \mathbf{L}}\equiv w_{\boldsymbol{i}_{1} \cdot \boldsymbol{k}_{1}, \boldsymbol{j}_{1} \cdot \boldsymbol{l}_{1}} w_{\boldsymbol{i}_{2} \cdot \boldsymbol{k}_{2}, \boldsymbol{j}_{2} \cdot \boldsymbol{l}_{2}} w_{\boldsymbol{i}_{1}^{\prime} \cdot \boldsymbol{k}_{1}^{\prime}, \boldsymbol{j}_{1}^{\prime} \cdot \boldsymbol{l}_{1}^{\prime}}^{*} w_{\boldsymbol{i}_{2}^{\prime} \cdot \boldsymbol{k}_{2}^{\prime}, \boldsymbol{j}_{2}^{\prime} \cdot \boldsymbol{l}_{2}^{\prime}}^{*}$. The four Kronecker deltas arise from the identity operators in $\Omega$, which eliminate four indices. Applying the second-moment Haar integral formula Eq. (\ref{A7}) introduces another four delta functions for each term
\begin{equation}\label{26} 
    \begin{aligned}
        \int d\mu(W) \text{Tr}[\Omega]
        =& \sum_{\mathbf{I}\cdot \mathbf{J},\mathbf{K}\cdot \mathbf{L}} \langle \boldsymbol{l}_{1} | A | \boldsymbol{l}_{1}^{\prime} \rangle \langle \boldsymbol{i}_{1}^{\prime} | B | \boldsymbol{i}_{1} \rangle \langle \boldsymbol{l}_{2} | C | \boldsymbol{l}_{2}^{\prime} \rangle \langle \boldsymbol{i}_{2}^{\prime} | D | \boldsymbol{i}_{2} \rangle \delta_{\boldsymbol{j}_{1}, \boldsymbol{j}_{2}^{\prime}} \delta_{\boldsymbol{j}_{2}, \boldsymbol{j}_{1}^{\prime}} \delta_{\boldsymbol{k}_{1}^{\prime}, \boldsymbol{k}_{1}} \delta_{\boldsymbol{k}_{2}^{\prime}, \boldsymbol{k}_{2}}
        \\
        &\times  \left[ \frac{1}{(d^{2})^{2} - 1} \left( \delta_{\substack{  
        \boldsymbol{i}_{1}, \boldsymbol{i}_{1}^{\prime} \\  
        \boldsymbol{k}_{1}, \boldsymbol{k}_{1}^{\prime}  
        }} \delta_{\substack{  
        \boldsymbol{i}_{2}, \boldsymbol{i}_{2}^{\prime} \\  
        \boldsymbol{k}_{2}, \boldsymbol{k}_{2}^{\prime}  
        }} \delta_{\substack{  
        \boldsymbol{j}_{1}, \boldsymbol{j}_{1}^{\prime} \\  
        \boldsymbol{l}_{1}, \boldsymbol{l}_{1}^{\prime}  
        }} \delta_{\substack{  
        \boldsymbol{j}_{2}, \boldsymbol{j}_{2}^{\prime} \\  
        \boldsymbol{l}_{2}, \boldsymbol{l}_{2}^{\prime}  
        }} + \delta_{\substack{  
        \boldsymbol{i}_{1}, \boldsymbol{i}_{2}^{\prime} \\  
        \boldsymbol{k}_{1}, \boldsymbol{k}_{2}^{\prime}  
        }} \delta_{\substack{  
        \boldsymbol{i}_{2}, \boldsymbol{i}_{1}^{\prime} \\  
        \boldsymbol{k}_{2}, \boldsymbol{k}_{1}^{\prime}  
        }} \delta_{\substack{  
        \boldsymbol{j}_{1}, \boldsymbol{j}_{2}^{\prime} \\  
        \boldsymbol{l}_{1}, \boldsymbol{l}_{2}^{\prime}  
        }} \delta_{\substack{  
        \boldsymbol{j}_{2}, \boldsymbol{j}_{1}^{\prime} \\  
        \boldsymbol{l}_{2}, \boldsymbol{l}_{1}^{\prime}  
        }} \right) \right.
        \\
        &\left.- \frac{1}{d^{2}[(d^{2})^{2} - 1]} \left( \delta_{\substack{  
        \boldsymbol{i}_{1}, \boldsymbol{i}_{1}^{\prime} \\  
        \boldsymbol{k}_{1}, \boldsymbol{k}_{1}^{\prime}  
        }} \delta_{\substack{  
        \boldsymbol{i}_{2}, \boldsymbol{i}_{2}^{\prime} \\  
        \boldsymbol{k}_{2}, \boldsymbol{k}_{2}^{\prime}  
        }} \delta_{\substack{  
        \boldsymbol{j}_{1}, \boldsymbol{j}_{2}^{\prime} \\  
        \boldsymbol{l}_{1}, \boldsymbol{l}_{2}^{\prime}  
        }} \delta_{\substack{  
        \boldsymbol{j}_{2}, \boldsymbol{j}_{1}^{\prime} \\  
        \boldsymbol{l}_{2}, \boldsymbol{l}_{1}^{\prime}  
        }} + \delta_{\substack{  
        \boldsymbol{i}_{1}, \boldsymbol{i}_{2}^{\prime} \\  
        \boldsymbol{k}_{1}, \boldsymbol{k}_{2}^{\prime}  
        }} \delta_{\substack{  
        \boldsymbol{i}_{2}, \boldsymbol{i}_{1}^{\prime} \\  
        \boldsymbol{k}_{2}, \boldsymbol{k}_{1}^{\prime}  
        }} \delta_{\substack{  
        \boldsymbol{j}_{1}, \boldsymbol{j}_{1}^{\prime} \\  
        \boldsymbol{l}_{1}, \boldsymbol{l}_{1}^{\prime}  
        }} \delta_{\substack{  
        \boldsymbol{j}_{2}, \boldsymbol{j}_{2}^{\prime} \\  
        \boldsymbol{l}_{2}, \boldsymbol{l}_{2}^{\prime}  
        }} \right) \right],
    \end{aligned}
\end{equation}  
by applying the delta functions to eliminate the corresponding indices, we obtain
\begin{equation}\label{27}
    \begin{split}
        \int d\mu(W) \text{Tr}[\Omega]
        = \frac{1}{d^{4} - 1} \left( \sum_{\substack{  
        \boldsymbol{i}_{1}, \boldsymbol{i}_{2}, \boldsymbol{l}_{1}, \boldsymbol{l}_{2} \\  
        \boldsymbol{k}_{1}, \boldsymbol{k}_{2}, \boldsymbol{j}_{1}  
        }} \langle \boldsymbol{l}_{1} | A | \boldsymbol{l}_{1} \rangle \langle \boldsymbol{i}_{1} | B | \boldsymbol{i}_{1} \rangle \langle \boldsymbol{l}_{2} | C | \boldsymbol{l}_{2} \rangle \langle \boldsymbol{i}_{2} | D | \boldsymbol{i}_{2} \rangle + \sum_{\substack{     \boldsymbol{i}_{1}, \boldsymbol{i}_{2}, \boldsymbol{l}_{1}, \boldsymbol{l}_{2} \\  
        \boldsymbol{k}_{1}, \boldsymbol{j}_{1}, \boldsymbol{j}_{2}  
        }} \langle \boldsymbol{l}_{1} | A | \boldsymbol{l}_{2} \rangle \langle \boldsymbol{i}_{2} | B | \boldsymbol{i}_{1} \rangle \langle \boldsymbol{l}_{2} | C | \boldsymbol{l}_{1} \rangle \langle \boldsymbol{i}_{1} | D | \boldsymbol{i}_{2} \rangle \right)\\
        - \frac{1}{d^{2}(d^{4} - 1)} \left( \sum_{\substack{  
        \boldsymbol{i}_{1}, \boldsymbol{i}_{2}, \boldsymbol{l}_{1}, \boldsymbol{l}_{2} \\  
        \boldsymbol{k}_{1}, \boldsymbol{k}_{2}, \boldsymbol{j}_{1}, \boldsymbol{j}_{2}  
        }} \langle \boldsymbol{l}_{1} | A | \boldsymbol{l}_{2} \rangle \langle \boldsymbol{i}_{1} | B | \boldsymbol{i}_{1} \rangle \langle \boldsymbol{l}_{2} | C | \boldsymbol{l}_{1} \rangle \langle \boldsymbol{i}_{2} | D | \boldsymbol{i}_{2} \rangle + \sum_{\substack{  
        \boldsymbol{i}_{1}, \boldsymbol{i}_{2}, \boldsymbol{l}_{1}, \boldsymbol{l}_{2} \\  
        \boldsymbol{k}_{1}, \boldsymbol{j}_{2}  
        }} \langle \boldsymbol{l}_{1} | A | \boldsymbol{l}_{1} \rangle \langle \boldsymbol{i}_{2} | B | \boldsymbol{i}_{1} \rangle \langle \boldsymbol{l}_{2} | C | \boldsymbol{l}_{2} \rangle \langle \boldsymbol{i}_{1} | D | \boldsymbol{i}_{2} \rangle \right),
    \end{split}
\end{equation} 
where the indices in the second row of the summation operator are free and contribute a multiplier factor of \(d\). After collecting the traces, we obtain
\begin{equation}
    \begin{split}
        \int d\mu(W) \text{Tr}[\Omega]
        = \frac{1}{d^{4} - 1} &\left( \text{Tr}[A] \text{Tr}[B] \text{Tr}[C] \text{Tr}[D] \cdot d^{3} + \text{Tr}[AC] \text{Tr}[BD] \cdot d^{3} \right)\\
        - \frac{1}{d^{2}(d^{4} - 1)} &\left( \text{Tr}[AC] \text{Tr}[B] \text{Tr}[D] \cdot d^{4} + \text{Tr}[A] \text{Tr}[C] \text{Tr}[BD] \cdot d^{2} \right)\\
        = \frac{1}{d^{4} - 1} ( \text{Tr}[A] \text{Tr}[B] \text{Tr}[C] \text{Tr}[D] d^{3} &+ \text{Tr}[AC] \text{Tr}[BD] d^{3} - \text{Tr}[A] \text{Tr}[C] \text{Tr}[BD] - \text{Tr}[AC] \text{Tr}[B] \text{Tr}[D] \cdot d^{2} ).
    \end{split}
\end{equation}
\hfill $\square$

\noindent{\textbf{Lemma 6.}} Let the composite Hilbert space $\mathcal{H} = \mathcal{H}_{1} \otimes \mathcal{H}_{2}$ have dimension $d \times d = d^{2}$ (both subspaces have dimension $d$), and let $\{ W_{y} \}_{y \in Y} \subset \mathcal{U}(d^{2})$ form a unitary $t$-design with $t \geq 2$, acting on the composite space $\mathcal{H}$. For any linear operators $A, B, C, D: \mathcal{H}_{2} \rightarrow \mathcal{H}_{2}$, define the operator  
\begin{equation}
    \Omega = \text{Tr}_{2}\left[ \begin{matrix}  
    \mathbb{I} \\  \otimes \\  A  \end{matrix} W^{\dagger} \begin{matrix}  \mathbb{I} \\  
    \otimes \\  B  \end{matrix} W \right] \text{Tr}_{2}\left[ \begin{matrix}  \mathbb{I} \\   
    \otimes \\ C \end{matrix} W^{\dagger} \begin{matrix}  \mathbb{I} \\  \otimes \\  D  \end{matrix} W \right],
\end{equation}  
for the operator $\Omega$, the following equality holds
\begin{equation}
    \begin{split}
        \int d\mu(W) \text{Tr}[\Omega] 
        =& \frac{1}{d^{4} - 1} \left( \text{Tr}[A] \text{Tr}[B] \text{Tr}[C] \text{Tr}[D] \cdot d^{3} + \text{Tr}[AC] \text{Tr}[BD] \cdot d^{3} \right)\\
        &- \frac{1}{d^{2}(d^{4} - 1)} \left( \text{Tr}[AC] \text{Tr}[B] \text{Tr}[D] \cdot d^{4} + \text{Tr}[A] \text{Tr}[C] \text{Tr}[BD] \cdot d^{2} \right).
    \end{split}
\end{equation}  

\noindent{\textit{Proof.}} The proof follows similarly to that of \noindent{\textbf{Lemma 5}}. We can apply Eq. (\ref{A24}) to expand the spectral decomposition of \(W\) and \(W^{\dagger}\) in the composite space, and then use this decomposition in the integral for the operator \(\Omega\) to obtain
\begin{equation}\label{A31}
    \begin{split}
        \int d\mu(W) \text{Tr}[\Omega] 
        &= \int d\mu(W) \text{Tr}\left\{ \text{Tr}_{2}\left[ \begin{matrix}  \mathbb{I} \\  \otimes \\  A  \end{matrix} W^{\dagger} \begin{matrix}  \mathbb{I} \\  \otimes \\  B  \end{matrix} W \right] \text{Tr}_{1}\left[ \begin{matrix}  \mathbb{I} \\  \otimes \\  C  \end{matrix} W^{\dagger} \begin{matrix} \mathbb{I} \\  \otimes \\  D  \end{matrix} W \right] \right\}
        \\
        &= \int d\mu(W) \sum_{\mathbf{I}\cdot \mathbf{J},\mathbf{K}\cdot \mathbf{L}} w_{\mathbf{I}\cdot \mathbf{J},\mathbf{K}\cdot \mathbf{L}}
        \text{Tr}\left\{ \text{Tr}_{2}\left[ \begin{matrix}  
        \mathbb{I} \\  \otimes \\  A  \end{matrix} \begin{matrix}  
        | \boldsymbol{j}_{1}^{\prime} \rangle \langle \boldsymbol{i}_{1}^{\prime} | \\  \otimes \\  | \boldsymbol{l}_{1}^{\prime} \rangle \langle \boldsymbol{k}_{1}^{\prime} |  \end{matrix} \begin{matrix}  \mathbb{I} \\  \otimes \\  B  \end{matrix} \begin{matrix}  | \boldsymbol{i}_{1} \rangle \langle \boldsymbol{j}_{1} | \\  \otimes \\  | \boldsymbol{k}_{1} \rangle \langle \boldsymbol{l}_{1} |  \end{matrix} \right] \text{Tr}_{2}\left[ \begin{matrix}  \mathbb{I} \\  \otimes \\  C  \end{matrix} \begin{matrix}  | \boldsymbol{j}_{2}^{\prime} \rangle \langle \boldsymbol{i}_{2}^{\prime} | \\  \otimes \\  | \boldsymbol{l}_{2}^{\prime} \rangle \langle \boldsymbol{k}_{2}^{\prime} |  \end{matrix} \begin{matrix}  \mathbb{I} \\  \otimes \\  D  \end{matrix} \begin{matrix}  | \boldsymbol{i}_{2} \rangle \langle \boldsymbol{j}_{2} | \\  \otimes \\  | \boldsymbol{k}_{2} \rangle \langle \boldsymbol{l}_{2} |  \end{matrix} \right] \right\}
        \\
        &= \sum_{\mathbf{I}\cdot \mathbf{J},\mathbf{K}\cdot \mathbf{L}} \int d\mu(W) w_{\mathbf{I}\cdot \mathbf{J},\mathbf{K}\cdot \mathbf{L}}
        \langle \boldsymbol{l}_{1} | A | \boldsymbol{l}_{1}^{\prime} \rangle \langle \boldsymbol{k}_{1}^{\prime} | B | \boldsymbol{k}_{1} \rangle \langle \boldsymbol{l}_{2} | C | \boldsymbol{l}_{2}^{\prime} \rangle \langle \boldsymbol{k}_{2}^{\prime} | D | \boldsymbol{k}_{2} \rangle \delta_{\boldsymbol{j}_{1}, \boldsymbol{j}_{2}^{\prime}} \delta_{\boldsymbol{j}_{2}, \boldsymbol{j}_{1}^{\prime}} \delta_{\boldsymbol{i}_{1}^{\prime}, \boldsymbol{i}_{1}} \delta_{\boldsymbol{i}_{2}^{\prime}, \boldsymbol{i}_{2}}.
    \end{split}
\end{equation} 
where the definitions of $\mathbf{I}\cdot \mathbf{J},\mathbf{K}\cdot \mathbf{L}$ and $w_{\mathbf{I}\cdot \mathbf{J},\mathbf{K}\cdot \mathbf{L}}$ are the same as those in \textbf{Lemma 5}. Four delta function originate from the unit operators in $\Omega$, which can remove four indexes. By using integral Eq. (\ref{A7}), another four delta functions can be introduced for each item
\begin{equation}\label{26} 
    \begin{aligned}
        \int d\mu(W) \text{Tr}[\Omega]
        =& \sum_{\mathbf{I}\cdot \mathbf{J},\mathbf{K}\cdot \mathbf{L}} \langle \boldsymbol{l}_{1} | A | \boldsymbol{l}_{1}^{\prime} \rangle \langle \boldsymbol{k}_{1}^{\prime} | B | \boldsymbol{k}_{1} \rangle \langle \boldsymbol{l}_{2} | C | \boldsymbol{l}_{2}^{\prime} \rangle \langle \boldsymbol{k}_{2}^{\prime} | D | \boldsymbol{k}_{2} \rangle \delta_{\boldsymbol{j}_{1}, \boldsymbol{j}_{2}^{\prime}} \delta_{\boldsymbol{j}_{2}, \boldsymbol{j}_{1}^{\prime}} \delta_{\boldsymbol{i}_{1}^{\prime}, \boldsymbol{i}_{1}} \delta_{\boldsymbol{i}_{2}^{\prime}, \boldsymbol{i}_{2}}
        \\
        &\times  \left[ \frac{1}{(d^{2})^{2} - 1} \left( \delta_{\substack{  
        \boldsymbol{i}_{1}, \boldsymbol{i}_{1}^{\prime} \\  
        \boldsymbol{k}_{1}, \boldsymbol{k}_{1}^{\prime}  
        }} \delta_{\substack{  
        \boldsymbol{i}_{2}, \boldsymbol{i}_{2}^{\prime} \\  
        \boldsymbol{k}_{2}, \boldsymbol{k}_{2}^{\prime}  
        }} \delta_{\substack{  
        \boldsymbol{j}_{1}, \boldsymbol{j}_{1}^{\prime} \\  
        \boldsymbol{l}_{1}, \boldsymbol{l}_{1}^{\prime}  
        }} \delta_{\substack{  
        \boldsymbol{j}_{2}, \boldsymbol{j}_{2}^{\prime} \\  
        \boldsymbol{l}_{2}, \boldsymbol{l}_{2}^{\prime}  
        }} + \delta_{\substack{  
        \boldsymbol{i}_{1}, \boldsymbol{i}_{2}^{\prime} \\  
        \boldsymbol{k}_{1}, \boldsymbol{k}_{2}^{\prime}  
        }} \delta_{\substack{  
        \boldsymbol{i}_{2}, \boldsymbol{i}_{1}^{\prime} \\  
        \boldsymbol{k}_{2}, \boldsymbol{k}_{1}^{\prime}  
        }} \delta_{\substack{  
        \boldsymbol{j}_{1}, \boldsymbol{j}_{2}^{\prime} \\  
        \boldsymbol{l}_{1}, \boldsymbol{l}_{2}^{\prime}  
        }} \delta_{\substack{  
        \boldsymbol{j}_{2}, \boldsymbol{j}_{1}^{\prime} \\  
        \boldsymbol{l}_{2}, \boldsymbol{l}_{1}^{\prime}  
        }} \right) \right.
        \\
        &\left.- \frac{1}{d^{2}[(d^{2})^{2} - 1]} \left( \delta_{\substack{  
        \boldsymbol{i}_{1}, \boldsymbol{i}_{1}^{\prime} \\  
        \boldsymbol{k}_{1}, \boldsymbol{k}_{1}^{\prime}  
        }} \delta_{\substack{  
        \boldsymbol{i}_{2}, \boldsymbol{i}_{2}^{\prime} \\  
        \boldsymbol{k}_{2}, \boldsymbol{k}_{2}^{\prime}  
        }} \delta_{\substack{  
        \boldsymbol{j}_{1}, \boldsymbol{j}_{2}^{\prime} \\  
        \boldsymbol{l}_{1}, \boldsymbol{l}_{2}^{\prime}  
        }} \delta_{\substack{  
        \boldsymbol{j}_{2}, \boldsymbol{j}_{1}^{\prime} \\  
        \boldsymbol{l}_{2}, \boldsymbol{l}_{1}^{\prime}  
        }} + \delta_{\substack{  
        \boldsymbol{i}_{1}, \boldsymbol{i}_{2}^{\prime} \\  
        \boldsymbol{k}_{1}, \boldsymbol{k}_{2}^{\prime}  
        }} \delta_{\substack{  
        \boldsymbol{i}_{2}, \boldsymbol{i}_{1}^{\prime} \\  
        \boldsymbol{k}_{2}, \boldsymbol{k}_{1}^{\prime}  
        }} \delta_{\substack{  
        \boldsymbol{j}_{1}, \boldsymbol{j}_{1}^{\prime} \\  
        \boldsymbol{l}_{1}, \boldsymbol{l}_{1}^{\prime}  
        }} \delta_{\substack{  
        \boldsymbol{j}_{2}, \boldsymbol{j}_{2}^{\prime} \\  
        \boldsymbol{l}_{2}, \boldsymbol{l}_{2}^{\prime}  
        }} \right) \right],
    \end{aligned}
\end{equation}  
by applying the delta functions to eliminate the corresponding indices, we obtain
\begin{equation}\label{27}
    \begin{split}
        \int d\mu(W) \text{Tr}[\Omega]
        = \frac{1}{d^{4} - 1} \left( \sum_{\substack{  
        \boldsymbol{k}_{1}, \boldsymbol{k}_{2}, \boldsymbol{l}_{1}, \boldsymbol{l}_{2} \\  
        \boldsymbol{i}_{1}, \boldsymbol{i}_{2}, \boldsymbol{j}_{1}  
        }} \langle \boldsymbol{l}_{1} | A | \boldsymbol{l}_{1} \rangle \langle \boldsymbol{k}_{1} | B | \boldsymbol{k}_{1} \rangle \langle \boldsymbol{l}_{2} | C | \boldsymbol{l}_{2} \rangle \langle \boldsymbol{k}_{2} | D | \boldsymbol{k}_{2} \rangle + \sum_{\substack{     \boldsymbol{k}_{1}, \boldsymbol{k}_{2}, \boldsymbol{l}_{1}, \boldsymbol{l}_{2} \\  
        \boldsymbol{i}_{1}, \boldsymbol{j}_{1}, \boldsymbol{j}_{2}  
        }} \langle \boldsymbol{l}_{1} | A | \boldsymbol{l}_{2} \rangle \langle \boldsymbol{k}_{2} | B | \boldsymbol{k}_{1} \rangle \langle \boldsymbol{l}_{2} | C | \boldsymbol{l}_{1} \rangle \langle \boldsymbol{k}_{1} | D | \boldsymbol{k}_{2} \rangle \right)\\
        - \frac{1}{d^{2}(d^{4} - 1)} \left( \sum_{\substack{  
        \boldsymbol{k}_{1}, \boldsymbol{k}_{2}, \boldsymbol{l}_{1}, \boldsymbol{l}_{2} \\  
        \boldsymbol{i}_{1}, \boldsymbol{i}_{2}, \boldsymbol{j}_{1}, \boldsymbol{j}_{2}  
        }} \langle \boldsymbol{l}_{1} | A | \boldsymbol{l}_{2} \rangle \langle \boldsymbol{k}_{1} | B | \boldsymbol{k}_{1} \rangle \langle \boldsymbol{l}_{2} | C | \boldsymbol{l}_{1} \rangle \langle \boldsymbol{k}_{2} | D | \boldsymbol{k}_{2} \rangle + \sum_{\substack{  
        \boldsymbol{k}_{1}, \boldsymbol{k}_{2}, \boldsymbol{l}_{1}, \boldsymbol{l}_{2} \\  
        \boldsymbol{i}_{1}, \boldsymbol{j}_{2}  
        }} \langle \boldsymbol{l}_{1} | A | \boldsymbol{l}_{1} \rangle \langle \boldsymbol{k}_{2} | B | \boldsymbol{k}_{1} \rangle \langle \boldsymbol{l}_{2} | C | \boldsymbol{l}_{2} \rangle \langle \boldsymbol{k}_{1} | D | \boldsymbol{k}_{2} \rangle \right),
    \end{split}
\end{equation} 
where the indices in the second row of the summation operator are free and contribute a multiplier factor of \(d\). After collecting the traces, we obtain
\begin{equation}
    \begin{split}
        \int d\mu(W) \text{Tr}[\Omega]
        = \frac{1}{d^{4} - 1} &\left( \text{Tr}[A] \text{Tr}[B] \text{Tr}[C] \text{Tr}[D] \cdot d^{3} + \text{Tr}[AC] \text{Tr}[BD] \cdot d^{3} \right)\\
        - \frac{1}{d^{2}(d^{4} - 1)} &\left( \text{Tr}[AC] \text{Tr}[B] \text{Tr}[D] \cdot d^{4} + \text{Tr}[A] \text{Tr}[C] \text{Tr}[BD] \cdot d^{2} \right)\\
        = \frac{1}{d^{4} - 1} ( \text{Tr}[A] \text{Tr}[B] \text{Tr}[C] \text{Tr}[D] d^{3} &+ \text{Tr}[AC] \text{Tr}[BD] d^{3} - \text{Tr}[A] \text{Tr}[C] \text{Tr}[BD] - \text{Tr}[AC] \text{Tr}[B] \text{Tr}[D] \cdot d^{2} ).
    \end{split}
\end{equation}
\hfill $\square$  

\section{\label{B}Section B. Equivalent Model of QRNNs and the Formalism}
In this section, we present a detailed proof of the equivalence between the standard model $\mathcal Q_1$ and equivalent
model $\mathcal Q_2$ of QRNN introduced in the main text. We also introduce the formalism of the QRNN model that will be used in the subsequent theoretical analysis.

\subsection{\label{sec:level2}1. Equivalent Model of QRNNs}
\begin{figure}[h]
    \centering
    \includegraphics[width=4.2in]{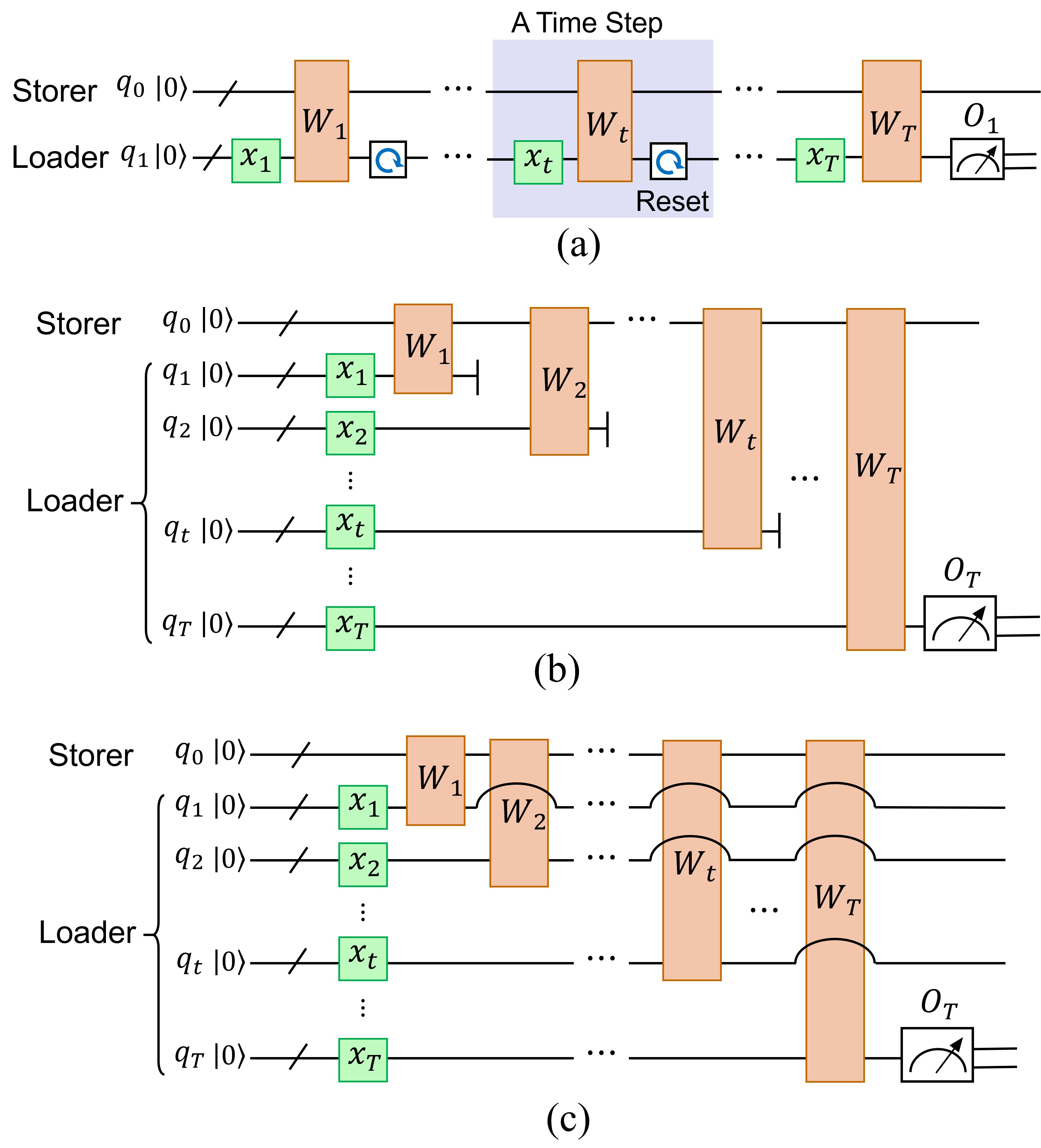}
    \caption{QRNN models. 
    (a) Standard QRNN model $\mathcal{Q}_{1}$ introduced in Refs.~\cite{bausch2020recurrent,takaki2021learning,li2023quantum,li2025quantum}. 
    (b) Transitional QRNN model ${\widetilde{\mathcal{Q}}}_{1}$, in which the sequential data loading on a single register is unfolded into simultaneous initial inputs distributed over multiple registers $q_{1}\sim q_{T}$. 
    (c) Equivalent QRNN model $\mathcal{Q}_{2}$, illustrating that the step-by-step partial trace operations on registers $q_{1}\sim q_{T-1}$ can be omitted without changing the final expectation value.}
    \label{FB1}
\end{figure}

QRNNs are quantum models designed for sequential learning tasks, including time-series prediction and related problems. A QRNN consists of three main components: data input modules, parameterization modules, and measurements. The standard model $\mathcal{Q}_{1}$ is illustrated in Fig.~\ref{FB1}(a). It employs two quantum registers: the first register $q_{0}$, referred to as the \emph{Storer}, is used for information storage and is not measured; the second register $q_{1}$, referred to as the \emph{Loader}, is used to encode the input data $x_{t}$ at different time steps $t$~\cite{li2023quantum}. At time step $t$, the input $x_t$ is embedded into the Loader register as a quantum state, given by
\begin{equation}\label{B0}
    \rho_{t} \equiv \rho_{x_t} = U_{in}(f(x_t))|\mathbf{0}\rangle\langle \mathbf{0}| U_{in}^\dagger(f(x_t)),
\end{equation}
where ${\left|\mathbf{0} \right \rangle \left\langle\mathbf{0}\right|}\equiv {\left|0 \right \rangle \left\langle0\right|}^{\otimes m}$ denotes the initial state of the register. This state is then combined with the hidden state
$\rho_{h_{t-1}}=\text{Tr}_L [W_t (\rho_{x_{t-1}} \otimes \rho_{h_{t-2}}) W_t^\dagger]$
stored in the Storer via a variational block $W_t$, which acts on the joint Storer$\otimes$Loader system. This separation naturally enables QRNNs to handle variable-length sequences and allows for qubit-efficient implementations by permitting the reset and reuse of the Loader register after each time step. After the final input $x_{T}$ is encoded, the Storer $q_{0}$ contains information accumulated from all previous time steps, while the Loader $q_{1}$ retains information associated with the final time step $T$. Following the coupling operation, the Loader $q_{1}$ is measured to obtain the final prediction, whose expectation value is related to the inputs at all time steps:
\begin{equation}\label{B1}
  \left\langle O \right\rangle_{std} = \text{Tr}\left\{ W_{T}\left( {\text{Tr}}_{T - 1}\left[ \cdots\left( {\text{Tr}}_{2}\left[ W_{2}\left( {\text{Tr}}_{1}\left[ W_{1}\rho_{0}\otimes\rho_{1}W_{1}^{\dagger} \right]\otimes\rho_{2} \right)W_{2}^{\dagger} \right] \right)\cdots \right]\otimes\rho_{T} \right)W_{T}^{\dagger}\left( \mathbb{I}_0\otimes O_1 \right) \right\}.
\end{equation}
Here, $\text{Tr}_{t}$ denotes the partial trace over the system introduced at time step $t$, $O_1$ is an observable acting on the Loader $q_{1}$, and $\mathbb{I}_0$ is the identity operator acting on the Storer $q_{0}$.

However, one can equivalently use multiple Loader registers $q_{1}\sim q_{T}$ to input data, as shown in Fig.~\ref{FB1}(b), to replace the repeated use of a single Loader $q_{1}$ in Fig.~\ref{FB1}(a), thereby obtaining the transitional model ${\widetilde{\mathcal{Q}}}_{1}$, whose equivalence is straightforward. Furthermore, all data input modules $U_{x_{t}}$ can be naturally placed before the first partial trace operation, as indicated in Fig.~\ref{FB1}(b). This is because the data input at any time step does not affect the partial trace operations associated with previous time steps. Accordingly, the output of the transitional model ${\widetilde{\mathcal{Q}}}_{1}$ is given by
\begin{equation}
  \begin{gathered}
  \left\langle {O} \right\rangle_{tr} 
  = \text{Tr}\left\{ W_{T0}\otimes\mathbb{I}_{\overline{T0}}\left( {\text{Tr}}_{T - 1}\left[ \cdots\left( {\text{Tr}}_{2}\left[ W_{20}\otimes\mathbb{I}_{\overline{20}} \times
  \right. \right. \right. \right. \right.  \\ \left. \left. \left. \left. \left. 
  \left( {\text{Tr}}_{1} \left[ W_{10}\otimes\mathbb{I}_{\overline{10}}\rho_{all}W_{10}^{\dagger}\otimes\mathbb{I}_{\overline{10}} \right] \right)W_{20}\otimes\mathbb{I}_{\overline{20}} \right] \right)\cdots \right] \right)W_{T}^{\dagger}\otimes\mathbb{I}_{\overline{T0}}\left( \mathbb{I}_{\overline{T}}\otimes O_{T} \right) \right\},
  \end{gathered}
\end{equation}
where $\rho_{all} \equiv \otimes_{t = 0}^{T}\rho_{t}$ denotes the quantum state obtained after inputting all data. After the system expansion shown in Fig.~\ref{FB1}(b), the variational quantum circuit $W_{t}$ is extended to $W_{t0}\otimes\mathbb{I}_{\overline{t0}}$, where $\mathbb{I}_{\overline{t0}}$ denotes the identity operator acting on all systems except $q_{0}$ and $q_{t}$.

Next, we construct the equivalent model $\mathcal{Q}_{2}$, as shown in Fig.~\ref{FB1}(c), in which the partial trace operations at each time step are further discarded and replaced by identity operators. As a result, the output expression becomes
\begin{equation}
  \begin{split}
  \left\langle O \right\rangle_{eq} = \text{Tr}\left\{ \left( \prod_{t = 1}^{T}W_{t0}\otimes\mathbb{I}_{\overline{t0}} \right)\left( \bigotimes_{i = 0}^{T}\rho_{t} \right)\left( \prod_{t = 1}^{T}W_{t0}^{\dagger}\otimes\mathbb{I}_{\overline{t0}} \right)O_{T} \right\}.
  \end{split}
\end{equation}
This output is equivalent to the output $\left\langle O \right\rangle_{std}$ of the standard model $\mathcal{Q}_{1}$, and a detailed proof is provided below.

Since the equivalence between the standard model $\mathcal{Q}_{1}$ and the transitional model ${\widetilde{\mathcal{Q}}}_{1}$ is straightforward, we directly derive the output of the equivalent model $\mathcal{Q}_{2}$ from the output expression of the transitional model ${\widetilde{\mathcal{Q}}}_{1}$:
\begin{equation}
  \begin{split}
  \left\langle O \right\rangle_{std} 
  \equiv 
  \left\langle {O} \right\rangle_{tr}
  = \text{Tr}\left\{ \left[ \begin{matrix}
  \begin{matrix}
  {\hat{W}}_{T0} \\
  1_{1}
  \end{matrix} \\
  \begin{matrix}
  1_{2} \\
   \vdots \\
  {\check{W}}_{T0}
  \end{matrix}
  \end{matrix}\left( {\text{Tr}}_{T - 1}\left[ \cdots\left( {\text{Tr}}_{2}\left[ \begin{matrix}
  \begin{matrix}
  {\hat{W}}_{20} \\
  1_{1}
  \end{matrix} \\
  \begin{matrix}
  {\check{W}}_{20} \\
   \vdots \\
  \mathbb{I}_{T}
  \end{matrix}
  \end{matrix}\left( {\text{Tr}}_{1}\begin{bmatrix}
  {\hat{W}}_{10} & \rho_{0} & {\hat{W}}_{10}^{\dagger} \\
  {\check{W}}_{10} & \rho_{1} & {\check{W}}_{10}^{\dagger} \\
  \mathbb{I}_{2} & \rho_{2} & \mathbb{I}_{2} \\
   \vdots & \vdots & \vdots \\
  \mathbb{I}_{T} & \rho_{T} & \mathbb{I}_{T}
  \end{bmatrix} \right)\begin{matrix}
  \begin{matrix}
  {\hat{W}}_{20}^{\dagger} \\
  1_{1}
  \end{matrix} \\
  \begin{matrix}
  {\check{W}}_{20}^{\dagger} \\
   \vdots \\
  \mathbb{I}_{T}
  \end{matrix}
  \end{matrix} \right] \right)\cdots \right] \right)\begin{matrix}
  \begin{matrix}
  {\hat{W}}_{T0}^{\dagger} \\
  1_{1}
  \end{matrix} \\
  \begin{matrix}
  1_{1} \\
   \vdots \\
  {\check{W}}_{T0}^{\dagger}
  \end{matrix}
  \end{matrix} \right]\begin{matrix}
  \begin{matrix}
  \mathbb{I}_{0} \\
  1_{1}
  \end{matrix} \\
  \begin{matrix}
  1_{2} \\
   \vdots \\
  O_{T}
  \end{matrix}
  \end{matrix} \right\}.
  \end{split}
\end{equation}
Here we adopt the \emph{tensor-product tableau notation} introduced in \textbf{Lemma~5}, in which tensor-product operations are arranged in a stacked layout to clarify their actions on different subsystems. We denote ${\hat{W}}_{t0}$ and ${\check{W}}_{t0}$ as the \emph{restricted representations} of the parameterized circuit $W_{t0}$ acting on subsystems $q_{0}$ and $q_{t}$, respectively. It is important to emphasize that these symbols serve purely as notational devices and do not imply that $W_{t0}$ factorizes as a tensor product of operators acting on the two subsystems. Owing to entanglement, $W_{t0}$ generally \textbf{cannot} be written as ${\hat{W}}_{t0} \otimes {\check{W}}_{t0}$; nevertheless, it still exhibits a tensor-product structure with respect to the remaining subsystems.

Additionally, the quantities $1_{1}$ and $1_{2}$ represent scalar extensions arising from the partial trace operations over systems $q_{1}$ and $q_{2}$. To preserve a consistent correspondence between subsystems in the tensor-product tableau notation, we may always tensor a $1$-dimensional identity operator $1$ as a visual placeholder. Based on the definition of the partial trace operation,
$\text{Tr}_{1}[\rho] = \sum_{\boldsymbol{i}}{\left \langle \boldsymbol{i} \right |_{1}\otimes\mathbb{I}_{\overline{1}}\rho \left| \boldsymbol{i}\right\rangle_{1}\otimes\mathbb{I}_{\overline{1}}}$,
we have
\begin{equation}\label{B6}
    \left\langle {O} \right\rangle_{tr} =
\end{equation} 
\centerline{$\displaystyle \text{Tr}\left\{ \left[ 
          \begin{matrix}
          {\hat{W}}_{T0} \\
          1_{1} \\
          1_{2} \\
           \vdots \\
          {\check{W}}_{T0}
          \end{matrix} \\
          \left( \sum_{\boldsymbol{i}_{T - 1}}^{}{
          \begin{matrix}
          \mathbb{I}_{0} \\
          \mathbb{I}_{1}\\
          \vdots \\
          \left \langle \boldsymbol{i}_{T-1} \right |  \\
          \mathbb{I}_{T}
          \end{matrix}\left[ \cdots\left( \sum_{\boldsymbol{i}_{2}}^{}{
          \begin{matrix}
          \mathbb{I}_{0} \\
          \mathbb{I}_{1} \\
          \left \langle \boldsymbol{i}_{2} \right |  \\
           \vdots \\
          \mathbb{I}_{T}
          \end{matrix}\left[ 
          \begin{matrix}
          {\hat{W}}_{20} \\
          1_{1}\\
          {\check{W}}_{20} \\
           \vdots \\
          \mathbb{I}_{T}
          \end{matrix}\left( \sum_{\boldsymbol{i}_{1}}^{}{
          \begin{matrix}
          \mathbb{I}_{0} \\
          \left \langle \boldsymbol{i}_{1} \right | \\
          \mathbb{I}_{2} \\
           \vdots \\
          \mathbb{I}_{T}
          \end{matrix}
          \begin{bmatrix}
          {\hat{W}}_{10} & \rho_{0} & {\hat{W}}_{10}^{\dagger} \\
          {\check{W}}_{10} & \rho_{1} & {\check{W}}_{10}^{\dagger} \\
          \mathbb{I}_{2} & \rho_{2} & \mathbb{I}_{2} \\
           \vdots & \vdots & \vdots \\
          \mathbb{I}_{T} & \rho_{T} & \mathbb{I}_{T}
          \end{bmatrix}\begin{matrix}
          \begin{matrix}
          \mathbb{I}_{0} \\
          \left | \boldsymbol{i} \right \rangle _{1}
          \end{matrix} \\
          \begin{matrix}
          \mathbb{I}_{2} \\
           \vdots \\
          \mathbb{I}_{T}
          \end{matrix}
          \end{matrix}} \right)\begin{matrix}
          \begin{matrix}
          {\hat{W}}_{20}^{\dagger} \\
          1_{1}
          \end{matrix} \\
          \begin{matrix}
          {\check{W}}_{20}^{\dagger} \\
           \vdots \\
          \mathbb{I}_{T}
          \end{matrix}
          \end{matrix} \right]\begin{matrix}
          \begin{matrix}
          \mathbb{I}_{0} \\
          \mathbb{I}_{1}
          \end{matrix} \\
          \begin{matrix}
          \left | \boldsymbol{i} \right \rangle _{2} \\
           \vdots \\
          \mathbb{I}_{T}
          \end{matrix}
          \end{matrix}} \right)\cdots \right]\begin{matrix}
          \begin{matrix}
          \mathbb{I}_{0} \\
          \mathbb{I}_{1}
          \end{matrix} \\
          \begin{matrix}
           \vdots \\
          \left | \boldsymbol{i} \right \rangle _{T-1} \\
          \mathbb{I}_{T}
          \end{matrix}
          \end{matrix}} \right)\begin{matrix}
          \begin{matrix}
          {\hat{W}}_{T0}^{\dagger} \\
          1_{1}
          \end{matrix} \\
          \begin{matrix}
          1_{2} \\
           \vdots \\
          {\check{W}}_{T0}^{\dagger}
          \end{matrix}
          \end{matrix} \right]\begin{matrix}
          \begin{matrix}
          \mathbb{I}_{0} \\
          1_{1}
          \end{matrix} \\
          \begin{matrix}
          1_{2} \\
           \vdots \\
          O_{T}
          \end{matrix}
          \end{matrix} \right\}
$}
\vspace{2ex} 
\centerline{$\displaystyle 
= \sum_{k = 1}^{T - 1}{\sum_{\boldsymbol{i}_{k}}^{}{\text{Tr}\left\{ \left[ \begin{matrix}
  \begin{matrix}
  {\hat{W}}_{T0} \\
  1_{1}
  \end{matrix} \\
  \begin{matrix}
  1_{2} \\
   \vdots \\
  {\check{W}}_{T0}
  \end{matrix}
  \end{matrix}\left( \begin{matrix}
  \begin{matrix}
  \mathbb{I}_{0} \\
  1_{1}
  \end{matrix} \\
  \begin{matrix}
   \vdots \\
  \left \langle \boldsymbol{i} \right |_{T-1}  \\
  \mathbb{I}_{T}
  \end{matrix}
  \end{matrix}\left[ \cdots\left( \begin{matrix}
  \begin{matrix}
  \mathbb{I}_{0} \\
  1_{1}
  \end{matrix} \\
  \begin{matrix}
  \left \langle \boldsymbol{i} \right |_{2}  \\
   \vdots \\
  \mathbb{I}_{T}
  \end{matrix}
  \end{matrix}\left[ \begin{matrix}
  \begin{matrix}
  {\hat{W}}_{20} \\
  1_{1}
  \end{matrix} \\
  \begin{matrix}
  {\check{W}}_{20} \\
   \vdots \\
  \mathbb{I}_{T}
  \end{matrix}
  \end{matrix}\left( \begin{matrix}
  \begin{matrix}
  \mathbb{I}_{0} \\
  \left \langle \boldsymbol{i} \right |_{1} 
  \end{matrix} \\
  \begin{matrix}
  \mathbb{I}_{2} \\
   \vdots \\
  \mathbb{I}_{T}
  \end{matrix}
  \end{matrix}
  \begin{bmatrix}
  {\hat{W}}_{10} & \rho_{0} & {\hat{W}}_{10}^{\dagger} \\
  {\check{W}}_{10} & \rho_{1} & {\check{W}}_{10}^{\dagger} \\
  \mathbb{I}_{2} & \rho_{2} & \mathbb{I}_{2} \\
   \vdots & \vdots & \vdots \\
  \mathbb{I}_{T} & \rho_{T} & \mathbb{I}_{T}
  \end{bmatrix}
  \begin{matrix}
  \begin{matrix}
  \mathbb{I}_{0} \\
  \left | \boldsymbol{i} \right \rangle _{1}
  \end{matrix} \\
  \begin{matrix}
  \mathbb{I}_{2} \\
   \vdots \\
  \mathbb{I}_{T}
  \end{matrix}
  \end{matrix} \right)\begin{matrix}
  \begin{matrix}
  {\hat{W}}_{20}^{\dagger} \\
  1_{1}
  \end{matrix} \\
  \begin{matrix}
  {\check{W}}_{20}^{\dagger} \\
   \vdots \\
  \mathbb{I}_{T}
  \end{matrix}
  \end{matrix} \right]\begin{matrix}
  \begin{matrix}
  \mathbb{I}_{0} \\
  1_{1}
  \end{matrix} \\
  \begin{matrix}
  \left | \boldsymbol{i} \right \rangle _{2} \\
   \vdots \\
  \mathbb{I}_{T}
  \end{matrix}
  \end{matrix} \right)\cdots \right]\begin{matrix}
  \begin{matrix}
  \mathbb{I}_{0} \\
  1_{1}
  \end{matrix} \\
  \begin{matrix}
   \vdots \\
  \left | \boldsymbol{i} \right \rangle _{T-1} \\
  \mathbb{I}_{T}
  \end{matrix}
  \end{matrix} \right)\begin{matrix}
  \begin{matrix}
  {\hat{W}}_{T0}^{\dagger} \\
  1_{1}
  \end{matrix} \\
  \begin{matrix}
  1_{2} \\
   \vdots \\
  {\check{W}}_{T0}^{\dagger}
  \end{matrix}
  \end{matrix} \right]\begin{matrix}
  \begin{matrix}
  \mathbb{I}_{0} \\
  1_{1}
  \end{matrix} \\
  \begin{matrix}
  1_{2} \\
   \vdots \\
  O_{T}
  \end{matrix}
  \end{matrix} \right\}}.}
$}.
\vspace{2ex}

\noindent{}By applying the key identity $1\langle\boldsymbol{i}|=\langle\boldsymbol{i}| \mathbb{I},|\boldsymbol{i}\rangle 1=\mathbb{I}|\boldsymbol{i}\rangle$, the bra and ket vectors can be systematically moved outward across the tensor structure. This allows us to rearrange the expression into the form

\vspace{2ex}
\centerline{$\displaystyle \sum_{k = 1}^{T - 1}{\sum_{\boldsymbol{i}_{k}}^{}{\text{Tr}\left\{ \left[ \begin{matrix}
  \begin{matrix}
  {\hat{W}}_{T0} \\
  \left \langle \boldsymbol{i_{1}} \right |
  \end{matrix} \\
  \begin{matrix}
   \vdots \\
  \left \langle \boldsymbol{i_{T-1}} \right | \\
  {\check{W}}_{T0}
  \end{matrix}
  \end{matrix}\left( \begin{matrix}
  \begin{matrix}
  \mathbb{I}_{0} \\
  \mathbb{I}_{1}
  \end{matrix} \\
  \begin{matrix}
   \vdots \\
  \mathbb{I}_{T - 1} \\
  \mathbb{I}_{T}
  \end{matrix}
  \end{matrix}\left[ \cdots\left( \begin{matrix}
  \begin{matrix}
  \mathbb{I}_{0} \\
  \mathbb{I}_{1}
  \end{matrix} \\
  \begin{matrix}
  \mathbb{I}_{2} \\
   \vdots \\
  \mathbb{I}_{T}
  \end{matrix}
  \end{matrix}\left[ \begin{matrix}
  \begin{matrix}
  {\hat{W}}_{20} \\
  \mathbb{I}_{1}
  \end{matrix} \\
  \begin{matrix}
  {\check{W}}_{20} \\
   \vdots \\
  \mathbb{I}_{T}
  \end{matrix}
  \end{matrix}\left( \begin{matrix}
  \begin{matrix}
  \mathbb{I}_{0} \\
  \mathbb{I}_{1}
  \end{matrix} \\
  \begin{matrix}
  \mathbb{I}_{2} \\
   \vdots \\
  \mathbb{I}_{T}
  \end{matrix}
  \end{matrix}\begin{bmatrix}
  {\hat{W}}_{10} & \rho_{0} & {\hat{W}}_{10}^{\dagger} \\
  {\check{W}}_{10} & \rho_{1} & {\check{W}}_{10}^{\dagger} \\
  \mathbb{I}_{2} & \rho_{2} & \mathbb{I}_{2} \\
   \vdots & \vdots & \vdots \\
  \mathbb{I}_{T} & \rho_{T} & \mathbb{I}_{T}
  \end{bmatrix}\begin{matrix}
  \begin{matrix}
  \mathbb{I}_{0} \\
  \mathbb{I}_{1}
  \end{matrix} \\
  \begin{matrix}
  \mathbb{I}_{2} \\
   \vdots \\
  \mathbb{I}_{T}
  \end{matrix}
  \end{matrix} \right)\begin{matrix}
  \begin{matrix}
  {\hat{W}}_{20}^{\dagger} \\
  \mathbb{I}_{1}
  \end{matrix} \\
  \begin{matrix}
  {\check{W}}_{20}^{\dagger} \\
   \vdots \\
  \mathbb{I}_{T}
  \end{matrix}
  \end{matrix} \right]\begin{matrix}
  \begin{matrix}
  \mathbb{I}_{0} \\
  \mathbb{I}_{1}
  \end{matrix} \\
  \begin{matrix}
  \mathbb{I}_{2} \\
   \vdots \\
  \mathbb{I}_{T}
  \end{matrix}
  \end{matrix} \right)\cdots \right]\begin{matrix}
  \begin{matrix}
  \mathbb{I}_{0} \\
  \mathbb{I}_{1}
  \end{matrix} \\
  \begin{matrix}
   \vdots \\
  \mathbb{I}_{T - 1} \\
  \mathbb{I}_{T}
  \end{matrix}
  \end{matrix} \right)\begin{matrix}
  \begin{matrix}
  {\hat{W}}_{T0}^{\dagger} \\
  \mathbb{I}_{1}
  \end{matrix} \\
  \begin{matrix}
   \vdots \\
  \mathbb{I}_{T - 1} \\
  {\check{W}}_{T0}^{\dagger}
  \end{matrix}
  \end{matrix} \right]\begin{matrix}
  \begin{matrix}
  \mathbb{I}_{0} \\
  \left | \boldsymbol{i_{1}} \right \rangle
  \end{matrix} \\
  \begin{matrix}
   \vdots \\
  \left | \boldsymbol{i_{T-1}} \right \rangle \\
  O_{T}
  \end{matrix}
  \end{matrix} \right\}}}
$}
\vspace{2ex}
\centerline{$\displaystyle = \sum_{k = 1}^{T - 1}{\sum_{\boldsymbol{i}_{k}}^{}{\text{Tr}\left\{ \left[ \begin{matrix}
  \begin{matrix}
  \mathbb{I}_{0} \\
  \left | \boldsymbol{i_{1}} \right \rangle
  \end{matrix} \\
  \begin{matrix}
   \vdots \\
  \left | \boldsymbol{i_{T-1}} \right \rangle \\
  O_{T}
  \end{matrix}
  \end{matrix}\begin{matrix}
  \begin{matrix}
  {\hat{W}}_{T0} \\
  \left \langle \boldsymbol{i_{1}} \right |
  \end{matrix} \\
  \begin{matrix}
   \vdots \\
  \left \langle \boldsymbol{i_{T-1}} \right |  \\
  {\check{W}}_{T0}
  \end{matrix}
  \end{matrix}\left( \left[ \cdots\left( \left[ \begin{matrix}
  \begin{matrix}
  {\hat{W}}_{20} \\
  \mathbb{I}_{1}
  \end{matrix} \\
  \begin{matrix}
  {\check{W}}_{20} \\
   \vdots \\
  \mathbb{I}_{T}
  \end{matrix}
  \end{matrix}\left( \begin{bmatrix}
  {\hat{W}}_{10} & \rho_{0} & {\hat{W}}_{10}^{\dagger} \\
  {\check{W}}_{10} & \rho_{1} & {\check{W}}_{10}^{\dagger} \\
  \mathbb{I}_{2} & \rho_{2} & \mathbb{I}_{2} \\
   \vdots & \vdots & \vdots \\
  \mathbb{I}_{T} & \rho_{T} & \mathbb{I}_{T}
  \end{bmatrix} \right)\begin{matrix}
  \begin{matrix}
  {\hat{W}}_{20}^{\dagger} \\
  \mathbb{I}_{1}
  \end{matrix} \\
  \begin{matrix}
  {\check{W}}_{20}^{\dagger} \\
   \vdots \\
  \mathbb{I}_{T}
  \end{matrix}
  \end{matrix} \right] \right)\cdots \right] \right)\begin{matrix}
  \begin{matrix}
  {\hat{W}}_{T0}^{\dagger} \\
  \mathbb{I}_{1}
  \end{matrix} \\
  \begin{matrix}
   \vdots \\
  \mathbb{I}_{T - 1} \\
  {\check{W}}_{T0}^{\dagger}
  \end{matrix}
  \end{matrix} \right] \right\}}}
$}
\vspace{2ex}
\centerline{$\displaystyle = \text{Tr}\left\{ \left[ \begin{matrix}
  \begin{matrix}
  \mathbb{I}_{0} \\
  \mathbb{I}_{1}
  \end{matrix} \\
  \begin{matrix}
   \vdots \\
  \mathbb{I}_{T - 1} \\
  O_{T}
  \end{matrix}
  \end{matrix}\begin{matrix}
  \begin{matrix}
  {\hat{W}}_{T0} \\
  \mathbb{I}_{1}
  \end{matrix} \\
  \begin{matrix}
   \vdots \\
  \mathbb{I}_{T - 1} \\
  {\check{W}}_{T0}
  \end{matrix}
  \end{matrix}\left( \left[ \cdots\left( \left[ \begin{matrix}
  \begin{matrix}
  {\hat{W}}_{20} \\
  \mathbb{I}_{1}
  \end{matrix} \\
  \begin{matrix}
  {\check{W}}_{20} \\
   \vdots \\
  \mathbb{I}_{T}
  \end{matrix}
  \end{matrix}\left( \begin{bmatrix}
  {\hat{W}}_{10} & \rho_{0} & {\hat{W}}_{10}^{\dagger} \\
  {\check{W}}_{10} & \rho_{1} & {\check{W}}_{10}^{\dagger} \\
  \mathbb{I}_{2} & \rho_{2} & \mathbb{I}_{2} \\
   \vdots & \vdots & \vdots \\
  \mathbb{I}_{T} & \rho_{T} & \mathbb{I}_{T}
  \end{bmatrix} \right)\begin{matrix}
  \begin{matrix}
  {\hat{W}}_{20}^{\dagger} \\
  \mathbb{I}_{1}
  \end{matrix} \\
  \begin{matrix}
  {\check{W}}_{20}^{\dagger} \\
   \vdots \\
  \mathbb{I}_{T}
  \end{matrix}
  \end{matrix} \right] \right)\cdots \right] \right)\begin{matrix}
  \begin{matrix}
  {\hat{W}}_{T0}^{\dagger} \\
  \mathbb{I}_{1}
  \end{matrix} \\
  \begin{matrix}
   \vdots \\
  \mathbb{I}_{T - 1} \\
  {\check{W}}_{T0}^{\dagger}
  \end{matrix}
  \end{matrix} \right] \right\}
$}
\begin{equation}
  = \text{Tr}
  \begin{Bmatrix}
  \mathbb{I}_{0} & {\hat{W}}_{T0} & \cdots & {\hat{W}}_{20} & {\hat{W}}_{10} & \rho_{0} & {\hat{W}}_{10}^{\dagger} & {\hat{W}}_{20}^{\dagger} & \cdots & {\hat{W}}_{T0}^{\dagger} \\
  \mathbb{I}_{1} & \mathbb{I}_{1} & \cdots & \mathbb{I}_{1} & {\check{W}}_{10} & \rho_{1} & {\check{W}}_{10}^{\dagger} & \mathbb{I}_{1} & \cdots & \mathbb{I}_{1} \\
  \mathbb{I}_{2} & \mathbb{I}_{2} & \cdots & {\check{W}}_{20} & \mathbb{I}_{2} & \rho_{2} & \mathbb{I}_{2} & {\check{W}}_{20}^{\dagger} & \cdots & \mathbb{I}_{2} \\
   \vdots & \vdots & \iddots & \vdots & \vdots & \vdots & \vdots & \vdots & \ddots & \vdots \\
  O_{T} & {\check{W}}_{T0} & \cdots & \mathbb{I}_{T} & \mathbb{I}_{T} & \rho_{T} & \mathbb{I}_{T} & \mathbb{I}_{T} & \cdots & {\check{W}}_{T0}^{\dagger}
  \end{Bmatrix} = \left\langle O \right\rangle_{eq}.
\end{equation}
Finally, we arrive at the output expression of the equivalent model $\mathcal{Q}_{2}$, which coincides with that of the standard QRNN model. Therefore, the equivalence is proven.

\hfill $\square$  

To illustrate the equivalence more transparently, we present the proof for the simple case of $T = 3$. In this case, the output of the transitional model $\widetilde{\mathcal{Q}}_1$ is given by

\vspace{2ex}
\centerline{$\displaystyle 
    \left\langle {O} \right\rangle_{tr} = \text{Tr}\left\{ W_{30}\otimes\mathbb{I}_{\overline{30}}\left( \text{Tr}_{2}\left[ W_{20}\otimes\mathbb{I}_{\overline{20}}\left( \text{Tr}_{1}\left[ W_{10}\otimes\mathbb{I}_{\overline{10}}\rho_{all}W_{10}^{\dagger}\otimes\mathbb{I}_{\overline{10}} \right] \right)W_{20}\otimes\mathbb{I}_{\overline{20}} \right] \right)W_{3}^{\dagger}\otimes\mathbb{I}_{\overline{30}}\left( \mathbb{I}_{\overline{3}}\otimes O_{3} \right) \right\} $}
\vspace{2ex}
\centerline{$\displaystyle 
    = \text{Tr}\left\{ \left[ 
    \begin{matrix}
    W_{30} \\
    1_{1} \\
    1_{2} \\
    W_{30}
    \end{matrix}
    \left( \text{Tr}_{2}\left[ 
    \begin{matrix}
    W_{20} \\
    1_{1} \\d
    W_{20} \\
    \mathbb{I}_{3}
    \end{matrix}
    \left(\text{Tr}_{1}
        \begin{bmatrix}
        {W}_{10} & \rho_{0} & {W}_{10}^{\dagger} \\
        {W}_{10} & \rho_{1} & {W}_{10}^{\dagger} \\
        \mathbb{I}_{2} & \rho_{2} & \mathbb{I}_{2} \\
        \mathbb{I}_{3} & \rho_{3} & \mathbb{I}_{3}
        \end{bmatrix}
    \right)
    \begin{matrix}
    \begin{matrix}
    W_{20}^{\dagger} \\
    1_{1}
    \end{matrix} \\
    \begin{matrix}
    W_{20}^{\dagger} \\
    \mathbb{I}_{3}
    \end{matrix}
    \end{matrix} \right] \right)\begin{matrix}
    \begin{matrix}
    W_{30}^{\dagger} \\
    1_{1}
    \end{matrix} \\
    \begin{matrix}
    1_{2} \\
    W_{30}^{\dagger}
    \end{matrix}
    \end{matrix} \right]\begin{matrix}
    \begin{matrix}
    \mathbb{I}_{0} \\
    1_{1}
    \end{matrix} \\
    \begin{matrix}
    1_{2} \\
    O_{3}
    \end{matrix}
    \end{matrix} \right\}
$}
\vspace{2ex}
\centerline{$\displaystyle 
    = \text{Tr}\left\{ \left[ \begin{matrix}
    \begin{matrix}
    W_{30} \\
    1_{1}
    \end{matrix} \\
    \begin{matrix}
    1_{2} \\
    W_{30}
    \end{matrix}
    \end{matrix}\left( \sum_{\boldsymbol{j}}^{}{\begin{matrix}
    \begin{matrix}
    \mathbb{I}_{0} \\
    1_{1}
    \end{matrix} \\
    \begin{matrix}
    \left \langle \boldsymbol{j} \right |   \\
    \mathbb{I}_{3}
    \end{matrix}
    \end{matrix}\left[ \begin{matrix}
    \begin{matrix}
    W_{20} \\
    1_{1}
    \end{matrix} \\
    \begin{matrix}
    W_{20} \\
    \mathbb{I}_{3}
    \end{matrix}
    \end{matrix}\left( \sum_{\boldsymbol{i}}^{}{\begin{matrix}
    \begin{matrix}
    \mathbb{I}_{0} \\
    \left \langle \boldsymbol{i} \right |  
    \end{matrix} \\
    \begin{matrix}
    \mathbb{I}_{2} \\
    \mathbb{I}_{3}
    \end{matrix}
    \end{matrix}
        \begin{bmatrix}
        {W}_{10} & \rho_{0} & {W}_{10}^{\dagger} \\
        {W}_{10} & \rho_{1} & {W}_{10}^{\dagger} \\
        \mathbb{I}_{2} & \rho_{2} & \mathbb{I}_{2} \\
        \mathbb{I}_{3} & \rho_{3} & \mathbb{I}_{3}
        \end{bmatrix}
    \begin{matrix}
    \begin{matrix}
    \mathbb{I}_{0} \\
    \left | \boldsymbol{i} \right \rangle 
    \end{matrix} \\
    \begin{matrix}
    \mathbb{I}_{2} \\
    \mathbb{I}_{3}
    \end{matrix}
    \end{matrix}} \right)\begin{matrix}
    \begin{matrix}
    W_{20}^{\dagger} \\
    1_{1}
    \end{matrix} \\
    \begin{matrix}
    W_{20}^{\dagger} \\
    \mathbb{I}_{3}
    \end{matrix}
    \end{matrix} \right]\begin{matrix}
    \begin{matrix}
    \mathbb{I}_{0} \\
    1_{1}
    \end{matrix} \\
    \begin{matrix}
    \left | \boldsymbol{j} \right \rangle  \\
    \mathbb{I}_{3}
    \end{matrix}
    \end{matrix}} \right)\begin{matrix}
    \begin{matrix}
    W_{30}^{\dagger} \\
    1_{1}
    \end{matrix} \\
    \begin{matrix}
    1_{2} \\
    W_{30}^{\dagger}
    \end{matrix}
    \end{matrix} \right]\begin{matrix}
    \begin{matrix}
    \mathbb{I}_{0} \\
    1_{1}
    \end{matrix} \\
    \begin{matrix}
    1_{2} \\
    O_{3}
    \end{matrix}
    \end{matrix} \right\}
$}
\vspace{2ex}
\centerline{$\displaystyle 
    = \sum_{\boldsymbol{i}}^{}{\sum_{\boldsymbol{j}}^{}{\text{Tr}\left\{ \left[ \begin{matrix}
    \begin{matrix}
    W_{30} \\
    1_{1}
    \end{matrix} \\
    \begin{matrix}
    1_{2} \\
    W_{30}
    \end{matrix}
    \end{matrix}\left( \begin{matrix}
    \begin{matrix}
    \mathbb{I}_{0} \\
    1_{1}
    \end{matrix} \\
    \begin{matrix}
    \left \langle \boldsymbol{j} \right |   \\
    \mathbb{I}_{3}
    \end{matrix}
    \end{matrix}\left[ \begin{matrix}
    \begin{matrix}
    W_{20} \\
    1_{1}
    \end{matrix} \\
    \begin{matrix}
    W_{20} \\
    \mathbb{I}_{3}
    \end{matrix}
    \end{matrix}\left( \begin{matrix}
    \begin{matrix}
    \mathbb{I}_{0} \\
    \left \langle \boldsymbol{i} \right |  
    \end{matrix} \\
    \begin{matrix}
    \mathbb{I}_{2} \\
    \mathbb{I}_{3}
    \end{matrix}
    \end{matrix}
        \begin{bmatrix}
        {W}_{10} & \rho_{0} & {W}_{10}^{\dagger} \\
        {W}_{10} & \rho_{1} & {W}_{10}^{\dagger} \\
        \mathbb{I}_{2} & \rho_{2} & \mathbb{I}_{2} \\
        \mathbb{I}_{3} & \rho_{3} & \mathbb{I}_{3}
        \end{bmatrix}
    \begin{matrix}
    \begin{matrix}
    \mathbb{I}_{0} \\
    \left | \boldsymbol{i} \right \rangle 
    \end{matrix} \\
    \begin{matrix}
    \mathbb{I}_{2} \\
    \mathbb{I}_{3}
    \end{matrix}
    \end{matrix} \right)\begin{matrix}
    \begin{matrix}
    W_{20}^{\dagger} \\
    1_{1}
    \end{matrix} \\
    \begin{matrix}
    W_{20}^{\dagger} \\
    \mathbb{I}_{3}
    \end{matrix}
    \end{matrix} \right]\begin{matrix}
    \begin{matrix}
    \mathbb{I}_{0} \\
    1_{1}
    \end{matrix} \\
    \begin{matrix}
    \left | \boldsymbol{j} \right \rangle  \\
    \mathbb{I}_{3}
    \end{matrix}
    \end{matrix} \right)\begin{matrix}
    \begin{matrix}
    W_{30}^{\dagger} \\
    1_{1}
    \end{matrix} \\
    \begin{matrix}
    1_{2} \\
    W_{30}^{\dagger}
    \end{matrix}
    \end{matrix} \right]\begin{matrix}
    \begin{matrix}
    \mathbb{I}_{0} \\
    1_{1}
    \end{matrix} \\
    \begin{matrix}
    1_{2} \\
    O_{3}
    \end{matrix}
    \end{matrix} \right\}}}
$}
\vspace{2ex}
\centerline{$\displaystyle 
    = \sum_{\boldsymbol{i}}^{}{\sum_{\boldsymbol{j}}^{}{\text{Tr}\left\{ \left[ \begin{matrix}
    \begin{matrix}
    W_{30} \\
    \left \langle \boldsymbol{i} \right | 
    \end{matrix} \\
    \begin{matrix}
    \left \langle \boldsymbol{j} \right |  \\
    W_{30}
    \end{matrix}
    \end{matrix}\left( \begin{matrix}
    \begin{matrix}
    \mathbb{I}_{0} \\
    \mathbb{I}_{1}
    \end{matrix} \\
    \begin{matrix}
    \mathbb{I}_{2} \\
    \mathbb{I}_{3}
    \end{matrix}
    \end{matrix}\left[ \begin{matrix}
    \begin{matrix}
    W_{20} \\
    \mathbb{I}_{1}
    \end{matrix} \\
    \begin{matrix}
    W_{20} \\
    \mathbb{I}_{3}
    \end{matrix}
    \end{matrix}\left( \begin{matrix}
    \begin{matrix}
    \mathbb{I}_{0} \\
    \mathbb{I}_{1}
    \end{matrix} \\
    \begin{matrix}
    \mathbb{I}_{2} \\
    \mathbb{I}_{3}
    \end{matrix}
    \end{matrix}
        \begin{bmatrix}
        {W}_{10} & \rho_{0} & {W}_{10}^{\dagger} \\
        {W}_{10} & \rho_{1} & {W}_{10}^{\dagger} \\
        \mathbb{I}_{2} & \rho_{2} & \mathbb{I}_{2} \\
        \mathbb{I}_{3} & \rho_{3} & \mathbb{I}_{3}
        \end{bmatrix}
    \begin{matrix}
    \begin{matrix}
    \mathbb{I}_{0} \\
    \mathbb{I}_{1}
    \end{matrix} \\
    \begin{matrix}
    \mathbb{I}_{2} \\
    \mathbb{I}_{3}
    \end{matrix}
    \end{matrix} \right)\begin{matrix}
    \begin{matrix}
    W_{20}^{\dagger} \\
    \mathbb{I}_{1}
    \end{matrix} \\
    \begin{matrix}
    W_{20}^{\dagger} \\
    \mathbb{I}_{3}
    \end{matrix}
    \end{matrix} \right]\begin{matrix}
    \begin{matrix}
    \mathbb{I}_{0} \\
    \mathbb{I}_{1}
    \end{matrix} \\
    \begin{matrix}
    \mathbb{I}_{2} \\
    \mathbb{I}_{3}
    \end{matrix}
    \end{matrix} \right)\begin{matrix}
    \begin{matrix}
    W_{30}^{\dagger} \\
    \mathbb{I}_{1}
    \end{matrix} \\
    \begin{matrix}
    \mathbb{I}_{2} \\
    W_{30}^{\dagger}
    \end{matrix}
    \end{matrix} \right]\begin{matrix}
    \begin{matrix}
    \mathbb{I}_{0} \\
    \left | \boldsymbol{i} \right \rangle
    \end{matrix} \\
    \begin{matrix}
    \left | \boldsymbol{j} \right \rangle \\
    O_{3}
    \end{matrix}
    \end{matrix} \right\}}}
$}
\vspace{2ex}
\centerline{$\displaystyle 
    = \sum_{\boldsymbol{i}}^{}{\sum_{\boldsymbol{j}}^{}{\text{Tr}\left\{ \begin{matrix}
    \begin{matrix}
    \mathbb{I}_{0} \\
    \left | \boldsymbol{i} \right \rangle
    \end{matrix} \\
    \begin{matrix}
    \left | \boldsymbol{j} \right \rangle \\
    O_{3}
    \end{matrix}
    \end{matrix}\begin{matrix}
    \begin{matrix}
    W_{30} \\
    \left \langle \boldsymbol{i} \right | 
    \end{matrix} \\
    \begin{matrix}
    \left \langle \boldsymbol{j} \right |  \\
    W_{30}
    \end{matrix}
    \end{matrix}\left( \begin{matrix}
    \begin{matrix}
    \mathbb{I}_{0} \\
    \mathbb{I}_{1}
    \end{matrix} \\
    \begin{matrix}
    \mathbb{I}_{2} \\
    \mathbb{I}_{3}
    \end{matrix}
    \end{matrix}\left[ \begin{matrix}
    \begin{matrix}
    W_{20} \\
    \mathbb{I}_{1}
    \end{matrix} \\
    \begin{matrix}
    W_{20} \\
    \mathbb{I}_{3}
    \end{matrix}
    \end{matrix}\left( \begin{matrix}
    \begin{matrix}
    \mathbb{I}_{0} \\
    \mathbb{I}_{1}
    \end{matrix} \\
    \begin{matrix}
    \mathbb{I}_{2} \\
    \mathbb{I}_{3}
    \end{matrix}
    \end{matrix}
        \begin{bmatrix}
        {W}_{10} & \rho_{0} & {W}_{10}^{\dagger} \\
        {W}_{10} & \rho_{1} & {W}_{10}^{\dagger} \\
        \mathbb{I}_{2} & \rho_{2} & \mathbb{I}_{2} \\
        \mathbb{I}_{3} & \rho_{3} & \mathbb{I}_{3}
        \end{bmatrix}
    \begin{matrix}
    \begin{matrix}
    \mathbb{I}_{0} \\
    \mathbb{I}_{1}
    \end{matrix} \\
    \begin{matrix}
    \mathbb{I}_{2} \\
    \mathbb{I}_{3}
    \end{matrix}
    \end{matrix} \right)\begin{matrix}
    \begin{matrix}
    W_{20}^{\dagger} \\
    \mathbb{I}_{1}
    \end{matrix} \\
    \begin{matrix}
    W_{20}^{\dagger} \\
    \mathbb{I}_{3}
    \end{matrix}
    \end{matrix} \right]\begin{matrix}
    \begin{matrix}
    \mathbb{I}_{0} \\
    \mathbb{I}_{1}
    \end{matrix} \\
    \begin{matrix}
    \mathbb{I}_{2} \\
    \mathbb{I}_{3}
    \end{matrix}
    \end{matrix} \right)\begin{matrix}
    \begin{matrix}
    W_{30}^{\dagger} \\
    \mathbb{I}_{1}
    \end{matrix} \\
    \begin{matrix}
    \mathbb{I}_{2} \\
    W_{30}^{\dagger}
    \end{matrix}
    \end{matrix} \right\}}}
$}
\begin{equation}
    = \text{Tr}\begin{Bmatrix}
    \mathbb{I}_{0} & W_{30} & W_{20} & W_{10} & \rho_{0} & W_{10}^{\dagger} & W_{20}^{\dagger} & W_{30}^{\dagger} \\
    \mathbb{I}_{1} & \mathbb{I}_{1} & \mathbb{I}_{1} & W_{10} & \rho_{1} & W_{10}^{\dagger} & \mathbb{I}_{1} & \mathbb{I}_{1} \\
    \mathbb{I}_{2} & \mathbb{I}_{2} & W_{20} & \mathbb{I}_{2} & \rho_{2} & \mathbb{I}_{2} & W_{20}^{\dagger} & \mathbb{I}_{2} \\
    O_{3} & W_{30} & \mathbb{I}_{3} & \mathbb{I}_{3} & \rho_{3} & \mathbb{I}_{3} & \mathbb{I}_{3} & W_{30}^{\dagger}
    \end{Bmatrix} = \left\langle O \right\rangle_{eq}.
\end{equation}
This completes the proof of equivalence for the case $T = 3$. For notational simplicity, we have omitted the explicit marking of the \emph{restricted representations} above.

\subsection{\label{sec:level2}2. Formalism of the QRNN Model}

\begin{figure}[h]
  \centering
  \includegraphics[width=4.5in]{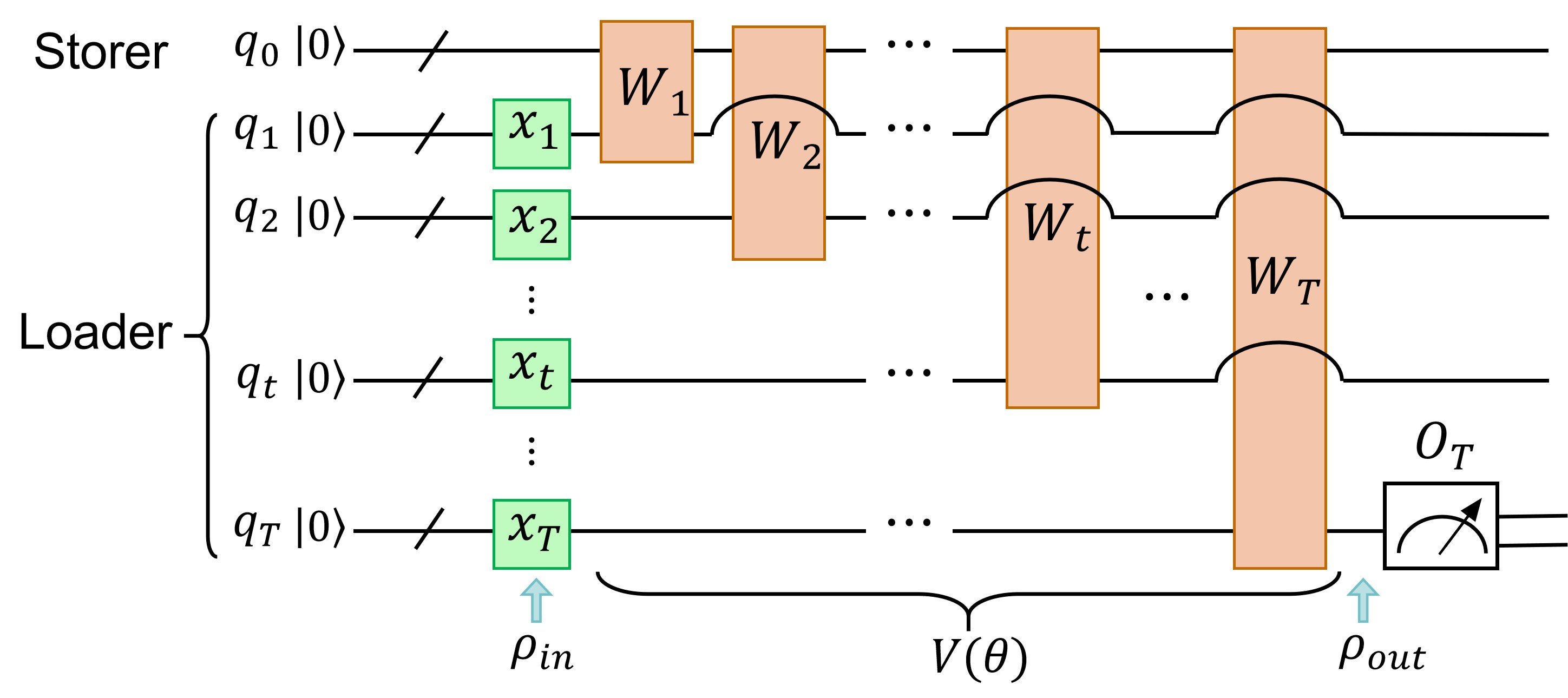}
  \caption{Equivalent model $\mathcal{Q}_{2}$ of QRNNs}
  \label{FB2}
\end{figure}

Here we present the formalism used to describe the QRNN model based on the architecture shown in Fig.~\ref{FB1}(c), which serves as the starting point for the subsequent theoretical analysis. For clarity, we reproduce the corresponding circuit diagram in Fig.~\ref{FB2} for the discussion below.

We first introduce a symbolic representation of the basic modules in the model $\mathcal{Q}_{2}$: All qubits are initially prepared in the state $\left | 0 \right \rangle$, such that the initial state of each register is given by $\rho_{t} = {(\left | 0 \right \rangle \left \langle 0 \right |)}^{\otimes m}$. The green modules shown in Fig.~\ref{FB2} can be combined into a single unitary operator,
\begin{equation}
    U_{T} = \left[ \otimes_{t = 1}^{T}U_{\text{in}}(f(x_{t})) \right] \otimes \mathbb{I}_{0},
\end{equation}
where $f(x_{t})$ processes the input data $x_{t}$ such that $f(x_{t}) \in [0,2\pi]$. The operator $U_{\text{in}}(\cdot)$ corresponds to the green module in the circuit and encodes the processed data $f(x_{t})$ into the quantum state. After the data input stage, the circuit state becomes $\rho_{\text{in}} = U_{T}\left | \boldsymbol{0} \right \rangle \left \langle \boldsymbol{0} \right | U_{T}^{\dagger}$, which subsequently evolves under the parameterized quantum circuit (PQC)
\begin{equation}\label{B10}
    V(\boldsymbol{\theta}) = \prod_{t = 1}^{T}W_{t} = \prod_{t = 1}^{T}{\left( W_{t0}\otimes\mathbb{I}_{\overline{t0}} \right)},
\end{equation}
where $W_{t0}$ denotes the parameterized unitary at time step $t$, acting non-trivially only on subsystems $q_{0}$ and $q_{t}$. Accordingly, $W_{t} = W_{t0}\otimes\mathbb{I}_{\overline{t0}}$ represents the corresponding operator acting on the full system.

For the case of shared parameters, the operators $W_{t0}$ share the same circuit structure and parameters across different time steps, and can be expressed as
\begin{equation}\label{B11}
    W_{t0}(\boldsymbol{\theta}) = G_{\zeta}(\theta^{\zeta})\cdots G_{v}(\theta^{\nu})\cdots G_{1}(\theta^{1}).
\end{equation}
In contrast, for the case of non-sharing parameters, we allow $W_{t0}$ to have different circuit structures and independent parameters associated with each time step $t$, namely
\begin{equation}\label{B12}
    W_{t0}(\boldsymbol{\theta}) = G_{\zeta_{t0}}(\theta_{t0}^{\zeta_{t0}})\cdots G_{v}(\theta_{t0}^{\nu})\cdots G_{1}(\theta_{t0}^{1}),
\end{equation}
where $\theta_{t0}^{\nu}$ denotes a continuously varying parameter, and the total number of parameters $\zeta_{t0}$ depends on the specific structure of $W_{t0}$. We adopt the same assumptions on PQCs as in Ref.~\cite{cerezo2021cost}. Specifically, each gate takes the form $G_{v}(\theta_{t0}^{\nu}) = R_{\nu}(\theta_{t0}^{\nu})Q_{v}$, where $Q_{v}$ is a parameter-free quantum gate and $R_{\nu}(\theta_{t0}^{\nu}) = \exp( - i\theta_{t0}^{\nu}\sigma_{\nu}/2)$. Here, $\sigma_{\nu}:\mathcal{H}_{t0} \rightarrow \mathcal{H}_{t0}$ is a Pauli operator acting on the Hilbert space $\mathcal{H}_{t0}$, and acts non-trivially on a single qubit $j$ within $q_{0}$ or $q_{t}$, i.e., $\sigma_{\nu} = {(\sigma_{\nu})}_{j} \otimes \mathbb{I}_{\overline{j}}$, where ${(\sigma_{\nu})}_{j} \in \{\sigma_{x},\sigma_{y},\sigma_{z}\}$.

After the variational quantum circuit $V(\boldsymbol{\theta})$, the quantum state evolves to $\rho_{\text{out}} = V(\boldsymbol{\theta})\rho_{\text{in}}{V(\boldsymbol{\theta})}^{\dagger}$. By measuring the qubit group $q_{T}$, we obtain the expectation value of the operator $O_T$, given by
\begin{equation}\label{B13}
    C 
    \equiv \langle O \rangle 
    = \text{Tr}\left[ O V(\boldsymbol{\theta})\rho_{\text{in}}{V(\boldsymbol{\theta})}^{\dagger} \right]
    = \text{Tr}\left[ O_T\otimes\mathbb{I}_{\overline{T}} \, V(\boldsymbol{\theta})\rho_{\text{in}}{V(\boldsymbol{\theta})}^{\dagger} \right],
\end{equation}
which is typically used as the predicted value of the model. The loss function defined over $N$ data points $\{\boldsymbol{x}^{(1)}, \ldots, \boldsymbol{x}^{(N)}\}$ is given by
\begin{equation}
  \begin{split}
    \mathcal{L}_{N} 
    = \frac{1}{N} \sum_{i=1}^{N} \mathcal{L}(\langle O \rangle^{(i)}) 
    = \frac{1}{N} \sum_{i=1}^{N} \mathcal{L} \left( \text{Tr}\left[ O V(\boldsymbol{\theta})\rho_{\text{in}}^{(i)}{V(\boldsymbol{\theta})}^{\dagger} \right] \right),
  \end{split}
\end{equation}
where $\mathcal{L}(\cdot)$ denotes a specific loss function, such as the mean squared error. In practice, we often choose $O_T = y^{(i)}\mathbb{I}_T - M_T$, where $y^{(i)}$ denotes the corresponding label and $M_T$ is a measurement operator. When $M_T$ takes the form $c\left | \boldsymbol{0} \right \rangle \left \langle \boldsymbol{0} \right |_T$, $O_T$ corresponds to a global observable; when it takes the form $\frac{1}{m}\sum_{j=1}^m c_j \left | 0 \right \rangle \left \langle 0 \right |_j \otimes \mathbb{I}_j$, it corresponds to a local observable, where $c$ and $c_j$ are real coefficients. A detailed discussion of these observables is provided at the end of Section~D.

To optimize the loss function, the model parameters are typically updated by evaluating gradients. However, when the gradient magnitudes become too small, parameter updates are negligible, rendering the training process ineffective. Consequently, the scale of the loss function gradient plays a crucial role in the optimization process. The gradient of $\mathcal{L}$ can be decomposed into two factors: $\partial \mathcal{L} / \partial \langle O \rangle$ and $\partial \langle O \rangle / \partial \theta_t^{\nu}$. The first factor depends solely on the specific form of the loss function and can be appropriately bounded by design. Therefore, our primary focus lies on the second factor, which involves the expectation value of the observable defined in Eq.~(\ref{B13}). Without loss of generality, we directly refer to $C=\langle O \rangle$ as the loss function in the following discussions.

As will be shown later, when the parameter matrices follow the Haar distribution, the expected gradient vanishes. In this regime, if the variance of the gradient also approaches zero, the loss landscape develops a flat region known as a barren plateau.

\section{\label{C}Section C. Gradient Expectation of QRNNs without Parameter Sharing}

In this section, we compute the expectation of the partial derivative of the QRNN loss function with respect to the parameter $\theta_{t0}^{\nu}$ under the assumption that all parameters are independent, i.e., no parameter sharing is employed. Based on the discussion in Section~B, we start from the measurement outcome
\begin{equation}
    C = \text{Tr}\left[ OV(\boldsymbol{\theta})\rho_{\text{in}}V(\boldsymbol{\theta})^{\dagger} \right],
\end{equation}
and proceed to evaluate the partial derivative of the loss function with respect to $\theta_{t0}^{\nu}$. We note that the parameter $\theta_{t0}^{\nu}$ appears only in the corresponding parameter matrix $W_{t0}$. For notational simplicity, we hereafter denote this matrix $W_{t0}$ by $W$ and the parameter $\theta_{t0}^{\nu}$ by $\theta^{\nu}$. With respect to the parameter $\theta^{\nu}$, the matrix $W$ can be decomposed into two parts as $W = W_{B}W_{A}$, where
\begin{equation}
    W_{A} = \prod_{\mu = 1}^{v - 1}{G_{\mu}(\theta_{t0}^{\mu})}, \quad \quad \quad
    W_{B} = \prod_{\mu = v}^{\zeta_{t0}}{G_{\mu}(\theta_{t0}^{\mu})}.
\end{equation}
From this decomposition, we obtain the first-order partial derivatives of $W$ and $W^{\dagger}$ with respect to the parameter $\theta^{\nu}$ as
\begin{equation}
    \partial_{\nu}W = \frac{\partial W}{\partial\theta^{\nu}} = W_{B}\left( - \frac{i}{2}\sigma_{\nu}\right)W_{A}, \quad
    \partial_{\nu}W^{\dagger} = \frac{\partial W^{\dagger}}{\partial\theta^{\nu}} = W_{A}^{\dagger}\left(\frac{i}{2}\sigma_{\nu}\right)W_{B}^{\dagger}.
\end{equation}
Furthermore, the full parameterized quantum circuit $V(\boldsymbol{\theta})$ can be expressed as
\begin{equation}
    V(\boldsymbol{\theta}) = V_{\mathcal{R}}(W_{t0} \otimes \mathbb{I}_{\overline{t0}})V_{\mathcal{L}},
\end{equation}
where $V_{\mathcal{L}}$ denotes the collection of parameter matrices located to the left of $W_{t0}$, and $V_{\mathcal{R}}$ denotes those located to its right. The first-order partial derivative of the measurement outcome $C$ with respect to the parameter $\theta^{\nu}$ in $W_{t0}$ is then given by
\begin{equation}\label{C5}
    \begin{split}
        \frac{\partial C}{\partial\theta^{\nu}} 
        =& \text{Tr}\left[ OV_{\mathcal{R}}\left(\frac{\partial W}{\partial\theta^{\nu}} \otimes \mathbb{I}_{\overline{t0}}\right)V_{\mathcal{L}}\rho_{\text{in}}V_{\mathcal{L}}^{\dagger}(W^{\dagger} \otimes \mathbb{I}_{\overline{t0}})V_{\mathcal{R}}^{\dagger} \right] \\
        &+ \text{Tr}\left[ OV_{\mathcal{R}}(W \otimes \mathbb{I}_{\overline{t0}})V_{\mathcal{L}}\rho_{\text{in}}V_{\mathcal{L}}^{\dagger}\left(\frac{\partial W^{\dagger}}{\partial\theta^{\nu}} \otimes \mathbb{I}_{\overline{t0}}\right)V_{\mathcal{R}}^{\dagger} \right] \\
        =& \text{Tr}\left[ OV_{\mathcal{R}}\left(W_{B}\left( - \frac{i}{2}\sigma_{\nu}\right)W_{A} \otimes \mathbb{I}_{\overline{t0}}\right)V_{\mathcal{L}}\rho_{\text{in}}V_{\mathcal{L}}^{\dagger}(W_{A}^{\dagger}W_{B}^{\dagger} \otimes \mathbb{I}_{\overline{t0}})V_{\mathcal{R}}^{\dagger} \right] \\
        &+ \text{Tr}\left[ OV_{\mathcal{R}}(W_{B}W_{A} \otimes \mathbb{I}_{\overline{t0}})V_{\mathcal{L}}\rho_{\text{in}}V_{\mathcal{L}}^{\dagger}\left(W_{A}^{\dagger}\left(\frac{i}{2}\sigma_{\nu}\right)W_{B}^{\dagger} \otimes \mathbb{I}_{\overline{t0}}\right)V_{\mathcal{R}}^{\dagger} \right] \\
        =& -\text{Tr}\left\{ (W_{B}^{\dagger} \otimes \mathbb{I}_{\overline{t0}})V_{\mathcal{R}}^{\dagger}OV_{\mathcal{R}}(W_{B} \otimes \mathbb{I}_{\overline{t0}})\left[\frac{i}{2}\sigma_{\nu} \otimes \mathbb{I}_{\overline{t0}}, (W_{A} \otimes \mathbb{I}_{\overline{t0}})V_{\mathcal{L}}\rho_{\text{in}}V_{\mathcal{L}}^{\dagger}(W_{A}^{\dagger} \otimes \mathbb{I}_{\overline{t0}})\right] \right\}.
    \end{split}
\end{equation}
To further evaluate the expectation $\left\langle \partial_{\nu}C \right\rangle_{V}$, we consider three distinct cases: 

(1) only $\{W_{A}\}$ form a $1$-design;  

(2) only $\{W_{B}\}$ form a $1$-design;  

(3) both $\{W_{A}\}$ and $\{W_{B}\}$ form a $1$-design.  

We first consider case (1). Since the matrices $W_{A}$, $W_{B}$, $V_{\mathcal{R}}$, and $V_{\mathcal{L}}$ are independent, the expectation $\left\langle \partial_{\nu}C \right\rangle_{V}$ can be evaluated sequentially over the four modules. In particular, we have $\left\langle \partial_{\nu}C \right\rangle_{V} = \left\langle \left\langle \partial_{\nu}C \right\rangle_{W_{A}} \right\rangle_{W_{B},V_{\mathcal{R}},V_{\mathcal{L}}}.$ According to the definition of a $1$-design in Eq.~(\ref{A6}), we obtain
\begin{equation}
\begin{split}
\left\langle \partial_{\nu}C \right\rangle_{W_{A}}
&= \frac{i}{2}\text{Tr}\left\{ (W_{B}^{\dagger} \otimes \mathbb{I}_{\overline{t0}})V_{\mathcal{R}}^{\dagger}OV_{\mathcal{R}}(W_{B} \otimes \mathbb{I}_{\overline{t0}})
\left[\sigma_{\nu} \otimes \mathbb{I}_{\overline{t0}}, 
\int d\mu(W_{A})(W_{A} \otimes \mathbb{I}_{\overline{t0}})
V_{\mathcal{L}}\rho_{\text{in}}V_{\mathcal{L}}^{\dagger}
(W_{A}^{\dagger} \otimes \mathbb{I}_{\overline{t0}})
\right] \right\} \\
&= \frac{i}{2}\text{Tr}\left\{ (W_{B}^{\dagger} \otimes \mathbb{I}_{\overline{t0}})V_{\mathcal{R}}^{\dagger}OV_{\mathcal{R}}(W_{B} \otimes \mathbb{I}_{\overline{t0}})
\left[\sigma_{\nu} \otimes \mathbb{I}_{\overline{t0}},
\frac{\text{Tr}_{t0}\!\left[ V_{\mathcal{L}}\rho_{\text{in}}V_{\mathcal{L}}^{\dagger}\right]
\otimes \mathbb{I}_{t0}}{2^{2m}}
\right] \right\} = 0,
\end{split}
\end{equation}
where we have used \textbf{Lemma 4} and the fact that the commutator
$\left[\begin{matrix}A \\\otimes \\\mathbb{I}\end{matrix}, \begin{matrix}\mathbb{I} \\\otimes \\B\end{matrix}\right] = 0$. We now turn to case (2), where only $\{W_{B}\}$ form a $1$-design. In this case, we have
\begin{equation}
\left\langle \partial_{\nu}C \right\rangle_{W_{B}}
= \frac{i}{2}\int d\mu(W_{B})\text{Tr}\left\{
(W_{B}^{\dagger} \otimes \mathbb{I})V_{\mathcal{R}}^{\dagger}OV_{\mathcal{R}}
(W_{B} \otimes \mathbb{I})\, C_A
\right\}
= \frac{1}{2^{2m}}\text{Tr}\left\{
\text{Tr}_{t0}\!\left[ V_{\mathcal{R}}^{\dagger}OV_{\mathcal{R}}\right]
\text{Tr}_{t0}\!\left[ C_A\right]
\right\},
\end{equation}
where \textbf{Lemma 4} is applied, and the operator $C_A$ is defined as
$C_A = \left[\sigma_{\nu} \otimes \mathbb{I}, (W_{A} \otimes \mathbb{I})\rho^{(\mathcal{L})} (W_{A}^{\dagger} \otimes \mathbb{I})
\right],$ with $\rho^{(\mathcal{L})} = V_{\mathcal{L}}\rho_{\text{in}}V_{\mathcal{L}}^{\dagger}$. For notational simplicity, we abbreviate $\mathbb{I}_{\overline{t0}}$ as $\mathbb{I}$. 

The density operator $\rho^{(\mathcal{L})}$ admits the spectral decomposition
\begin{equation}
\rho^{(\mathcal{L})}
= \sum_{\boldsymbol{i},\boldsymbol{j}} p_{i,j}
\left| \boldsymbol{i} \right\rangle \left| \boldsymbol{j} \right\rangle
\left\langle \boldsymbol{i} \right|
\left\langle \boldsymbol{j} \right|
= \sum_{\boldsymbol{i},\boldsymbol{j}} p_{i,j}
\begin{matrix}
{\left| \boldsymbol{i} \right\rangle \left\langle \boldsymbol{i} \right|}_{t0} \\
{\left| \boldsymbol{j} \right\rangle \left\langle \boldsymbol{j} \right|}_{\overline{t0}}
\end{matrix},
\end{equation}
where $\left| \boldsymbol{i} \right\rangle$ and $\left| \boldsymbol{j} \right\rangle$ denote computational basis states of the subsystems $\{q_{0},q_{t}\}$ and its complement, respectively. Consequently, $\text{Tr}_{t0}[C_A]$ can be evaluated as
\begin{equation}
    \begin{split}
        \text{Tr}_{t0}\left[ C_A\right] 
        &= \sum_{\boldsymbol{i},\boldsymbol{j}} p_{\boldsymbol{i},\boldsymbol{j}}\left\{ \text{Tr}_{t0}\left[ \begin{matrix}
        \sigma_{\nu} \\
        \otimes \\
        \mathbb{I}
        \end{matrix} \begin{matrix}
        W_{A} \\
        \otimes \\
        \mathbb{I}
        \end{matrix} \begin{matrix}
        \left| \boldsymbol{i} \right\rangle \left\langle \boldsymbol{i} \right| \\
        \otimes \\
        \left| \boldsymbol{j} \right\rangle \left\langle \boldsymbol{j} \right| 
        \end{matrix} \begin{matrix}
        W_{A}^{\dagger} \\
        \otimes \\
        \mathbb{I}
        \end{matrix} \right] - \text{Tr}_{t0}\left[ \begin{matrix}
        W_{A} \\
        \otimes \\
        \mathbb{I}
        \end{matrix} \begin{matrix}
        \left| \boldsymbol{i} \right\rangle \left\langle \boldsymbol{i} \right| \\
        \otimes \\
        \left| \boldsymbol{j} \right\rangle \left\langle \boldsymbol{j} \right| 
        \end{matrix} \begin{matrix}
        W_{A}^{\dagger} \\
        \otimes \\
        \mathbb{I}
        \end{matrix} \begin{matrix}
        \sigma_{\nu} \\
        \otimes \\
        \mathbb{I}
        \end{matrix} \right] \right\}\\
        &= \sum_{\boldsymbol{i},\boldsymbol{j}} p_{\boldsymbol{i},\boldsymbol{j}}\left| \boldsymbol{j} \right\rangle \left\langle \boldsymbol{j} \right| \text{Tr}\left[ \sigma_{\nu}W_{A}\left| \boldsymbol{i} \right\rangle \left\langle \boldsymbol{i} \right| W_{A}^{\dagger} - W_{A}\left| \boldsymbol{i} \right\rangle \left\langle \boldsymbol{i} \right| W_{A}^{\dagger}\sigma_{\nu} \right] = 0,
    \end{split}
\end{equation}
As a result,
\begin{equation}
    \left\langle \partial_{\nu}C \right\rangle_{W_{B}}
    = \frac{1}{2^{2m}}\text{Tr}\left\{
    \text{Tr}_{t0}\!\left[ V_{\mathcal{R}}^{\dagger}OV_{\mathcal{R}}\right]
    \text{Tr}_{t0}\!\left[ C\right]
    \right\}
    = 0.
\end{equation}
In summary, as long as either $\{W_{A}\}$ or $\{W_{B}\}$ form a $1$-design, we have
$\left\langle \partial_{\nu}C \right\rangle = 0$.
This result also holds trivially when both $W_{A}$ and $W_{B}$ form a $1$-design.

\section{\label{D}Section D. Gradient Variance of QRNNs without Parameter Sharing}
This section evaluates the variance of the partial derivative $\partial_\nu C$ of the QRNN loss function under the assumption that all parameters are independent (i.e., without parameter sharing). As shown in Section C, the expectation value of the gradient vanishes. Consequently, the gradient variance reduces to the expectation of the squared gradient,
$\left\langle \left( \partial_{\nu}C - \left\langle \partial_{\nu}C \right\rangle \right)^{2} \right\rangle_{V(\boldsymbol{\theta})}
= \left\langle \left( \partial_{\nu}C \right)^{2} \right\rangle_{V(\boldsymbol{\theta})}$.
The primary objective of this section is therefore to evaluate
$\left\langle \left( \partial_{\nu}C \right)^{2} \right\rangle_{V(\boldsymbol{\theta})}$
by performing integrations over the modules $W_{A}$, $W_{B}$, $V_{\mathcal{R}}$, and $V_{\mathcal{L}}$, thereby establishing Theorem 1 in the main text.

According to Ref.~\cite{cerezo2021cost} and Eq.~(\ref{C5}), the variance of the gradient is given by
\begin{equation}\label{DE1}
    \left\langle \left( \partial_{\nu}C \right)^{2} \right\rangle_{V(\boldsymbol{\theta})} = \frac{2^{2m - 1}\text{Tr}\left[ \sigma_{\nu}^{2} \right]}{{(2^{4m} - 1)}^{2}}
    \sum_{\substack{\boldsymbol{p},\boldsymbol{q} \\
    \boldsymbol{p'},\boldsymbol{q'}}}
    {\left\langle {\Delta\Omega}_{\boldsymbol{pq}}^{\boldsymbol{p'q'}} \right\rangle_{V_{\mathcal{R}}}\left\langle {\Delta\Psi}_{\boldsymbol{pq}}^{\boldsymbol{p'q'}} \right\rangle_{V_{\mathcal{L}}}},
\end{equation}
where
\begin{equation}\label{ED2}
    \begin{split}
        {\Delta\Omega}_{\boldsymbol{pq}}^{\boldsymbol{p'q'}} 
        &= \text{Tr}\left[ \Omega_{\boldsymbol{qp}}\Omega_{\boldsymbol{q'p'}} \right] - \frac{\text{Tr}\left[ \Omega_{\boldsymbol{qp}} \right] \text{Tr}\left[ \Omega_{\boldsymbol{q'p'}} \right]}{2^{2m}},\\
        {\Delta\Psi}_{\boldsymbol{pq}}^{\boldsymbol{p'q'}} 
        &= \text{Tr}\left[ \Psi_{\boldsymbol{pq}}\Psi_{\boldsymbol{p'q'}} \right] - \frac{\text{Tr}\left[ \Psi_{\boldsymbol{pq}} \right] \text{Tr}\left[ \Psi_{\boldsymbol{p'q'}} \right]}{2^{2m}},
    \end{split}
\end{equation}
and
\begin{equation}
    \begin{split}
        \Omega_{\boldsymbol{qp}} 
        &= {\text{Tr}}_{\overline{t0}}\left[ (\left | \boldsymbol{p} \right \rangle \left \langle \boldsymbol{q} \right | \otimes \mathbb{I}_{t0})V_{\mathcal{R}}^{\dagger}OV_{\mathcal{R}} \right],\\
        \Psi_{\boldsymbol{pq}} 
        &= {\text{Tr}}_{\overline{t0}}\left[ (\left | \boldsymbol{q} \right \rangle \left \langle \boldsymbol{p} \right | \otimes \mathbb{I}_{t0})V_{\mathcal{L}}\rho_{in}V_{\mathcal{L}}^{\dagger} \right].
    \end{split}
\end{equation}
Note that Eq.~(\ref{DE1}) involves integration over both matrices $W_{A}$ and $W_{B}$, which each form a unitary $2$-design.

\subsection{\label{sec:level2}1. Calculation for $V_{\mathcal{R}}$}
In this subsection, we evaluate the contribution of $V_{\mathcal{R}}$ to Eq.~(\ref{DE1}). In the QRNN architecture, only the $q_{T}$ register is measured at the end of the circuit, such that
\begin{equation}
    O = O_{T} \otimes \mathbb{I}_{\overline{T}}.
\end{equation}

\begin{figure}[h]
        \centering
        \includegraphics[width=7in]{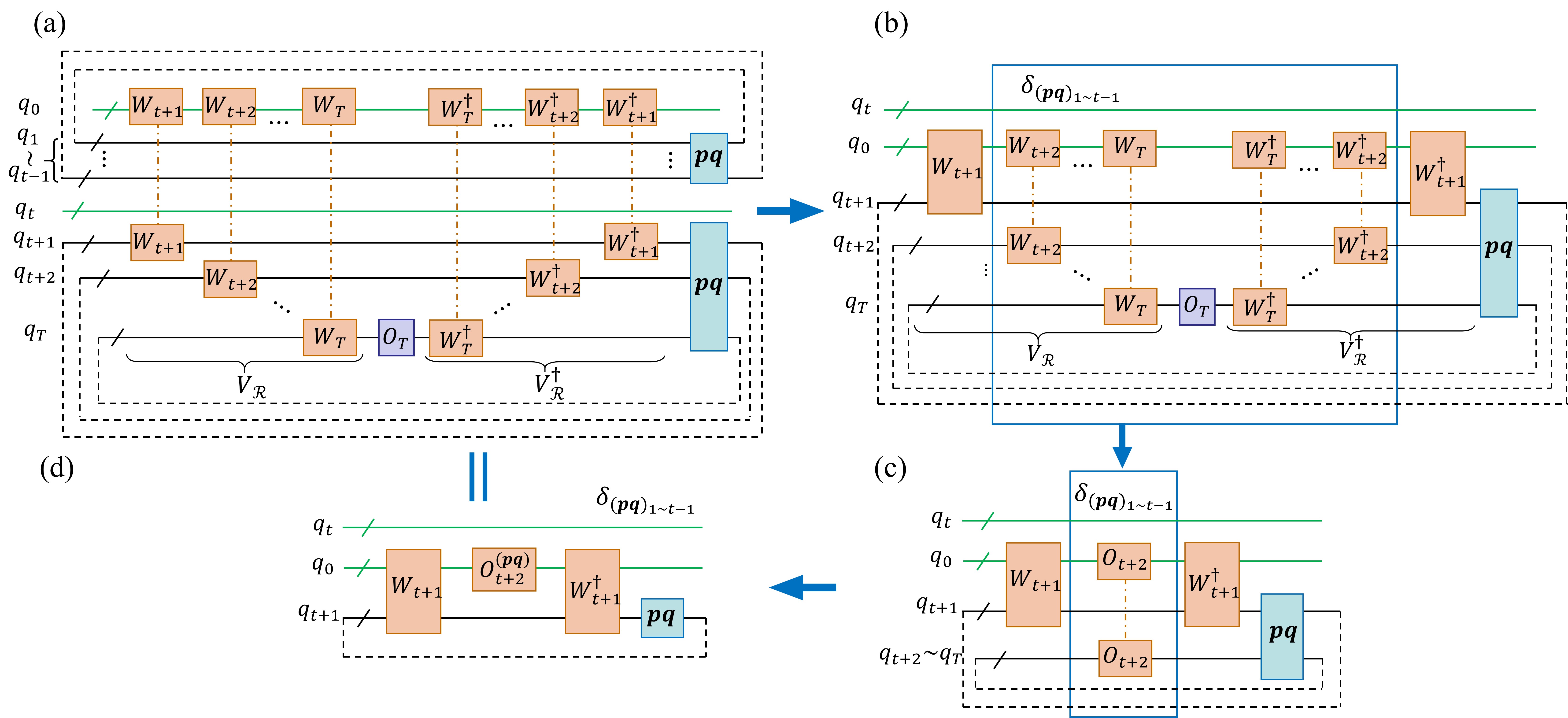}
        \caption{\label{fig:2} Reduction of the operator $\Omega_{\bm{qp}}$. (a) Tensor network representation of $\Omega_{\boldsymbol{q}\boldsymbol{p}}=\operatorname{Tr}_{\bar{t0}}\left[\left(|\boldsymbol{p}\rangle\langle\boldsymbol{q}| \otimes \mathbb{I}_{t 0}\right) V_{\mathcal{R}}^{\dagger} O_T V_{\mathcal{R}}\right]$. (b) Partial trace of $\left|\boldsymbol{p}\right\rangle\left\langle\boldsymbol{q}\right|$ over the subsystems $q_{1}\sim q_{t-1}$, yielding the factor $\delta_{\boldsymbol{(pq)}_{1\sim t-1}}$. (c) All operations outside the time step $t+1$ are grouped into a single effective operator $O_{t+2}$ acting on registers $q_0$, $q_t$, and $q_{t+2}\sim q_T$. (d) After tracing out the subsystems $q_{t+2}\sim q_T$, the remaining operator reduces to an equivalent operator $O_{t+2}^{\boldsymbol{(pq)}}$ acting solely on register $q_0$.}
        \label{FD1}
\end{figure}
\noindent{}The operator $V_{\mathcal{R}}$ consists of $(T - t)$ circuit modules, $W_{t + 1}\sim W_{T}$. Assuming that each parameterized module forms a unitary $2$-design, the integration over $V_{\mathcal{R}}$ can be performed sequentially from left to right, module by module. Specifically,
$\left\langle {\Delta\Omega}_{\boldsymbol{pq}}^{\boldsymbol{p'q'}} \right\rangle_{V_{\mathcal{R}}}
= \left\langle \left\langle \left\langle {\Delta\Omega}_{\boldsymbol{pq}}^{\boldsymbol{p'q'}} \right\rangle_{W_{t + 1}} \right\rangle_{W_{t + 2}}\cdots \right\rangle_{W_{T}}$. We begin with the first module $W_{t + 1}$, Fig.~\ref{FD1} illustrates the tensor network representation of the operator $\Omega_{\boldsymbol{qp}}$, and the successive contraction and simplification steps proceed as follows.

(a) Tensor network representation of $\Omega_{\boldsymbol{q} \boldsymbol{p}}=\operatorname{Tr}_{\bar{t0}}\left[\left(|\boldsymbol{p}\rangle\langle\boldsymbol{q}| \otimes \mathbb{I}_{t 0}\right) V_{\mathcal{R}}^{\dagger} O_T V_{\mathcal{R}}\right]$, where $|\boldsymbol{p}\rangle$ and $|\boldsymbol{q}\rangle$ denote computational basis states on all qubits except those in registers $q_0$ and $q_t$. The dashed lines represent partial trace operations, while the two green lines indicate the subsystems on which the operator $\Omega_{\boldsymbol{qp}}$ acts after the partial trace. Paired squares connected by dash-dot lines represent a full circuit module (e.g., $W_T$ denotes the circuit module acting on registers $q_0$ and $q_T$, as shown in Fig.~\ref{FB2}).

(b) The partial trace operation of the operator $\left | \boldsymbol{p} \right \rangle \left \langle \boldsymbol{q} \right |$ over the subsystems $q_{1}\sim q_{t - 1}$ is simplified to $\delta_{\boldsymbol{(pq)}_{1\sim t - 1}}$.

(c) The matrix in the boxed region of panel (b) is absorbed into a single operator $O_{t + 2}$, which acts on the registers $q_0$, $q_t$, and $q_{t+2}\sim q_{T}$.

\begin{figure}[h]
    \centering
    \includegraphics[width=7in]{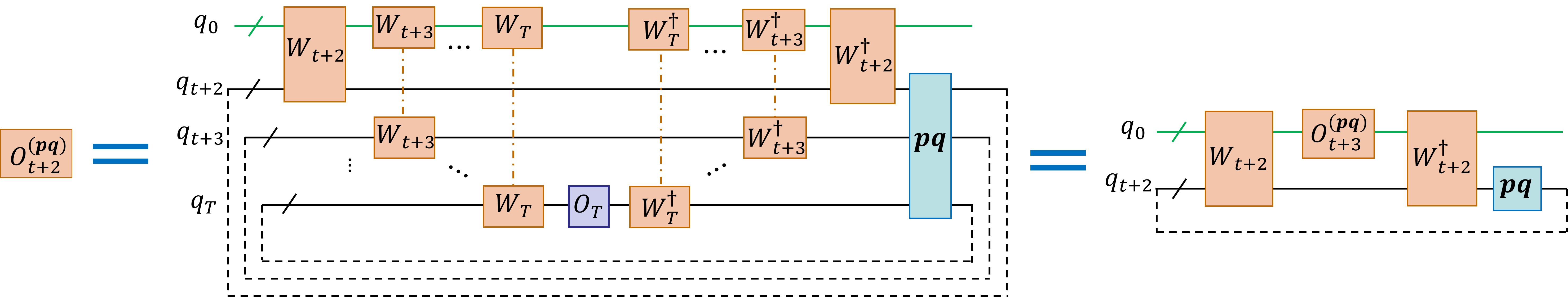}
    \caption{Tensor network representation of $O_{t + 2}^{\boldsymbol{(pq)}}$.}
    \label{FD2}
\end{figure}

(d) The partial trace operations of $O_{t + 2}$ and $\left | \boldsymbol{p} \right \rangle \left \langle \boldsymbol{q} \right |$ over the subsystems $q_{t + 2}\sim q_{T}$ are further simplified. The remaining part of $O_{t + 2}$ is equivalent to an operator $O_{t + 2}^{\boldsymbol{(pq)}}$ acting on the register $q_{0}$, as shown in detail in Fig.~\ref{FD2}. The final equivalent circuit for the operator $\Omega_{\boldsymbol{qp}}$ is shown in Fig.~\ref{FD1}(d). The operator $O_{t+2}^{\boldsymbol{(pq)}}$ can be further reduced to $O_{t+3}^{\boldsymbol{(pq)}}$ by incorporating $W_{t+2}$, and this procedure continues recursively until reaching the final operators $O_T$ and $W_T$.

After integrating over the matrix $W_{t + 1}$, the remaining modules in the operator $O_{t + 2}^{\boldsymbol{(pq)}}$ can be treated in the same manner. As illustrated in Fig. ~\ref{FD1}, similar simplification procedures are applied to obtain the corresponding equivalent quantum circuits, and \textbf{Lemma 3} or \textbf{Lemma 5} can then be used to complete the integration over all remaining modules.

\begin{figure}[h]
    \centering
    \includegraphics[width=5.5in]{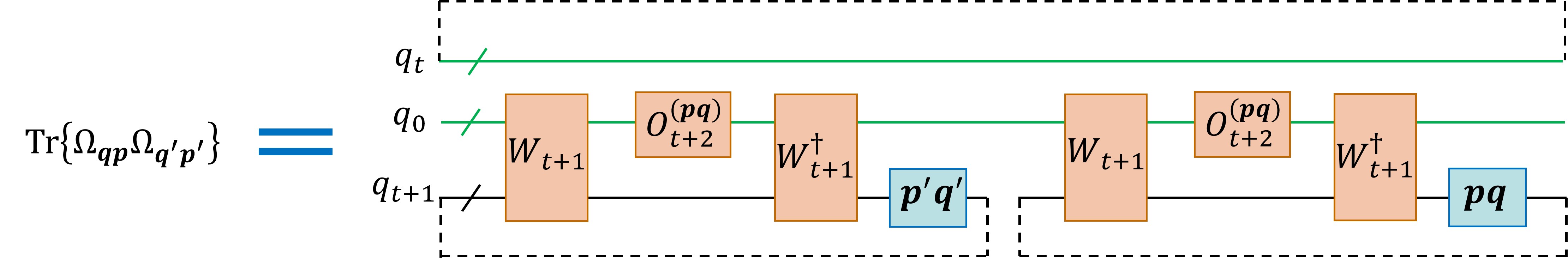}
    \caption{Tensor network representation of $\text{Tr}\left[ \Omega_{\boldsymbol{qp}}\Omega_{\boldsymbol{q'p'}} \right]$.}
    \label{FD3}
\end{figure}

\begin{figure}[h]
    \centering
    \includegraphics[width=3.5in]{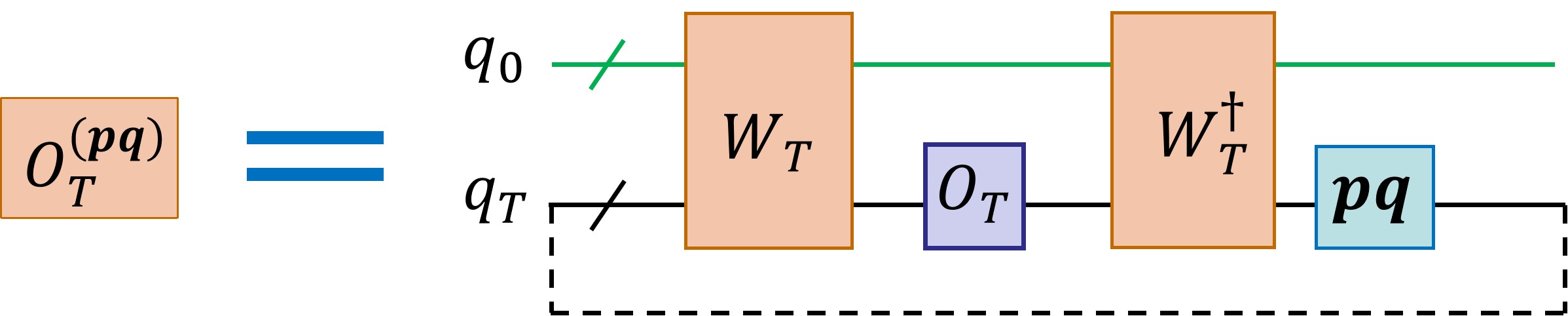}
    \caption{Tensor network representation of $O_{T}^{\boldsymbol{(pq)}}$.}
    \label{FD5}
\end{figure}

Here, we take the integration over the matrix $W_{t + 1}$ as an explicit example to illustrate the calculation procedure. We first evaluate the integration of the term $\text{Tr}\left[ \Omega_{\boldsymbol{qp}}\Omega_{\boldsymbol{q'p'}} \right]$ appearing in Eq.~(\ref{ED2}). According to its tensor network representation shown in Fig.~\ref{FD3}, we obtain
$$\int_{}^{}{d\mu \text{Tr}\left[ \Omega_{\boldsymbol{qp}}\Omega_{\boldsymbol{q'p'}} \right]} \\[-1em]$$
\begin{equation}\label{ED5}
    \begin{split}
        = \int_{}^{}{d\mu(W_{t + 1})\delta_{\boldsymbol{(pq)}_{< t}}\delta_{\boldsymbol{(p'q')}_{< t}}2^{m}\times\text{Tr}\left\{ {\text{Tr}}_{t + 1}\left[ \begin{matrix}
        \mathbb{I} \\
        \otimes \\
        \left | \boldsymbol{p} \right \rangle \left \langle \boldsymbol{q} \right |
        \end{matrix}W_{t + 1}^{\dagger}\begin{matrix}
        O_{t + 2}^{\boldsymbol{(pq)}} \\
        \otimes \\
        \mathbb{I}
        \end{matrix}W_{t + 1} \right]{\text{Tr}}_{t + 1}\left[ \begin{matrix}
        \mathbb{I} \\
        \otimes \\
        \left | \boldsymbol{p'} \right \rangle \left \langle \boldsymbol{q'} \right |
        \end{matrix}W_{t + 1}^{\dagger}\begin{matrix}
        O_{t + 2}^{\boldsymbol{(p'q')}} \\
        \otimes \\
        \mathbb{I}
        \end{matrix}W_{t + 1} \right] \right\}}\\
        \\[-1em]
        = \delta_{\boldsymbol{(pq)}_{< t}}\delta_{\boldsymbol{(p'q')}_{< t}}\frac{1}{2^{4m} - 1} \{ \delta_{\boldsymbol{(pq)}_{t + 1}}\delta_{\boldsymbol{(p'q')}_{t + 1}}\text{Tr}[ O_{t + 2}^{\boldsymbol{(pq)}}] \text{Tr}[ O_{t + 2}^{\boldsymbol{(p'q')}}] \cdot 2^{4m} 
        + \delta_{\boldsymbol{(pq')}_{t + 1}}\delta_{\boldsymbol{(p'q)}_{t + 1}}\text{Tr}[ O_{t + 2}^{\boldsymbol{(pq)}} O_{t + 2}^{\boldsymbol{(p'q')}}] \cdot 2^{4m}\\
        \\[-1em]
        - \delta_{\boldsymbol{(pq')}_{t + 1}}\delta_{\boldsymbol{(p'q)}_{t + 1}}\text{Tr}[ O_{t + 2}^{\boldsymbol{(pq)}}] \text{Tr}[ O_{t + 2}^{\boldsymbol{(p'q')}}] \cdot 2^{3m}
        - \delta_{\boldsymbol{(pq)}_{t + 1}}\delta_{\boldsymbol{(p'q')}_{t + 1}}\text{Tr}[ O_{t + 2}^{\boldsymbol{(pq)}} O_{t + 2}^{\boldsymbol{(p'q')}}] \cdot 2^{m} \},
    \end{split}
\end{equation}
\begin{figure}[h]
    \centering
    \includegraphics[width=5.5in]{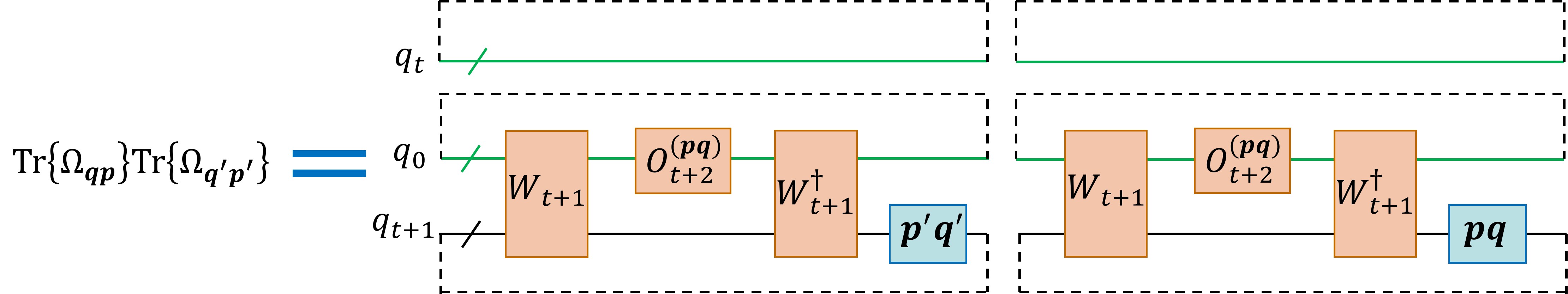}
    \caption{Tensor network representation of  $\text{Tr}\left[ \Omega_{\boldsymbol{qp}} \right] \text{Tr}\left[ \Omega_{\boldsymbol{q'p'}} \right]$.}
    \label{FD4}
\end{figure}
where $\delta_{\boldsymbol{(pq)}_{< t}}$ corresponds to the factor $\delta_{\boldsymbol{(pq)}_{1\sim t - 1}}$ shown in Fig.~\ref{FD1}(b). The coefficient $2^{m}$ in the first equation arises from taking the trace of the identity operator over the Register $q_{t}$, while the second equation follows directly from \textbf{Lemma 5}. In addition, according to \textbf{Lemma 3} and Fig.~\ref{FD4}, the integral corresponding to the second term in Eq.~(\ref{ED2}) holds
\begin{equation}\label{ED6}
    \begin{aligned}
        &\quad \quad \quad \quad \quad \quad \quad \quad \quad \quad \quad \quad \quad \quad \quad \quad \quad \quad 
        \int_{}^{}{d\mu\frac{\text{Tr}\left[ \Omega_{\boldsymbol{qp}} \right] \text{Tr}\left[ \Omega_{\boldsymbol{q'p'}} \right]}{2^{2m}}}\\
        \\[-1em]
        &= \int_{}^{}{d\mu(W_{t + 1})\delta_{\boldsymbol{(pq)}_{< t}}\delta_{\boldsymbol{(p'q')}_{< t}}\text{Tr}\left[ \begin{matrix}
        \mathbb{I} \\
        \otimes \\
        \left | \boldsymbol{p} \right \rangle \left \langle \boldsymbol{q} \right |
        \end{matrix}W_{t + 1}^{\dagger}\begin{matrix}
        O_{t + 2}^{\boldsymbol{(pq)}} \\
        \otimes \\
        \mathbb{I}
        \end{matrix}W_{t + 1} \right] \text{Tr}\left[ \begin{matrix}
        \mathbb{I} \\
        \otimes \\
        \left | \boldsymbol{p'} \right \rangle \left \langle \boldsymbol{q'} \right |
        \end{matrix}W_{t + 1}^{\dagger}\begin{matrix}
        O_{t + 2}^{\boldsymbol{(p'q')}} \\
        \otimes \\
        \mathbb{I}
        \end{matrix}W_{t + 1} \right]}\\
        \\[-1em]
        &= \delta_{\boldsymbol{(pq)}_{< t}}\delta_{\boldsymbol{(p'q')}_{< t}}\frac{1}{2^{4m} - 1}\{\delta_{\boldsymbol{(pq)}_{t + 1}}\delta_{\boldsymbol{(p'q')}_{t + 1}}\text{Tr}[ O_{t + 2}^{\boldsymbol{(pq)}}] \text{Tr}[ O_{t + 2}^{\boldsymbol{(p'q')}}] \cdot 2^{4m}
        + \delta_{\boldsymbol{(pq')}_{t + 1}}\delta_{\boldsymbol{(p'q)}_{t + 1}}\text{Tr}[ O_{t + 2}^{\boldsymbol{(pq)}} O_{t + 2}^{\boldsymbol{(p'q')}}] \cdot 2^{2m}\\
        \\[-1em]
        &\quad \quad \quad \quad \quad \quad \quad \quad \quad \quad \quad 
        - \delta_{\boldsymbol{(pq')}_{t + 1}}\delta_{\boldsymbol{(p'q)}_{t + 1}}\text{Tr}[ O_{t + 2}^{\boldsymbol{(pq)}}] \text{Tr}[ O_{t + 2}^{\boldsymbol{(p'q')}}] \cdot 2^{m}
        - \delta_{\boldsymbol{(pq)}_{t + 1}}\delta_{\boldsymbol{(p'q')}_{t + 1}}\text{Tr}[ O_{t + 2}^{\boldsymbol{(pq)}} O_{t + 2}^{\boldsymbol{(p'q')}}] \cdot 2^{m}\}.
    \end{aligned}
\end{equation}
Combining Eq.~(\ref{ED5}) and Eq.~(\ref{ED6}), we obtain
\begin{equation}
    \begin{split}
        \left\langle {\Delta\Omega}_{\boldsymbol{pq}}^{\boldsymbol{p'q'}} \right\rangle_{W_{t + 1}} 
        &= \int_{}^{}{d\mu(W_{t + 1})\left \{\text{Tr}\left[ \Omega_{\boldsymbol{qp}}\Omega_{\boldsymbol{q'p'}} \right] - \frac{\text{Tr}\left[ \Omega_{\boldsymbol{qp}} \right] \text{Tr}\left[ \Omega_{\boldsymbol{q'p'}} \right]}{2^{2m}} \right \}}\\
        &= \delta_{\boldsymbol{(pq)}_{< t}}\delta_{\boldsymbol{(p'q')}_{< t}}\delta_{\boldsymbol{(pq')}_{t + 1}}\delta_{\boldsymbol{(p'q)}_{t + 1}}\frac{2^{m}(2^{2m} - 1)}{2^{4m} - 1}\left \{ \text{Tr}[ O_{t + 2}^{\boldsymbol{(pq)}} O_{t + 2}^{\boldsymbol{(p'q')}}] - \frac{\text{Tr}[ O_{t + 2}^{\boldsymbol{(pq)}}]\text{Tr}[ O_{t + 2}^{\boldsymbol{(p'q')}}]}{{2^{m}}} \right \},
    \end{split}
\end{equation}
where the factor
$\left \{ \text{Tr}[ O_{t + 2}^{\boldsymbol{(pq)}} O_{t + 2}^{\boldsymbol{(p'q')}}]
- \text{Tr}[ O_{t + 2}^{\boldsymbol{(pq)}}] \text{Tr}[ O_{t + 2}^{\boldsymbol{(p'q')}}]/2^{m} \right \}$
can be simplified sequentially using the procedure illustrated in Figs. \ref{FD1} and \ref{FD2} . Then, by applying \textbf{Lemma 3} and \textbf{Lemma 5}, the integration over the matrix $W_{t + 2}$ can be carried out, yielding
\begin{equation}
    \begin{split}
        &\int d \mu\left(W_{t+2}\right)\left\{\operatorname{Tr}[O_{t+2}^{(\boldsymbol{p} \boldsymbol{q})} O_{t+2}^{\left(\boldsymbol{p}^{\prime} \boldsymbol{q}^{\prime}\right)}]-\frac{\operatorname{Tr}[O_{t+2}^{(\boldsymbol{p} \boldsymbol{q})}] \operatorname{Tr}[O_{t+2}^{\left(\boldsymbol{p}^{\prime} \boldsymbol{q}^{\prime}\right)}]}{2^m}\right\}\\
        &=\delta_{\left(\boldsymbol{p} \boldsymbol{q}^{\prime}\right)_{t+2}} \delta_{\left(\boldsymbol{p}^{\prime} \boldsymbol{q}\right)_{t+2}} \frac{2^m\left(2^{2 m}-1\right)}{2^{4 m}-1}\left\{\operatorname{Tr}[O_{t+3}^{(\boldsymbol{p} \boldsymbol{q})} O_{t+3}^{\left(\boldsymbol{p}^{\prime} \boldsymbol{q}^{\prime}\right)}]-\frac{\operatorname{Tr}[O_{t+3}^{(\boldsymbol{p} \boldsymbol{q})}] \operatorname{Tr}[O_{t+3}^{\left(\boldsymbol{p}^{\prime} \boldsymbol{q}^{\prime}\right)}]}{2^m}\right\}.
    \end{split}
\end{equation}
Repeating this procedure until the integration over the matrix $W_{T - 1}$ is completed, we obtain
\begin{equation}
    \begin{split}
        \left\langle {\Delta\Omega}_{\boldsymbol{pq}}^{\boldsymbol{p'q'}} \right\rangle_{W_{t + 1}\sim W_{T - 1}} 
        =& \delta_{\boldsymbol{(pq)}_{< t}}\delta_{\boldsymbol{(p'q')}_{< t}}\delta_{\boldsymbol{(pq')}_{> t}}\delta_{\boldsymbol{(p'q)}_{> t}}{\left(\frac{2^{m}}{2^{2m} + 1}\right)}^{T - 1 - t}\\
        &\times\left\{\operatorname{Tr}\left[O_T^{(\boldsymbol{p} \boldsymbol{q})} O_T^{\left(\boldsymbol{p}^{\prime} \boldsymbol{q}^{\prime}\right)}\right]-\frac{\operatorname{Tr}\left[O_T^{(\boldsymbol{p} \boldsymbol{q})}\right] \operatorname{Tr}\left[O_T^{\left(\boldsymbol{p}^{\prime} \boldsymbol{q}^{\prime}\right)}\right]}{2^m}\right\},
    \end{split}
\end{equation}
where $\delta_{\boldsymbol{(pq')}_{> t}}$ represents $\delta_{\boldsymbol{(pq')}_{t + 1\sim T - 1}}$. The final operator $O_{T}^{\boldsymbol{(pq)}}$ differs slightly from the previous cases, as illustrated in Fig.~\ref{FD5}. In particular, its circuit involves the operator $O_{T}$ acting on the register group $q_{T}$ rather than $q_{0}$. Consequently, the integration of $\text{Tr}\left [ O_{T}^{\boldsymbol{(pq)}} O_{T}^{\boldsymbol{(p'q')}}\right]$ over $W_{T}$ cannot be performed using \textbf{Lemma 5}, but instead requires \textbf{Lemma 6}, although the overall procedure is largely similar.

By applying \textbf{Lemma 3} and \textbf{Lemma 6} to integrate over all modules in $V_{\mathcal{R}}$, we obtain
\begin{equation}\label{ED-1}
    \begin{split}
        \left\langle {\Delta\Omega}_{\boldsymbol{pq}}^{\boldsymbol{p'q'}} \right\rangle_{V_{\mathcal{R}}} 
        &= \delta_{\boldsymbol{(pq)}_{< t}}\delta_{\boldsymbol{(p'q')}_{< t}}\delta_{\boldsymbol{(pq')}_{> t}}\delta_{\boldsymbol{(p'q)}_{> t}}{\left(\frac{2^{m}}{2^{2m} + 1}\right)}^{T - t}\{ \text{Tr}[ O_{T}^{2}] - \frac{\text{Tr}[ O_{T}] \text{Tr}[ O_{T}]}{2^{m}}\}\\
        &= \delta_{\boldsymbol{(pq)}_{< t}}\delta_{\boldsymbol{(p'q')}_{< t}}\delta_{\boldsymbol{(pq')}_{> t}}\delta_{\boldsymbol{(p'q)}_{> t}}{\left(\frac{2^{m}}{2^{2m} + 1}\right)}^{T - t}D_{HS}\left( O_{T}, \frac{\text{Tr}[ O_{T}]\mathbb{I}_{T}}{2^{m}} \right),
    \end{split}
\end{equation}
where the factor $\delta_{\boldsymbol{(pq')}_{> t}}$ is updated to $\delta_{\boldsymbol{(pq')}_{t + 1\sim T}}$, and
$D_{HS}(\rho,\sigma) = \text{Tr}\left[ (\rho - \sigma)^{2} \right]$
denotes the Hilbert--Schmidt distance. This result shows that the step-by-step integration of the matrices on the right-hand side ultimately depends only on the parameters $m$, $T$, and $t$.

\hfill $\square$  

\subsection{\label{sec:level2}2. Calculation for $V_{\mathcal{L}}$}
Next, we evaluate the expectation value associated with $V_{\mathcal{L}}$ appearing in Eq.~(\ref{ED2}). For convenience, we first restate the operator
\begin{equation}\label{64}
    {\Delta\Psi}_{\boldsymbol{pq}}^{\boldsymbol{p'q'}} = \text{Tr}\left[ \Psi_{\boldsymbol{pq}}\Psi_{\boldsymbol{p'q'}} \right] - \frac{\text{Tr}\left[ \Psi_{\boldsymbol{pq}} \right] \text{Tr}\left[ \Psi_{\boldsymbol{p'q'}} \right]}{2^{2m}},
\end{equation}
and
\begin{equation}
    \Psi_{\boldsymbol{pq}} = {\text{Tr}}_{\overline{t0}}\left[ (\left | \boldsymbol{q} \right \rangle \left \langle \boldsymbol{p} \right | \otimes \mathbb{I}_{t0})V_{\mathcal{L}}\rho_{in}V_{\mathcal{L}}^{\dagger} \right].
\end{equation}
Based on Eq.~(\ref{DE1}) and Eq.~(\ref{ED-1}), the variance can now be written as
\begin{equation}
    \left\langle (\partial_\nu C)^2 \right\rangle_{V} = \frac{2^{2m-1} \text{Tr}[\sigma_v^2]}{(2^{4m}-1)^2} \left( \frac{2^m}{2^{2m}+1} \right)^{T-t} D_{HS} \left( O_T, \frac{\text{Tr}[O_T] \mathbb{I}_T}{2^m} \right) \sum_{\substack{\boldsymbol{p},\boldsymbol{q} \\ \boldsymbol{p'},\boldsymbol{q'}}} \delta_{(\boldsymbol{pq})_{<t}} \delta_{(\boldsymbol{p'q'})_{<t}} \delta_{(\boldsymbol{pq'})_{>t}} \delta_{(\boldsymbol{p'q})_{>t}} \left\langle \Delta \Psi_{\boldsymbol{pq}}^{\boldsymbol{p'q'}} \right\rangle_{V_{\mathcal{L}}},
\end{equation}
where
\begin{equation}\label{ED14}
    \sum_{\substack{\boldsymbol{p},\boldsymbol{q} \\ \boldsymbol{p'},\boldsymbol{q'}}} \delta_{(\boldsymbol{pq})_{<t}} \delta_{(\boldsymbol{p'q'})_{<t}} \delta_{(\boldsymbol{pq'})_{>t}} \delta_{(\boldsymbol{p'q})_{>t}} \left\langle \Delta \Psi_{\boldsymbol{pq}}^{\boldsymbol{p'q'}} \right\rangle_{V_{\mathcal{L}}}
    = \sum_{\substack{(\boldsymbol{pq})_{>t} \\ (\boldsymbol{p'q'})_{>t}}} \delta_{(\boldsymbol{pq'})_{>t}} \delta_{(\boldsymbol{p'q})_{>t}}
    \sum_{\substack{(\boldsymbol{pq})_{<t} \\ (\boldsymbol{p'q'})_{<t}}} \delta_{(\boldsymbol{pq})_{<t}} \delta_{(\boldsymbol{p'q'})_{<t}} \left\langle \Delta \Psi_{\boldsymbol{pq}}^{\boldsymbol{p'q'}} \right\rangle_{V_{\mathcal{L}}}.
\end{equation}
We first evaluate the summation over the $(\boldsymbol{p},\boldsymbol{q},\boldsymbol{p'},\boldsymbol{q'})_{<t}$ sector in the above expression, where the combination of the factor $\delta_{(\boldsymbol{pq})_{<t}}$ and the operator $\Psi_{\boldsymbol{pq}}$ within $\left\langle \Delta \Psi_{\boldsymbol{pq}}^{\boldsymbol{p'q'}} \right\rangle_{V_{\mathcal{L}}}$ yields
\begin{equation}
    \begin{split}
        \sum_{\boldsymbol{(pq)}_{< t}}^{}{\delta_{\boldsymbol{(pq)}_{< t}}\Psi_{\boldsymbol{pq}}} 
        &= \sum_{\boldsymbol{(pq)}_{< t}}^{}{\delta_{\boldsymbol{(pq)}_{< t}}{\text{Tr}}_{\overline{t0}}\left[ \begin{pmatrix}
        {\left | \boldsymbol{q} \right \rangle \left \langle \boldsymbol{p} \right |}_{< t} \\
        \mathbb{I}_{t0} \\
        {\left | \boldsymbol{q} \right \rangle \left \langle \boldsymbol{p} \right |}_{> t}
        \end{pmatrix}V_{\mathcal{L}}\rho_{in}V_{\mathcal{L}}^{\dagger} \right]} 
        \\
        \\[-1em]
        &= \sum_{\boldsymbol{(p)}_{< t}}^{}{{\text{Tr}}_{\overline{t0}}\left[ \begin{pmatrix}
        \mathbb{I}_{< t} \\
        \mathbb{I}_{t0} \\
        {\left | \boldsymbol{q} \right \rangle \left \langle \boldsymbol{p} \right |}_{> t}
        \end{pmatrix}\begin{pmatrix}
        {\left | \boldsymbol{p} \right \rangle \left \langle \boldsymbol{p} \right |}_{< t} \\
        \mathbb{I}_{t0} \\
        \mathbb{I}_{> t}
        \end{pmatrix}V_{\mathcal{L}}\rho_{in}V_{\mathcal{L}}^{\dagger} \right],}
    \end{split}
\end{equation}
by invoking the definition of the partial trace,
${\text{Tr}}_{s}[ A] = \begin{pmatrix}\left \langle \boldsymbol{p} \right |_{s} \\    \mathbb{I}_{\overline{s}}\end{pmatrix}A\begin{pmatrix}    \left | \boldsymbol{p} \right \rangle_{s} \\    \mathbb{I}_{\overline{s}}\end{pmatrix}$,
we obtain
\begin{equation}\label{ED16}
    \begin{split}
        \sum_{\boldsymbol{(pq)}_{< t}}^{}{\delta_{\boldsymbol{(pq)}_{< t}}\Psi_{\boldsymbol{pq}}}
        &= \sum_{\boldsymbol{(p)}_{< t}}^{}{{\text{Tr}}_{> t}\left[ \sum_{\boldsymbol{(i)}_{< t}}^{}{\begin{pmatrix}
        \left \langle \boldsymbol{i} \right |_{< t} \\
        \mathbb{I}_{t0} \\
        \mathbb{I}_{> t}
        \end{pmatrix}\begin{pmatrix}
        \mathbb{I}_{< t} \\
        \mathbb{I}_{t0} \\
        {\left | \boldsymbol{q} \right \rangle \left \langle \boldsymbol{p} \right |}_{> t}
        \end{pmatrix}\begin{pmatrix}
        {\left | \boldsymbol{p} \right \rangle \left \langle \boldsymbol{p} \right |}_{< t} \\
        \mathbb{I}_{t0} \\
        \mathbb{I}_{> t}
        \end{pmatrix}V_{\mathcal{L}}\rho_{in}V_{\mathcal{L}}^{\dagger}\begin{pmatrix}
        \left | \boldsymbol{i} \right \rangle_{< t} \\
        \mathbb{I}_{t0} \\
        \mathbb{I}_{> t}
        \end{pmatrix}} \right]}
        \\
        \\[-1em]
        &= {\text{Tr}}_{> t}\left[ \begin{pmatrix}
        \mathbb{I}_{t0} \\
        {\left | \boldsymbol{q} \right \rangle \left \langle \boldsymbol{p} \right |}_{> t}
        \end{pmatrix}\sum_{\boldsymbol{(p)}_{< t}}^{}{\begin{pmatrix}
        \left \langle \boldsymbol{p} \right |_{< t} \\
        \mathbb{I}_{t0} \\
        \mathbb{I}_{> t}
        \end{pmatrix}V_{\mathcal{L}}\rho_{in}V_{\mathcal{L}}^{\dagger}\begin{pmatrix}
        \left | \boldsymbol{p} \right \rangle_{< t} \\
        \mathbb{I}_{t0} \\
        \mathbb{I}_{> t}
        \end{pmatrix}} \right]
        \\
        \\[-1em]
        &= {\text{Tr}}_{> t}\left[ \begin{pmatrix}
        \mathbb{I}_{t0} \\
        {\left | \boldsymbol{q} \right \rangle \left \langle \boldsymbol{p} \right |}_{> t}
        \end{pmatrix}{\widetilde{\rho}}_{\overline{< t}} \right] = {\text{Tr}}_{> t}\left[ \mathbb{I}_{t0} \otimes {\left | \boldsymbol{q} \right \rangle \left \langle \boldsymbol{p} \right |}_{> t}{\widetilde{\rho}}_{\overline{< t}} \right],
    \end{split}
\end{equation}
where
$\begin{pmatrix}
{\left | \boldsymbol{q} \right \rangle \left \langle \boldsymbol{p} \right |}_{< t} \\
\mathbb{I}_{t0} \\
{\left | \boldsymbol{q} \right \rangle \left \langle \boldsymbol{p} \right |}_{> t}
\end{pmatrix}$
represents
$\left( {\left | \boldsymbol{q} \right \rangle \left \langle \boldsymbol{p} \right |}_{< t} \otimes \mathbb{I}_{t0} \otimes {\left | \boldsymbol{q} \right \rangle \left \langle \boldsymbol{p} \right |}_{> t} \right)$, which is
written in the \emph{tensor-product tableau notation}. We further define
\begin{equation}
    {\widetilde{\rho}}_{\overline{< t}} = {\text{Tr}}_{< t}\left[ V_{\mathcal{L}}\rho_{in}V_{\mathcal{L}}^{\dagger} \right]
\end{equation}
as the reduced state obtained by the action of $V_{\mathcal{L}}$ on the input state over the registers $\{ q_{0}, q_{t\sim T} \}$, where ${\text{Tr}}_{< t}$ denotes the partial trace over the registers $q_{1\sim t - 1}$. 

Similar to Eq.~(\ref{ED16}), we obtain
\begin{equation}
    \sum_{(\boldsymbol{p'q'})_{<t}} \delta_{(\boldsymbol{p'q'})_{<t}} \Psi_{\boldsymbol{p'q'}}
    = \text{Tr}_{>t} \left[ \begin{pmatrix} \mathbb{I}_{t0} \\ {\left | \boldsymbol{q'} \right \rangle \left \langle \boldsymbol{p'} \right |}_{> t} \end{pmatrix} \widetilde{\rho}_{\overline{<t}} \right]
    = \text{Tr}_{>t} \left[ \mathbb{I}_{t0} \otimes {\left | \boldsymbol{q'} \right \rangle \left \langle \boldsymbol{p'} \right |}_{> t} \, \widetilde{\rho}_{\overline{<t}} \right]
    \equiv \Psi_{(\boldsymbol{p'q'})_{>t}}.
\end{equation}
It is evident that the numerical values resulting from the filtering by the factors $\delta_{(\boldsymbol{p'q'})_{<t}}$ and $\delta_{(\boldsymbol{pq'})_{<t}}$ depend only on the components $(\boldsymbol{pq})_{>t}$ and $(\boldsymbol{p'q'})_{>t}$. Accordingly, the relevant summation in Eq.~(\ref{ED14}) can be written as
\begin{equation}
    \sum_{(\boldsymbol{pq})_{<t}, (\boldsymbol{p'q'})_{<t}} \delta_{(\boldsymbol{pq})_{<t}} \delta_{(\boldsymbol{p'q'})_{<t}}
    \left\langle \Delta \Psi_{\boldsymbol{pq}}^{\boldsymbol{p'q'}} \right\rangle_{V_{\mathcal{L}}}
    = \left\langle
    \text{Tr}\!\left[\Psi_{(\boldsymbol{pq})_{>t}} \Psi_{(\boldsymbol{p'q'})_{>t}}\right]
    - \frac{\text{Tr}\!\left[\Psi_{(\boldsymbol{pq})_{>t}}\right]\text{Tr}\!\left[\Psi_{(\boldsymbol{p'q'})_{>t}}\right]}{2^{2m}}
    \right\rangle_{V_{\mathcal{L}}}
    \equiv \left\langle \Delta \Psi_{(\boldsymbol{pq})_{>t}}^{(\boldsymbol{p'q'})_{>t}} \right\rangle_{V_{\mathcal{L}}}.
\end{equation}
Furthermore, by incorporating the summation with the factors $\delta_{(\boldsymbol{pq'})_{>t}}$ and $\delta_{(\boldsymbol{p'q})_{>t}}$, we obtain
\begin{equation}
    \sum_{(\boldsymbol{pq})_{>t}, (\boldsymbol{p'q'})_{>t}} \delta_{(\boldsymbol{pq'})_{>t}} \delta_{(\boldsymbol{p'q})_{>t}}
    \left\langle \Delta \Psi_{(\boldsymbol{pq})_{>t}}^{(\boldsymbol{p'q'})_{>t}} \right\rangle_{V_{\mathcal{L}}}
    = \sum_{(\boldsymbol{pq})_{>t}} \text{Tr}\!\left[\Psi_{(\boldsymbol{pq})_{>t}} \Psi_{(\boldsymbol{qp})_{>t}}\right]
    - \sum_{(\boldsymbol{pq})_{>t}} \frac{\text{Tr}\!\left[\Psi_{(\boldsymbol{pq})_{>t}}\right]\text{Tr}\!\left[\Psi_{(\boldsymbol{qp})_{>t}}\right]}{2^{2m}}.
\end{equation}
where the first term in the above expression can be expanded as
\begin{equation}
    \begin{split}
        \sum_{\boldsymbol{(pq)}_{> t}}^{}{\text{Tr}\left[ \Psi_{\boldsymbol{pq}}\Psi_{\boldsymbol{qp}} \right]}
        &= \sum_{\boldsymbol{(pq)}_{> t}}^{}{\text{Tr}\left[ {\text{Tr}}_{> t}\left[ \begin{pmatrix}
        \mathbb{I}_{t0} \\
        {\left | \boldsymbol{q} \right \rangle \left \langle \boldsymbol{p} \right |}_{> t}
        \end{pmatrix}{\widetilde{\rho}}_{\overline{< t}} \right]{\text{Tr}}_{> t}\left[ \begin{pmatrix}
        \mathbb{I}_{t0} \\
        {\left | \boldsymbol{p} \right \rangle \left \langle \boldsymbol{q} \right |}_{> t}
        \end{pmatrix}{\widetilde{\rho}}_{\overline{< t}} \right] \right]}
        \\
        \\[-1em]
        &= \sum_{\boldsymbol{(pq)}_{> t}}^{}{\text{Tr}\left[ \sum_{\boldsymbol{(i)}_{> t}}^{}{\begin{pmatrix}
        \mathbb{I}_{t0} \\
        \left \langle \boldsymbol{i} \right |_{> t}
        \end{pmatrix}\begin{pmatrix}
        \mathbb{I}_{t0} \\
        {\left | \boldsymbol{q} \right \rangle \left \langle \boldsymbol{p} \right |}_{> t}
        \end{pmatrix}{\widetilde{\rho}}_{\overline{< t}}\begin{pmatrix}
        \mathbb{I}_{t0} \\
        \left | \boldsymbol{i} \right \rangle_{> t}
        \end{pmatrix}}\sum_{\boldsymbol{(j)}_{> t}}^{}{\begin{pmatrix}
        \mathbb{I}_{t0} \\
        \left \langle \boldsymbol{j} \right |_{> t}
        \end{pmatrix}\begin{pmatrix}
        \mathbb{I}_{t0} \\
        {\left | \boldsymbol{p} \right \rangle \left \langle \boldsymbol{q} \right |}_{> t}
        \end{pmatrix}{\widetilde{\rho}}_{\overline{< t}}\begin{pmatrix}
        \mathbb{I}_{t0} \\
        \left | \boldsymbol{j} \right \rangle_{> t}
        \end{pmatrix}} \right]}
        \\
        \\[-1em]
        &= \sum_{\boldsymbol{(pq)}_{> t}}^{}{\text{Tr}\left[ \begin{pmatrix}
        \mathbb{I}_{t0} \\
        \left \langle \boldsymbol{p} \right |_{> t}
        \end{pmatrix}{\widetilde{\rho}}_{\overline{< t}}\begin{pmatrix}
        \mathbb{I}_{t0} \\
        \left | \boldsymbol{q} \right \rangle_{> t}
        \end{pmatrix}\begin{pmatrix}
        \mathbb{I}_{t0} \\
        \left \langle \boldsymbol{q} \right |_{> t}
        \end{pmatrix}{\widetilde{\rho}}_{\overline{< t}}\begin{pmatrix}
        \mathbb{I}_{t0} \\
        \left | \boldsymbol{p} \right \rangle_{> t}
        \end{pmatrix} \right]}
        \\
        \\[-1em]
        &= \sum_{\boldsymbol{(pq)}_{> t}}^{}{\text{Tr}\left[ {\widetilde{\rho}}_{\overline{< t}}\begin{pmatrix}
        \mathbb{I}_{t0} \\
        {\left | \boldsymbol{q} \right \rangle \left \langle \boldsymbol{q} \right |}_{> t}
        \end{pmatrix}{\widetilde{\rho}}_{\overline{< t}}\begin{pmatrix}
        \mathbb{I}_{t0} \\
        {\left | \boldsymbol{p} \right \rangle \left \langle \boldsymbol{p} \right |}_{> t}
        \end{pmatrix} \right]}
        = \text{Tr}\left[ {\widetilde{\rho}}_{\overline{< t}}^{2} \right].
    \end{split}
\end{equation}
Similarly, the second term can be evaluated as
\begin{equation}
    \begin{split}
        \sum_{\boldsymbol{(pq)}_{> t}}^{}{\text{Tr}\left[ \Psi_{\boldsymbol{pq}} \right] \text{Tr}\left[ \Psi_{\boldsymbol{qp}} \right]}
        &= \sum_{\boldsymbol{(pq)}_{> t}}^{}{\text{Tr}\left[ \begin{pmatrix}
        \mathbb{I}_{t0} \\
        {\left | \boldsymbol{q} \right \rangle \left \langle \boldsymbol{p} \right |}_{> t}
        \end{pmatrix}{\widetilde{\rho}}_{\overline{< t}} \right] \text{Tr}\left[ \begin{pmatrix}
        \mathbb{I}_{t0} \\
        {\left | \boldsymbol{p} \right \rangle \left \langle \boldsymbol{q} \right |}_{> t}
        \end{pmatrix}{\widetilde{\rho}}_{\overline{< t}} \right]}
        \\
        \\[-1em]
        &= \sum_{\boldsymbol{(pq)}_{> t}}^{}{\sum_{\boldsymbol{(i)}_{t0, > t}}^{}{\begin{pmatrix}
        \left \langle \boldsymbol{i} \right |_{t0} \\
        \left \langle \boldsymbol{i} \right |_{> t}
        \end{pmatrix}\begin{pmatrix}
        \mathbb{I}_{t0} \\
        {\left | \boldsymbol{q} \right \rangle \left \langle \boldsymbol{p} \right |}_{> t}
        \end{pmatrix}{\widetilde{\rho}}_{\overline{< t}}\begin{pmatrix}
        \left | \boldsymbol{i} \right \rangle_{t0} \\
        \left | \boldsymbol{i} \right \rangle_{> t}
        \end{pmatrix}}\sum_{\boldsymbol{(j)}_{t0, > t}}^{}{\begin{pmatrix}
        \left \langle \boldsymbol{j} \right |_{t0} \\
        \left \langle \boldsymbol{j} \right |_{> t}
        \end{pmatrix}\begin{pmatrix}
        \mathbb{I}_{t0} \\
        {\left | \boldsymbol{p} \right \rangle \left \langle \boldsymbol{q} \right |}_{> t}
        \end{pmatrix}{\widetilde{\rho}}_{\overline{< t}}\begin{pmatrix}
        \left | \boldsymbol{j} \right \rangle_{t0} \\
        \left | \boldsymbol{j} \right \rangle_{> t}
        \end{pmatrix}}}
        \\
        \\[-1em]
        &= \sum_{\boldsymbol{(pq)}_{> t}}^{}{\sum_{\boldsymbol{(i)}_{t0}}^{}{\begin{pmatrix}
        \left \langle \boldsymbol{i} \right |_{t0} \\
        \left \langle \boldsymbol{p} \right |_{> t}
        \end{pmatrix}{\widetilde{\rho}}_{\overline{< t}}\begin{pmatrix}
        \left | \boldsymbol{i} \right \rangle_{t0} \\
        \left | \boldsymbol{q} \right \rangle_{> t}
        \end{pmatrix}}\sum_{\boldsymbol{(j)}_{t0}}^{}{\begin{pmatrix}
        \left \langle \boldsymbol{j} \right |_{t0} \\
        \left \langle \boldsymbol{q} \right |_{> t}
        \end{pmatrix}{\widetilde{\rho}}_{\overline{< t}}\begin{pmatrix}
        \left | \boldsymbol{j} \right \rangle_{t0} \\
        \left | \boldsymbol{p} \right \rangle_{> t}
        \end{pmatrix}}}
        \\
        \\[-1em]
        &= \sum_{\boldsymbol{(p)}_{> t}}^{}{\begin{pmatrix}
        \mathbb{I}_{t0} \\
        \left \langle \boldsymbol{p} \right |_{> t}
        \end{pmatrix}\left\{ \sum_{\boldsymbol{(i)}_{t0}}^{}{\begin{pmatrix}
        \left \langle \boldsymbol{i} \right |_{t0} \\
        \mathbb{I}_{> t}
        \end{pmatrix}{\widetilde{\rho}}_{\overline{< t}}\begin{pmatrix}
        \left | \boldsymbol{i} \right \rangle_{t0} \\
        \mathbb{I}_{> t}
        \end{pmatrix}} \right\}\left\{ \sum_{\boldsymbol{(j)}_{t0}}^{}{\begin{pmatrix}
        \left \langle \boldsymbol{j} \right |_{t0} \\
        \mathbb{I}_{> t}
        \end{pmatrix}{\widetilde{\rho}}_{\overline{< t}}\begin{pmatrix}
        \left | \boldsymbol{j} \right \rangle_{t0} \\
        \mathbb{I}_{> t}
        \end{pmatrix}} \right\}\begin{pmatrix}
        \mathbb{I}_{t0} \\
        \left | \boldsymbol{p} \right \rangle_{> t}
        \end{pmatrix}}
        = \text{Tr}\left[ {\widetilde{\rho}}_{> t}^{2} \right].
    \end{split}
\end{equation}
Noting that ${\widetilde{\rho}}_{> t} = \rho_{> t}^{(in)}$, and that the left parameterized quantum circuit of the QRNN does not affect the input data after time $t$, we further have
${\widetilde{\rho}}_{\overline{< t}} = {\widetilde{\rho}}_{t0} \otimes {\widetilde{\rho}}_{> t}$.
Thus, we obtain
\begin{equation}
    \begin{split}
        &\sum_{\boldsymbol{pq}}^{}{\delta_{\boldsymbol{(pq)}_{< t}}\delta_{\boldsymbol{(p'q')}_{< t}}\delta_{\boldsymbol{(pq')}_{> t}}\delta_{\boldsymbol{(p'q)}_{> t}}{\Delta\Psi}_{\boldsymbol{pq}}^{\boldsymbol{p'q'}}}\\
        &= \text{Tr}\left[ {\widetilde{\rho}}_{\overline{< t}}^{2} \right] - \frac{\text{Tr}\left[ {\widetilde{\rho}}_{> t}^{2} \right]}{2^{2m}} \\
        &= D_{HS}\left( {\widetilde{\rho}}_{\overline{< t}}, \frac{\mathbb{I}_{t0}}{2^{2m}} \otimes {\widetilde{\rho}}_{> t} \right)\\ 
        &= D_{HS}\left( {\widetilde{\rho}}_{t0}, \frac{\mathbb{I}_{t0}}{2^{2m}} \right) \text{Tr}\left[ {\widetilde{\rho}}_{> t}^{2} \right] \le D_{HS}\left( {\widetilde{\rho}}_{t0}, \frac{\mathbb{I}_{t0}}{2^{2m}} \right) \in [0,2],
    \end{split}
\end{equation}
where $D_{HS}(\rho,\sigma) = \text{Tr}\left[ (\rho - \sigma)^{2} \right]$ denotes the Hilbert--Schmidt distance, which measures the distance between two density operators in this context. The last equality follows from the identity
$D_{HS}(\rho \otimes \delta, \sigma \otimes \delta)
= \text{Tr}\left[ (\rho - \sigma)^{2} \right] \text{Tr}\left[ \delta^{2} \right]$.
It can be seen that when the reduced state ${\widetilde{\rho}}_{t0}$ on the registers $\{q_{0}, q_{t}\}$ after the action of $V_{\mathcal{L}}$ approaches the maximally mixed state, the gradient variance tends to zero, and the parameters in the variational circuit at time $t$ become increasingly difficult to train.

Putting everything together, we obtain
\begin{equation}
    \left\langle \left( \partial_{\nu}C \right)^{2} \right\rangle_{V}
    = \frac{2^{2m - 1}\text{Tr}\left[ \sigma_{\nu}^{2} \right]}{{(2^{4m} - 1)}^{2}}
    {\left(\frac{2^{m}}{2^{2m} + 1}\right)}^{T - t}
    D_{HS}\left( O_{T}, \frac{\text{Tr}[ O_{T}]\mathbb{I}_{T}}{2^{m}} \right)
    \left\langle D_{HS}\left( {\widetilde{\rho}}_{t0}, \frac{\mathbb{I}_{t0}}{2^{2m}} \right) \right\rangle_{V_{\mathcal{L}}}
    \text{Tr}\left[ {\widetilde{\rho}}_{> t}^{2} \right].
\end{equation}
We now provide a detailed analysis of the individual factors appearing in the above expression:

(a) The operator $\sigma_v$ acts on the registers $q_0$ and $q_t$ and is nontrivial only on a single qubit. Consequently, the factor $\text{Tr}[\sigma_v^2] = 2^{2m}$.

(b) The factor $\text{Tr}[\widetilde{\rho}_{>t}^2] < 1$ follows directly from the basic properties of density operators.

(c) For the expectation over $V_{\mathcal{L}}$, the quantity
$D_{HS}\left(\widetilde{\rho}_{t0}, \frac{\mathbb{I}_{t0}}{2^{2m}}\right) \in [0,2]$
quantifies the distance between two quantum states, with its specific value determined by the purity of the reduced state $\widetilde{\rho}_{t0}$.

(d) In contrast, the factor
$D_{HS}\left(O_T, \frac{\text{Tr}[O_T] \mathbb{I}_T}{2^m}\right)$
does not represent a distance between density operators, but instead involves an observable and therefore requires explicit evaluation. When
$O_T = \mathbb{I}_T - c\left | 0 \right \rangle \left \langle 0 \right |^{\otimes m}$
is a global observable, its computational complexity is constant, $\mathcal{O}(1)$. By contrast, when
$O_T = \mathbb{I}_T - \frac{1}{m}\sum_{j=1}^m c_j\left | 0 \right \rangle \left \langle 0 \right |_j \otimes \mathbb{I}_j$
is a local observable, the corresponding complexity scales as $\mathcal{O}(2^m/m)$. We note that the predicted value $y^{(i)}$ introduced in Section B has been omitted here, as it can always be absorbed into a global constant factor and has negligible impact on the complexity analysis.

These analyses lead directly to Theorem 1 stated in the main text.

\textit{Theorem 1.} (Upper Bound of Variance) For a QRNN with fully independent parameters, consider the parameter $\theta_t^v$ in the block $W_{t}$ shown in Fig.~\ref{FB2}. If $W_A$, $W_B$, and $W_{t0}(\forall t)$ form unitary $2$-designs, then the variance of the partial derivative in Eq.~(\ref{DE1}) admits the following upper bound:
\begin{equation}\label{D25}
    \text{Var}[\partial_\nu C] \leq g(T), \quad \text{with} \quad g(T) \in \mathcal{O} \left[ \left( \frac{1}{k} \right)^T \right],
\end{equation}
where $k = 2^m + 1/2^m \in [3/2, +\infty)$ and $m$ is constant. We also have
\begin{equation}\label{D26}
    \text{Var}[\partial_\nu C] \leq h(m), \quad \text{with} \quad h(m) \in \mathcal{O} \left[ \left( \frac{1}{2^c} \right)^m \right],
\end{equation}
where $c = 4 + T - t \in [4, +\infty)$ and $T$ is constant.

\hfill $\square$

Theorem 1 demonstrates that the gradient variance of a QRNN decays exponentially with both the number of time steps $T$ and the system size $m$, giving rise to the barren plateau (BP) phenomenon. This result indicates that, in the absence of mitigating strategies, training such models becomes increasingly difficult as their depth or size grows. The recurrent structure alone is therefore insufficient to prevent BPs in QRNNs, which is consistent with observations in other quantum neural network models, such as dissipative QNNs~\cite{sharma2022trainability,beer2020training}. Consequently, additional architectural strategies—such as parameter sharing—are required to preserve gradient magnitudes across many time steps and mitigate the barren plateau problem. These strategies will be explored further in Section~E.

\section{\label{E}Section E. Gradient Variance of QRNNs With Parameter Sharing}
In this section, we analyze the gradient variance of the loss function in QRNNs with parameters shared across recurrent blocks. We first quantify how parameter sharing amplifies the gradient variance along two key structural dimensions of the PQC, corresponding to the first and second examples in the main text. We then evaluate the gradient variance for a specific QRNN model under both parameter sharing and non-sharing configurations, corresponding to the third example in the main text. Finally, we discuss the general case and provide a proof of Theorem 2 presented in the main text.

\subsection{\label{E1}1. Gradient Variance along Two Structural Dimensions of the PQC}
We focus on the two fundamental structural dimensions of a PQC---circuit depth and qubit count---to quantify how parameter sharing amplifies the magnitude of the gradient variance. 

\begin{figure}[h]
    \centering
    \includegraphics[width=3.7in]{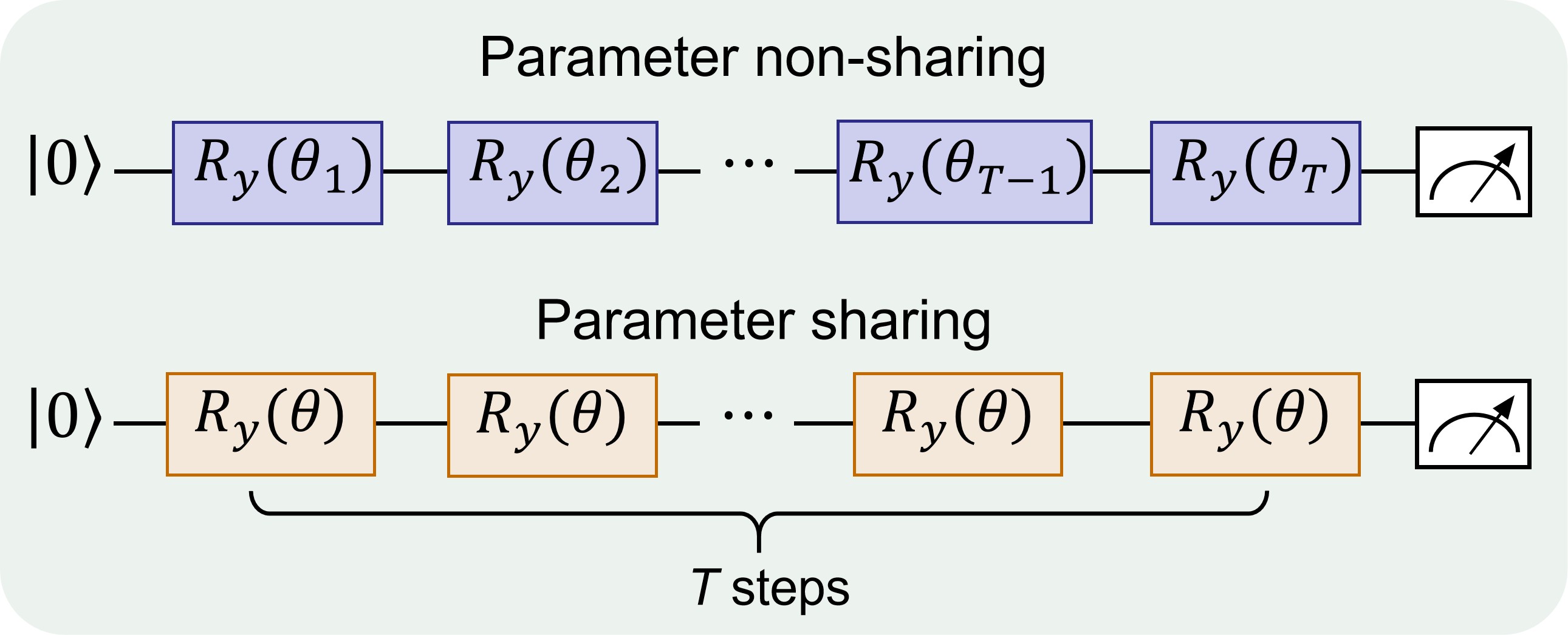}
     \caption{Ansatz circuits used to quantify how parameter sharing amplifies gradient variance along the circuit depth dimension (i.e., the longitudinal extension of the circuit).}
    \label{FE1}
\end{figure}

First, for the circuit depth dimension, we consider an ansatz of the form $V(\boldsymbol{\theta}) = \prod_{t=1}^T R_y(\theta_t)$, as shown in Fig. \ref{FE1}. In this circuit, a Pauli-Y rotation gate is applied $T$ times to a single qubit. In the non-sharing case, the parameters $\theta_{1},\cdots,\theta_{T}$ are independent, whereas under parameter sharing there is a single shared parameter $\theta$.

For the parameter non-sharing case, the loss function can be written as
\begin{equation}
    C = \text{Tr}\left[\left| \boldsymbol{0} \right\rangle \left\langle \boldsymbol{0} \right| R_y\left(\sum_{t = 1}^{T}\theta_{t}\right)\left| \boldsymbol{0} \right\rangle \left\langle \boldsymbol{0} \right| R_y^{\dagger}\left(\sum_{t = 1}^{T}\theta_{t}\right)\right] = \cos^{2}\left( \sum_{t = 1}^{T}\frac{\theta_{t}}{2} \right),
\end{equation}
where the initial state is chosen as $\rho = \left| \boldsymbol{0} \right\rangle \left\langle \boldsymbol{0} \right|$ and the measurement operator is $M = \left| \boldsymbol{0} \right\rangle \left\langle \boldsymbol{0} \right|$. The variance of the partial derivative of the loss function is then
\begin{equation}
    \text{Var}_{T}\left( \partial_{\theta_{t}}C \right) = \frac{1}{8}.
\end{equation}
Thus, the variance is independent of the time step and remains constant. 

For the parameter sharing case, the loss function becomes
\begin{equation}
    C = \text{Tr}\left[\left| \boldsymbol{0} \right\rangle \left\langle \boldsymbol{0} \right| R_y(T\theta)\left| \boldsymbol{0} \right\rangle \left\langle \boldsymbol{0} \right| R_y^{\dagger}(T\theta)\right] = \cos^{2}\left( \frac{T\theta}{2} \right),
\end{equation}
and the variance of the partial derivative is
\begin{equation}\label{E4}
    \text{Var}_{T}\left( \partial_{\theta}C \right) = \frac{T^{2}}{8}.
\end{equation}
Compared with the constant $1/8$ in the non-sharing case, this variance grows with the time step $T$. This behavior reflects the fact that, when parameters are independent, a small perturbation $\Delta\theta$ in a single module affects the output only locally. In contrast, under parameter sharing, the same perturbation in the shared parameter propagates through all $T$ modules, producing a collective effect of order $T\Delta\theta$. This leads to a multiplicative increase in the gradient variance.

\begin{figure}[h]
    \centering
    \includegraphics[width=3.7in]{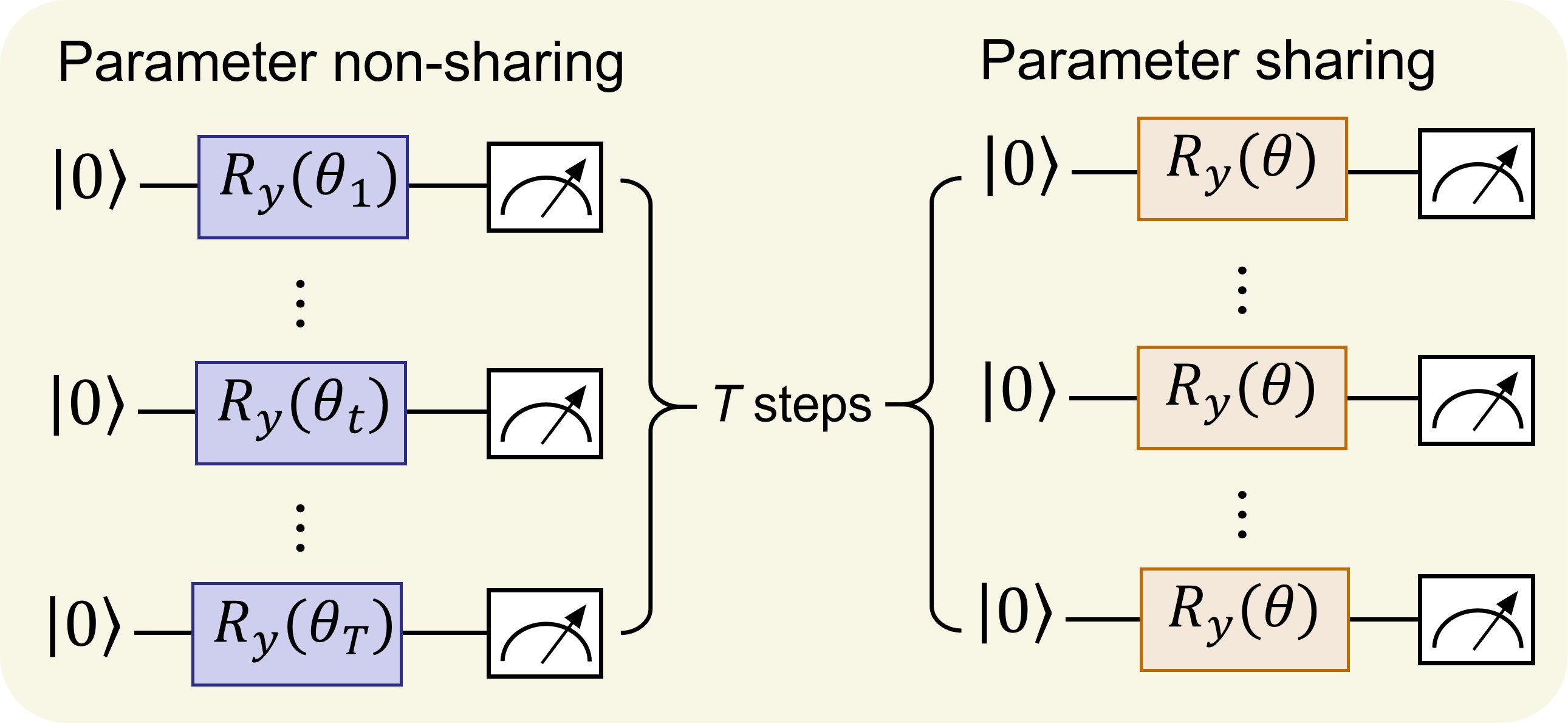}
    \caption{Ansatz circuits used to quantify how parameter sharing amplifies gradient variance along the qubit count dimension (i.e., the transverse extension of the circuit).}
    \label{FE2}
\end{figure}

Second, for the qubit count dimension, we consider an ansatz of the form $V(\boldsymbol{\theta}) = \bigotimes_{t=1}^T R_y(\theta_t)$, as shown in Fig. \ref{FE2}. For the parameter non-sharing case, the loss function can be written as 
\begin{equation}
    C = \text{Tr}\left[ M\left( \otimes_{t = 1}^{T}R_y(\theta_{t}) \right)\rho\left( \otimes_{t = 1}^{T}R_y^{\dagger}(\theta_{t}) \right)\right] = \prod_{t = 1}^{T}{\cos^{2}\left( \frac{\theta_{t}}{2} \right)},
\end{equation}
where the initial state is chosen as $\rho = \left( \left| \boldsymbol{0} \right\rangle \left\langle \boldsymbol{0} \right|  \right)^{\otimes T}$ and the measurement operator is $M = \left( \left| \boldsymbol{0} \right\rangle \left\langle \boldsymbol{0} \right|  \right)^{\otimes T}$. The variance of the partial derivative of the loss function is readily obtained as
\begin{equation}\label{E6}
\begin{aligned}
    \text{Var}_{T}\left( \partial_{\theta_{t}}C \right) &= \frac{1}{(2\pi)^{T}}\int_{0}^{2\pi}{\cos^{4}\left( \frac{\theta_{1}}{2} \right)d\theta_{1}}\cdots\int_{0}^{2\pi}{\cos^{2}\left( \frac{\theta_{t}}{2} \right)\sin^{2}\left( \frac{\theta_{t}}{2} \right)d\theta_{t}}\cdots\int_{0}^{2\pi}{\cos^{4}\left( \frac{\theta_{T}}{2} \right)d\theta_{T}} \\
    &= \frac{1}{(2\pi)^{T}}\left( \frac{3\pi}{4} \right)\cdots\left( \frac{\pi}{4} \right)\cdots\left( \frac{3\pi}{4} \right) \\
    &= \frac{1}{8}\left( \frac{3}{8} \right)^{T - 1}.
\end{aligned}
\end{equation}
Interestingly, even such a simple circuit can exhibit a BP phenomenon under global measurements, as also noted in Ref.~\cite{cerezo2021cost}. 

For the parameter sharing case, the loss function becomes
\begin{equation}
    C = \text{Tr}\left[ M\left( \otimes_{t = 1}^{T}R_y(\theta) \right)\rho\left( \otimes_{t = 1}^{T}R_y^{\dagger}(\theta) \right)\right] = \cos^{2T}\left( \frac{\theta}{2} \right),
\end{equation}
where the initial state is chosen as $\rho = \left( \left| \boldsymbol{0} \right\rangle \left\langle \boldsymbol{0} \right|  \right)^{\otimes n}$ and the measurement operator is $M = \left( \left| \boldsymbol{0} \right\rangle \left\langle \boldsymbol{0} \right|  \right)^{\otimes n}$. The variance of the partial derivative of the loss function is
\begin{equation}
    \left\langle \left( \frac{\partial C}{\partial\theta} \right)^{2} \right\rangle = \frac{T^{2}}{2\pi}\int_{0}^{2\pi}{\cos^{4T - 2}\left( \frac{\theta}{2} \right)}\sin^{2}\left( \frac{\theta}{2} \right)d\theta.
\end{equation}
Using the substitution $\varphi = \theta/2$ and the symmetry of the integrand about $\varphi = \pi/2$, we obtain
\begin{equation}
    \left\langle \left( \frac{\partial C}{\partial\theta} \right)^{2} \right\rangle = \frac{T^{2}}{\pi}\int_{0}^{\pi}{\cos^{4T - 2}(\varphi)}\sin^{2}(\varphi)d\varphi = \frac{2T^{2}}{\pi}\int_{0}^{\pi/2}{\cos^{4T - 2}(\varphi)}\sin^{2}(\varphi)d\varphi.
\end{equation}
Furthermore, using the trigonometric form of the Beta function
\begin{equation}
    B\left(\frac{p + 1}{2},\frac{q + 1}{2}\right) = 2\int_{0}^{\pi/2}{\sin^{p}(\varphi)}\cos^{q}(\varphi)d\varphi,
\end{equation}
and setting $p = 2$, $q = 4T - 2$, we have
\begin{equation}
    \left\langle \left( \frac{\partial C}{\partial\theta} \right)^{2} \right\rangle = \frac{T^{2}}{\pi}B\left( \frac{3}{2},\frac{4T - 1}{2} \right).
\end{equation}
Using the identity of Gamma function $B(p,q) = \Gamma(p)\Gamma(q)/\Gamma(p + q)$, it follows that
\begin{equation}
\begin{aligned}
    \left\langle \left( \frac{\partial C}{\partial\theta} \right)^{2} \right\rangle &= \frac{T^{2}}{\pi}\frac{\Gamma\left( \frac{3}{2} \right)\Gamma\left( \frac{4T - 1}{2} \right)}{\Gamma(2T + 1)} \\
    &= \frac{T^{2}}{\pi}\frac{\Gamma\left( \frac{3}{2} \right)\left( \frac{4T - 3}{2} \right)\left( \frac{4T - 5}{2} \right)\cdots\left( \frac{5}{2} \right)\left( \frac{3}{2} \right)\Gamma\left( \frac{3}{2} \right)}{2T!} \\
    &= \frac{T^{2}}{\pi}\frac{\Gamma\left( \frac{3}{2} \right)\left( \frac{4T - 2}{2} \right)\left( \frac{4T - 3}{2} \right)\left( \frac{4T - 4}{2} \right)\left( \frac{4T - 5}{2} \right)\cdots\left( \frac{5}{2} \right)\left( \frac{4}{2} \right)\left( \frac{3}{2} \right)\left( \frac{2}{2} \right)\Gamma\left( \frac{3}{2} \right)}{\left[ (2T - 1)(2T - 2)\cdots 2 \cdot 1 \right] 2T!} \\
    &= \frac{T^{2}}{\pi}\frac{\pi \cdot (4T - 2)!}{\left[ (2T - 1)! \right] \cdot (2T!) \cdot 2^{4T - 1}} \\
    &= \frac{{4T}^{3} \cdot (4T - 2)!}{(2T!)^{2} \cdot 2^{4T}} = \frac{T}{2^{4T}}\binom{4T - 2}{2T - 1},
\end{aligned}
\end{equation}
where we use $\Gamma(3/2) = \sqrt{\pi}$ and $\Gamma(n) = (n - 1)\Gamma(n - 1)$. Here, $\binom{4T - 2}{2T - 1} = \frac{(4T - 2)!}{(2T!)^{2}}$ denotes the binomial coefficient. 

To study how this variance scales with $T$, we consider the ratio of successive terms,
\begin{equation}
\begin{aligned}
    \frac{\text{Var}_{T + 1}}{\text{Var}_{T}}\  &= \frac{{4(T + 1)}^{3}}{{4(T)}^{3}}\frac{(4T + 2)!}{(4T - 2)!}\frac{(2T!)(2T!)}{(2T + 2)!(2T + 2)!}\frac{2^{4T}}{2^{4T + 4}} \\
    &= \frac{(T + 1)^{3}}{(T)^{3}}\frac{(4T + 2)(4T + 1)(4T)(4T - 1)}{(2T + 2)(2T + 2)(2T + 1)(2T + 1)}\frac{1}{2^{4}} \\
    &= \frac{(T + 1)^{3}}{(T)^{3}}\frac{(T + 1/2)(T + 1/4)(T)(T - 1/4)}{(T + 1)(T + 1)(T + 1/2)(T + 1/2)} \\
    &= \frac{(T + 1/4)}{(T)}\frac{(T + 1)}{(T + 1/2)}\frac{(T - 1/4)}{(T)} \\
    &= \frac{(T + 1/4)}{(T)}\frac{\left( T^{2} + T/2 \right) + (T - 1)/4}{\left( T^{2} + T/2 \right)} > 1.
\end{aligned}
\end{equation}
This implies that, for $T \geq 1$, the variance increases monotonically with $T$. Therefore, no BP occurs in this case. We next examine whether this growth admits an upper bound. Using Stirling's approximation, 
\begin{equation}
    n! \approx \sqrt{2\pi n}\left( \frac{n}{e} \right)^{n}.
\end{equation}
we can further simplify the variance as
\begin{equation}
    \text{Var}_{T}\left( \partial_{\theta}C \right) \approx \frac{{4T}^{3} \cdot \sqrt{2\pi(4T - 2)}\left( \frac{4T - 2}{e} \right)^{4T - 2}}{4\pi T \cdot \left( \frac{4T}{e} \right)^{4T}}.
\end{equation}
Taking the limit $T \rightarrow + \infty$, we obtain
\begin{equation}\label{E16}
    \lim_{T \rightarrow + \infty}{\text{Var}_{T}\left( \partial_{\theta}C \right)} \approx \frac{{4T}^{3} \cdot \sqrt{8\pi T}}{4\pi T \cdot (4T)^{2}} = \frac{\sqrt{2\pi T}}{8\pi} \in \mathcal{O}(\sqrt{T}).
\end{equation}
Thus, the variance is upper bounded by $\mathcal{O}(\sqrt{T})$, indicating that it does not lead to an exponential gradient explosion, as expected for QRNNs with CPTP forward transformations.    

Comparing Eq. (\ref{E6}) with Eq. (\ref{E16}), we can see that parameter sharing can substantially change the scaling behavior of the gradient variance along the qubit count dimension of the PQC, transitioning it from exponential decay to polynomial growth. Taken together, the two ansatz circuits above show that parameter sharing can amplify the gradient variance along the two key structural dimensions of the PQC, highlighting its potential to mitigate the BP phenomenon in QRNNs.

\subsection{\label{E2}2. Gradient Variance of a Specific QRNN Model}
\begin{figure}[h]
    \centering
    \includegraphics[width=5.2in]{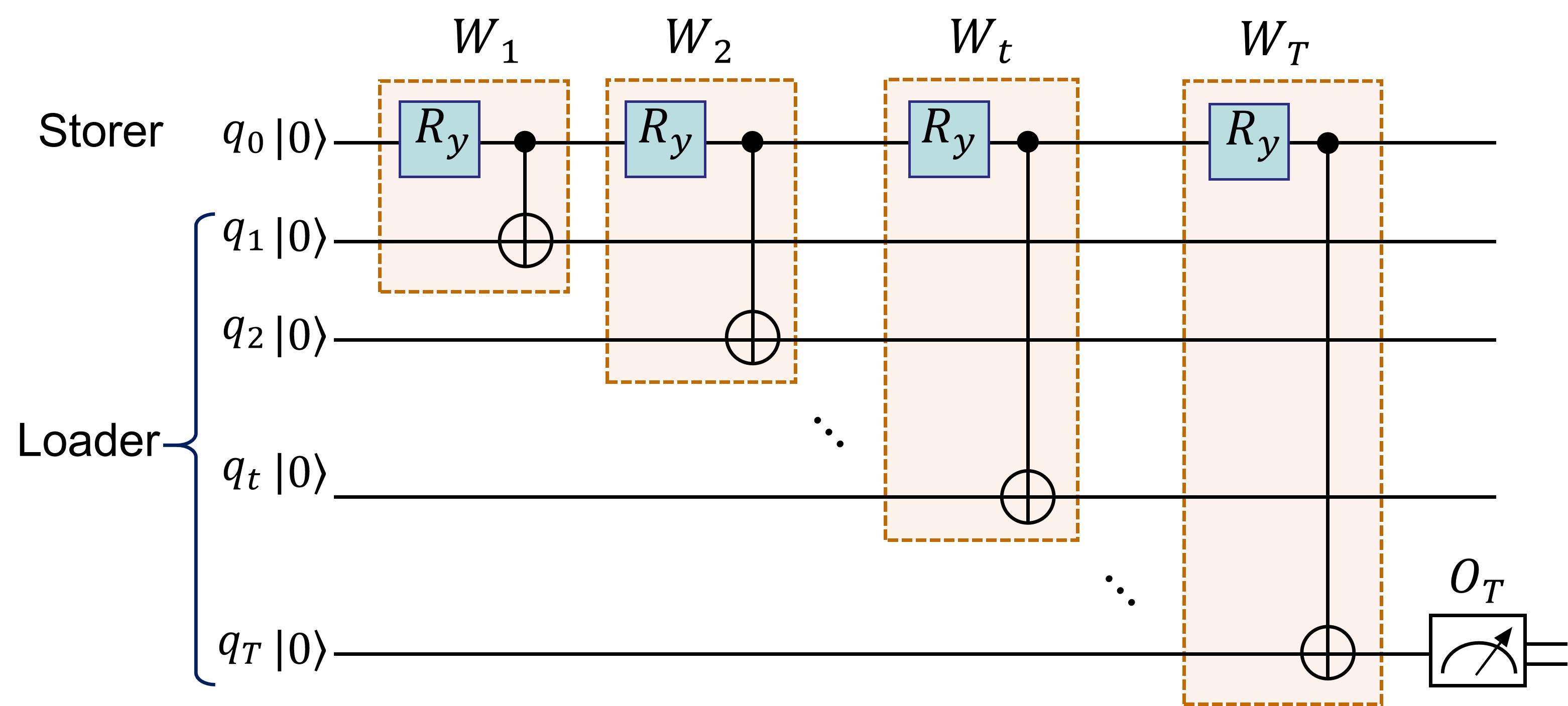}
    \caption{A specific QRNN architecture constructed from single-qubit $R_y$ gates and CNOT gates.}
    \label{FE3}
\end{figure}

By combining the longitudinal and transverse extension of the above ansatz circuits, we propose a specific QRNN model, as shown in Fig. \ref{FE3}. The circuit contains parameterized $R_y$ rotation gates and CNOT gates that entangle the Storer and Loader registers. We evaluate the gradient variance under both parameter-sharing and non-sharing settings. 

Without loss of generality, the initial quantum state is $\rho = \left( \left| \boldsymbol{0} \right\rangle \left\langle \boldsymbol{0} \right|  \right)^{\otimes(T + 1)}$, and the measurement operator is $M = \mathbb{I}_{\overline{T}}\otimes{\left| \boldsymbol{0} \right\rangle \left\langle \boldsymbol{0} \right| }_{T}$ acted on the last qubit $q_{T}$. When calculating the general form of the loss function, the entanglement in the quantum circuit causes the evolution matrix inside the trace operation to be unable to be expressed as a tensor product of subsystems. As the time step increases, the matrix dimension grows exponentially, greatly increasing computational complexity. We use mathematical induction to derive the loss function. 

First, for the parameter non-sharing case, the loss function for $T = 1,2,3,4,5$ can be expressed as
\begin{equation}\label{E17}
\begin{aligned}
    C_{1} &= \left( 1 + \cos\theta_{1} \right)/2, \\
    C_{2} &= \left( 1 + \cos\theta_{1}\cos\theta_{2} \right)/2, \\
    C_{3} &= \left( 1 + \cos\theta_{1}\cos\theta_{2}\cos\theta_{3} \right)/2, \\
    C_{4} &= \left( 1 + \cos\theta_{1}\cos\theta_{2}\cos\theta_{3}\cos\theta_{4} \right)/2, \\
    C_{5} &= \left( 1 + \cos\theta_{1}\cos\theta_{2}\cos\theta_{3}\cos\theta_{4}\cos\theta_{5} \right)/2.
\end{aligned}
\end{equation}
Given the regularity of the QRNN model with increasing time steps and the consistency of the output results, we conjecture that the general form of loss function is
\begin{equation}
    C_{T} = \frac{1}{2}\left( 1 + \prod_{t = 1}^{T}{\cos\theta_{t}} \right).
\end{equation}
We can use mathematical induction to prove the validity of this general formula.

\noindent{\textit{Proof.}} According to Eq. (\ref{E17}), the general formula holds for $T = 1$. Assume that for $T = k$, the general formula holds
\begin{equation}
    C_{k} = \frac{1}{2}\left( 1 + \prod_{t = 1}^{k}{\cos\theta_{t}} \right).
\end{equation}
After the action of $k$ Ansatz blocks, the quantum state of the entire system can be written as:
\begin{equation}
    \left| \boldsymbol{\psi_{k}} \right\rangle = \alpha_{k}\left| \boldsymbol{\phi_{0}} \right\rangle_{k} \otimes \left( \left| \boldsymbol{0} \right\rangle_{0}\left| \boldsymbol{0} \right\rangle_{k} \right) + \beta_{k}\left| \boldsymbol{\phi_{1}} \right\rangle_{k} \otimes \left( \left| \boldsymbol{1} \right\rangle_{0}\left| \boldsymbol{1} \right\rangle_{k} \right),
\end{equation}
where $\left| \boldsymbol{0} \right\rangle_{0}$ and $\left| \boldsymbol{1} \right\rangle_{0}$ represent the states of $q_{0}$, while $\left| \boldsymbol{0} \right\rangle_{k}$ and $\left| \boldsymbol{1} \right\rangle_{k}$ represent the states of $q_{k}$. Additionally, $\left| \boldsymbol{\phi_{0}} \right\rangle_{k}$ and $\left| \boldsymbol{\phi_{1}} \right\rangle_{k}$ describe the states of qubits $\{ q_{1},\ldots,q_{k - 1}\}$, respectively. Since the CNOT gates in the circuit only transfer the state of qubit $q_{0}$ to qubit $q_{k}$, and the initial state of qubit $q_{k}$ is $\left| \boldsymbol{0} \right\rangle_{k}$, only the all-zero or all-one states are possible. Because $C_{k} = \left| \alpha_{k} \right|^{2} = 1 - \left| \beta_{k} \right|^{2}$, the two coefficients can always be written as:
\begin{equation}
    \alpha_{k} = e^{i\gamma_{0}}\sqrt{\frac{1}{2}\left( 1 + \prod_{t = 1}^{k}{\cos\theta_{t}} \right)}, \quad \beta_{k} = e^{i\gamma_{1}}\sqrt{\frac{1}{2}\left( 1 - \prod_{t = 1}^{k}{\cos\theta_{t}} \right)},
\end{equation}
where $e^{i\gamma_{0}}$ and $e^{i\gamma_{1}}$ are the possible complex phase factors included in the two coefficients.

For $T = k + 1$, adding a new qubit $q_{k + 1}$, the quantum state after the action of the first $k$ Ansatz blocks becomes $\left| \boldsymbol{\psi_{k}} \right\rangle\left| \boldsymbol{0} \right\rangle_{k + 1}$. After passing through the $R_y$ gate and the CNOT gate, the quantum state becomes
\begin{equation}
\begin{aligned}
    \left| \boldsymbol{\psi_{k + 1}} \right\rangle =& \alpha_{k}\cos\frac{\theta_{k + 1}}{2}\left| \boldsymbol{0} \right\rangle_{0}(\left| \boldsymbol{\phi_{0}} \right\rangle_{k}\left| \boldsymbol{0} \right\rangle_{k})\left| \boldsymbol{0} \right\rangle_{k + 1} + \alpha_{k}\sin\frac{\theta_{k + 1}}{2}\left| \boldsymbol{1} \right\rangle_{0}(\left| \boldsymbol{\phi_{0}} \right\rangle_{k}\left| \boldsymbol{0} \right\rangle_{k})\left| \boldsymbol{1} \right\rangle_{k + 1} \\
    & - \beta_{k}\sin\frac{\theta_{k + 1}}{2}\left| \boldsymbol{0} \right\rangle_{0}(\left| \boldsymbol{\phi_{1}} \right\rangle_{k}\left| \boldsymbol{1} \right\rangle_{k})\left| \boldsymbol{0} \right\rangle_{k + 1} + \beta_{k}\cos\frac{\theta_{k + 1}}{2}\left| \boldsymbol{1} \right\rangle_{0}(\left| \boldsymbol{\phi_{1}} \right\rangle_{k}\left| \boldsymbol{1} \right\rangle_{k})\left| \boldsymbol{1} \right\rangle_{k + 1}.
\end{aligned}
\end{equation}
Due to the effect of the CNOT gates, the state of the last qubit is strongly correlated with the first qubit. Rewriting the quantum state based on the states of these two qubits
\begin{equation}
\begin{aligned}
    \left| \boldsymbol{\psi_{k + 1}} \right\rangle =& \left( \alpha_{k}\cos\frac{\theta_{k + 1}}{2}\left| \boldsymbol{\widetilde{\phi}_{0}} \right\rangle_{k} - \beta_{k}\sin\frac{\theta_{k + 1}}{2}\left| \boldsymbol{\widetilde{\phi}_{1}} \right\rangle_{k} \right)\left| \boldsymbol{0} \right\rangle_{0}\left| \boldsymbol{0} \right\rangle_{k + 1} \\
    &+ \left( \alpha_{k}\sin\frac{\theta_{k + 1}}{2}\left| \boldsymbol{\widetilde{\phi}_{0}} \right\rangle_{k} + \beta_{k}\cos\frac{\theta_{k + 1}}{2}\left| \boldsymbol{\widetilde{\phi}_{1}} \right\rangle_{k} \right)\left| \boldsymbol{1} \right\rangle_{0}\left| \boldsymbol{1} \right\rangle_{k + 1},
\end{aligned}
\end{equation}
where $\left| \boldsymbol{\widetilde{\phi}_{0}} \right\rangle_{k} = \left| \boldsymbol{\phi_{0}} \right\rangle_{k}\left| \boldsymbol{0} \right\rangle_{k}$, $\left| \boldsymbol{\widetilde{\phi}_{1}} \right\rangle_{k} = \left| \boldsymbol{\phi_{1}} \right\rangle_{k}\left| \boldsymbol{1} \right\rangle_{k}$. Further normalization gives,
\begin{equation}
\begin{aligned}
    \left| \boldsymbol{\psi_{k + 1}} \right\rangle =& \alpha_{k + 1}\left( \frac{\alpha_{k}}{\alpha_{k + 1}}\cos\frac{\theta_{k + 1}}{2}\left| \boldsymbol{\widetilde{\phi}_{0}} \right\rangle_{k} - \frac{\beta_{k}}{\alpha_{k + 1}}\sin\frac{\theta_{k + 1}}{2}\left| \boldsymbol{\widetilde{\phi}_{1}} \right\rangle_{k} \right)\left| \boldsymbol{0} \right\rangle_{0}\left| \boldsymbol{0} \right\rangle_{k + 1} \\
    &+ \beta_{k + 1}\left( \frac{\alpha_{k}}{\beta_{k + 1}}\sin\frac{\theta_{k + 1}}{2}\left| \boldsymbol{\widetilde{\phi}_{0}} \right\rangle_{k} + \frac{\beta_{k}}{\beta_{k + 1}}\cos\frac{\theta_{k + 1}}{2}\left| \boldsymbol{\widetilde{\phi}_{1}} \right\rangle_{k} \right)\left| \boldsymbol{1} \right\rangle_{0}\left| \boldsymbol{1} \right\rangle_{k + 1},
\end{aligned}
\end{equation}
where the normalization coefficients are
\begin{equation}
    \alpha_{k + 1} = \sqrt{\left| \alpha_{k}\cos\frac{\theta_{k + 1}}{2} \right|^{2} + \left| \beta_{k}\sin\frac{\theta_{k + 1}}{2} \right|^{2}}, \quad \beta_{k + 1} = \sqrt{\left| \alpha_{k}\sin\frac{\theta_{k + 1}}{2} \right|^{2} + \left| \beta_{k}\cos\frac{\theta_{k + 1}}{2} \right|^{2}}.
\end{equation}
Thus, we obtain the loss function
\begin{equation}
\begin{aligned}
    C_{k + 1} &= \left| \alpha_{k + 1} \right|^{2} = \left| \alpha_{k}\cos\frac{\theta_{k + 1}}{2} \right|^{2} + \left| \beta_{k}\sin\frac{\theta_{k + 1}}{2} \right|^{2} \\
    &= \frac{1}{2}\left( 1 + \prod_{t = 1}^{k}{\cos\theta_{t}} \right)\cos^{2}\frac{\theta_{k + 1}}{2} + \frac{1}{2}\left( 1 - \prod_{t = 1}^{k}{\cos\theta_{t}} \right)\sin^{2}\frac{\theta_{k + 1}}{2} \\
    &= \frac{1}{2}\left( 1 + \prod_{t = 1}^{k + 1}{\cos\theta_{t}} \right),
\end{aligned}
\end{equation}
which is consistent with the general formula. The proof is complete.

\hfill $\square$

From this result, by taking partial derivatives and integrating, we obtain the variance
\begin{equation}\label{E27}
\begin{aligned}
    \text{Var}_{T}\left( \partial_{\theta_{t}}C \right) &= \frac{1}{(2\pi)^{T}}\frac{1}{4}\int_{0}^{2\pi}{\cos^{2}\left( \theta_{1} \right)d\theta_{1}}\cdots\int_{0}^{2\pi}{\sin^{2}\left( \theta_{t} \right)d\theta_{t}}\cdots\int_{0}^{2\pi}{\cos^{2}\left( \theta_{T} \right)d\theta_{T}} \\
    &= \frac{1}{2^{T + 2}},
\end{aligned}
\end{equation}
which shows that the variance decreases exponentially, exhibiting the BP phenomenon.

Second, for the parameter sharing case, the loss function for $T = 1,2,3,4,5,6,7$ can be expressed as
\begin{equation}
\begin{aligned}
    C_{1} &= \cos^{2}\frac{\theta}{2}, \\
    C_{2} &= \cos^{4}\frac{\theta}{2} + \sin^{4}\frac{\theta}{2}, \\
    C_{3} &= \cos^{6}\frac{\theta}{2} + 3\cos^{2}\frac{\theta}{2}\sin^{4}\frac{\theta}{2}, \\
    C_{4} &= \cos^{8}\frac{\theta}{2} + 6\cos^{4}\frac{\theta}{2}\sin^{4}\frac{\theta}{2} + \sin^{8}\frac{\theta}{2}, \\
    C_{5} &= \cos^{10}\frac{\theta}{2} + 10\cos^{6}\frac{\theta}{2}\sin^{4}\frac{\theta}{2} + 5\cos^{2}\frac{\theta}{2}\sin^{8}\frac{\theta}{2}, \\
    C_{6} &= \cos^{12}\frac{\theta}{2} + 15\cos^{8}\frac{\theta}{2}\sin^{4}\frac{\theta}{2} + 15\cos^{4}\frac{\theta}{2}\sin^{8}\frac{\theta}{2} + \sin^{12}\frac{\theta}{2}, \\
    C_{7} &= \cos^{14}\frac{\theta}{2} + 21\cos^{10}\frac{\theta}{2}\sin^{4}\frac{\theta}{2} + 35\cos^{6}\frac{\theta}{2}\sin^{8}\frac{\theta}{2} + 7\cos^{2}\frac{\theta}{2}\sin^{12}\frac{\theta}{2}.
\end{aligned}
\end{equation}
From these results, the general formula can be summarized as
\begin{equation}\label{eq102}
    C_{T} = \sum_{k = 0}^{\left\lfloor \frac{T}{2} \right\rfloor}\binom{T}{2k}\cos^{2T - 4k}\left( \frac{\theta}{2} \right)\sin^{4k}\left( \frac{\theta}{2} \right),
\end{equation}
where the operation "$\left\lfloor \cdot \right\rfloor$" denotes the floor function. Note that the summation in Eq. (\ref{eq102}) corresponds to the even terms of a binomial expansion, $(a + b)^{n} = \sum_{k = 0}^{n}{\binom{n}{k}a^{n - k}b^{k}}$. Specifically, the even terms of a binomial expansion can be expressed as 
\begin{equation}
    \left[ (a + b)^{n} \right]_{\text{even terms}} = \frac{1}{2}\left[ (a + b)^{n} + (a - b)^{n} \right] = \sum_{k = 0}^{\left\lfloor \frac{n}{2} \right\rfloor}{\binom{n}{2k}a^{n - 2k}b^{2k}}.
\end{equation}
Let $a = \cos^{2}{(\theta/2)}$ and $b = \sin^{2}{(\theta/2)}$, and we can then derive the binomial expansion of Eq. (\ref{eq102}). By applying this method again, we obtain the general form of the loss function,
\begin{equation}\label{E31}
    C_{T} = \frac{1}{2}\left[ \left( \cos^{2}\frac{\theta}{2} + \sin^{2}\frac{\theta}{2} \right)^{T} + \left( \cos^{2}\frac{\theta}{2} - \sin^{2}\frac{\theta}{2} \right)^{T} \right] = \frac{1 + \cos^{T}\theta}{2}.
\end{equation}
The strict proof of Eq. (\ref{E31}) follows a similar mathematical induction process as that for the parameter non-sharing case. Differentiating this expression and squaring gives
\begin{equation}
    \left( \frac{dC_{T}}{d\theta} \right)^{2} = \frac{T^{2}}{4}\cos^{2T - 2}(\theta)\sin^{2}(\theta).
\end{equation}
Further integration yields the variance as follows:
\begin{equation}\label{E33}
\begin{aligned}
    \text{Var}_{T}\left( \partial_{\theta}C \right) &= \frac{T^{2}}{4}\frac{1}{2\pi}\int_{0}^{2\pi}{\cos^{2T - 2}(\theta)\sin^{2}(\theta)d\theta} \\
    &= \frac{T^{2}}{2\pi}\int_{0}^{\pi/2}{\cos^{2T - 2}(\theta)\sin^{2}(\theta)d\theta} \\
    &= \frac{T^{2}}{4\pi}B\left(\frac{3}{2},T - \frac{1}{2}\right) \\
    &= \frac{T^{2}}{4\pi}\frac{\Gamma\left( \frac{3}{2} \right)\Gamma\left( T - \frac{1}{2} \right)}{\Gamma(T + 1)} \\
    &= \frac{1}{2}\frac{T}{4^{T}}\frac{(2T - 2)!}{\left[ (T - 1)! \right]^{2}} = \frac{1}{2}\frac{T}{4^{T}}\binom{2T - 2}{T - 1},
\end{aligned}
\end{equation}
where we use the Beta function, the Gamma function, and their properties. 

To study how this variance scales with $T$, we also consider the ratio of successive terms,
\begin{equation}
    \frac{\text{Var}_{T + 1}}{\text{Var}_{T}}\  = \frac{(T + 1)(T - 1/2)}{T^{2}} = \frac{T^{2} + (T - 1)/2}{T^{2}} > 1.
\end{equation}
For $T > 1$, the ratio is greater than 1, indicating that the variance increases monotonically. Using Stirling's approximation for factorials, we can further examine the behavior (complexity) of the variance as $T \rightarrow + \infty$:
\begin{equation}\label{E35}
\begin{aligned}
    \lim_{T \rightarrow + \infty}{\text{Var}_{T}\left( \partial_{\theta}C \right)} &\approx \lim_{T \rightarrow + \infty}\frac{T \cdot \sqrt{2\pi(2T - 2)}\left( \frac{2T - 2}{e} \right)^{2T - 2}}{2 \cdot 4^{T} \cdot 2\pi(T - 1)\left( \frac{T - 1}{e} \right)^{2T - 2}} \\
    &= \frac{\sqrt{T}}{8\sqrt{\pi}} \in \mathcal{O}(\sqrt{T}),
\end{aligned}
\end{equation}
This shows that with parameter sharing, the variance increases with $T$ at a complexity of $\mathcal{O}(\sqrt{T})$, and neither barren plateau nor gradient explosion effects occur. 

By comparing Eq. (\ref{E27}) and Eq. (\ref{E35}), we can see that parameter sharing fundamentally alters the gradient scaling behavior in QRNNs, transitioning from exponential decay $\mathcal{O}(2^{-T})$ to polynomial growth $\mathcal{O}(\sqrt{T})$. 

\subsection{\label{E3}3. The General Case}
In this subsection, we focus on the general case and analyze the BP behavior of QRNNs with parameter sharing across recurrent blocks. To address the intractability of the Haar measure approach in the parameter sharing case, we compute the gradient variance by exploiting the trigonometric-function representation of the loss function, which allows us to establish Theorem 2 in the main text. As a complementary analysis, we conclude this subsection with a discussion based on the Haar-measure approach, which further supports our conclusions.

Based on the formalism of the QRNN model introduced in Section~B, the loss function can always be written as
\begin{equation}\label{E36-2}
    C
    = \mathrm{Tr}\left[
    O\, V(\theta)\,\rho\, V^{\dagger}(\theta)
    \right]
    = \mathrm{Tr}\left[
    O \left( \prod_{t=1}^{T} W_t \right)
    \rho
    \left( \prod_{t=1}^{T} W_t^{\dagger} \right)
    \right].
\end{equation}
Here we do not impose the constraint in Eq.~(\ref{B10}) that $W_t$ acts nontrivially only on subsystem $q_t$. As a result, the above expression can be viewed as a general loss function generated by a parameterized quantum circuit composed of $T$ ansatz blocks, and is therefore applicable to a broad class of variational quantum algorithms. Each block $W_t$ can be written as
\begin{equation}\label{E36-1}
    W_t
    = \prod_{i=1}^{\xi}
    \left[
    \cos\left(\frac{\theta_i^{(t)}}{2}\right)\mathbb{I}
    - i \sin\left(\frac{\theta_i^{(t)}}{2}\right)\sigma_i^{(t)}
    \right] Q_i^{(t)} ,
\end{equation}
where we adopt the same general ansatz expansion as in Eq.~(\ref{B12}), with a slight adjustment of index notation. Both $\mathbb{I}$ and $\sigma_i^{(t)}$ are operators defined on the full system Hilbert space. Specifically, $\sigma_i^{(t)}$ denotes a Pauli string that acts nontrivially only on a single qubit $j$, i.e., $\sigma_i^{(t)} = (\sigma_i^{(t)})_{j} \otimes \mathbb{I}_{\overline{j}}$ with $(\sigma_i^{(t)})_{j} \in \{\sigma_x,\sigma_y,\sigma_z\}$, and $Q_i^{(t)}$ denotes a parameter-free quantum gate. The superscript $(t)$ indicates that the parameters associated with different time steps are independent, corresponding to the non-sharing case; removing this superscript corresponds to the parameter-sharing case.

In the parameter non-sharing case, let the circuit contain $N$ independent parameters $\{\theta_i\}_{i=1}^N$. For a typical QRNN, the number of parameters scales linearly with the system size, i.e., $N = \xi T$, where $\xi$ denotes the number of parameters per PQC layer and is a constant. Since each parameterized gate (e.g., $R_x$, $R_y$, or $R_z$) introduces matrix elements proportional to $\cos(\theta_i/2)$ and $\sin(\theta_i/2)$, and the loss function involves both $U$ and $U^\dagger$, each independent parameter $\theta_i$ appears in the overall product only through two such trigonometric factors. By using standard double-angle identities, these factors combine into a single $\cos(\theta_i)$ or $\sin(\theta_i)$. Consequently, combining Eq.~(\ref{E36-2}) with the expansion in Eq.~(\ref{E36-1}), the loss function can always be written as
\begin{equation}\label{E36}
    C = \sum_k \alpha_k \prod_{i=1}^{\xi T} \cos^{a_{ki}}(\theta_i) \sin^{b_{ki}}(\theta_i),
\end{equation}
where for each $(k,i)$ the exponents satisfy $a_{ki}, b_{ki} \in \{0,1\}$ and $a_{ki}+b_{ki}\le1$. The coefficients $\alpha_k$ depend on the observable $O$, the density matrix $\rho$, and the parameter-free quantum gates $Q_i$.

After taking the derivative with respect to a specific parameter $\nu$, squaring, and performing a uniform average over all $\theta_i$ on $[0,2\pi)$, each independent parameter contributes an integration factor $1/(2\pi)$ together with a trigonometric integral $I_i \in [0,2\pi)$. Their product yields a constant factor $c_i \in (0,1)$ (for example, $\int_0^{2\pi}\cos^2\theta\,d\theta=\pi$ gives $c_i = 1/2$). Therefore, after integrating over all $N = \xi T$ independent parameters, the total variance acquires the factor
\begin{equation}\label{E37}
    \mathrm{Var}(\partial_\nu C_{\mathrm{non}})
    \lesssim
    \sum_k \beta_k \prod_{i=1}^{\xi T} c_{ki},
\end{equation}
where $\beta_k \ge 0$ collects the prefactors arising from differentiation and the coefficients $\alpha_k$.

We note that the number of distinct trigonometric monomials in Eq.~(\ref{E36}) can in principle be as large as $3^{\xi T}$, since each independent parameter may contribute through $1$, $\cos\theta_i$, or $\sin\theta_i$. Therefore, the relevant quantity in Eq.~(\ref{E37}) is not the bare combinatorial count of terms, but their total weighted contribution, encoded by the coefficients $\beta_k$. Although the formal expansion may contain exponentially many distinct monomials, these terms are not independent contributions with uniformly $O(1)$ weights: their coefficients are strongly constrained by the circuit structure, operator algebra, trace contractions, and cancellations induced by the observable. As a result, what competes with the multiplicative suppression $\prod_{i=1}^{\xi T} c_{ki}$ is the effective weighted term number $\sum_k \beta_k$, rather than the crude upper bound $3^{\xi T}$. Hence, whenever $\sum_k \beta_k$ grows more slowly than the inverse exponential suppression generated by the product factor, the total variance still decays exponentially with $T$, and a barren plateau emerges.

In writing Eq.~(\ref{E37}), we have implicitly focused on the situation where the observable $O$ is able to collect contributions from all independent parameters, so that the dominant terms in the cost function retain all $\xi T$ nontrivial factors. This is naturally the case for global observables, for which the backward-evolved operator $U^\dagger O U$ typically has extensive support. By contrast, for local observables, some parameters may lie outside the effective causal region of the measurement and therefore do not contribute to the cost function. In that case, the number of nontrivial factors in the product is effectively reduced from $\xi T$ to a smaller number, weakening the multiplicative suppression and potentially avoiding barren plateaus. This is consistent with the cost-function-dependent barren plateau picture reported in Ref.~\cite{cerezo2021cost}: local cost functions can remain trainable because they effectively involve fewer independent parameters, whereas global cost functions are more likely to exhibit exponentially vanishing gradients. From this viewpoint, the onset of barren plateaus is governed not only by the total number of trainable parameters, but more essentially by the effective parameter number seen by the observable.

We note that the above discussion based on the number of summation terms is only qualitative. The precise scaling is determined not by the bare combinatorial count of terms, but by their total weighted contribution together with the multiplicative suppression factor. Nevertheless, one point is clear: whenever a barren plateau arises in the non-sharing case, its exponential complexity originates from the product factor, leading to a scaling of the form $O(c^{\xi T})$ with $0<c<1$. As will be seen below, this is the fundamental distinction from the parameter sharing case.

In the parameter sharing case, the number of trainable parameters remains fixed and does not grow with $T$. Let there be $\xi$ shared parameters $\{\theta_j\}_{j=1}^\xi$. Because the same parameter is repeated $T$ times across the circuit, trigonometric functions of $\theta_j$ can appear with powers up to $T$. The loss function takes the general form
\begin{equation}\label{E38}
    C = \sum_k \beta_k \prod_{j=1}^{\xi} \cos^{a_{kj}}(\theta_j) \sin^{b_{kj}}(\theta_j),
\end{equation}
where for each $(k,j)$ the exponents satisfy $a_{kj}, b_{kj} \ge 0$ and $a_{kj}+b_{kj} \le T$, with the upper bound $T$ stemming from the $T$-fold repetition of each gate. And $\beta_k$ is a constant coefficient that depends mainly on the observable and other quantum gates independent of the parameters. Then, the variance of the gradient can be written as 
\begin{equation}\label{E38_Var}
    \mathrm{Var}\left(\partial_{\nu}C_{\mathrm{share}}\right)
    = \mathbb{E}\left[\left(\partial_{\nu}C\right)^2\right]
    \approx \sum_k \beta_k^2 \,\gamma_\nu(T)\,
           \prod_{j=1}^{\xi}
           \frac{1}{2\pi}\int_0^{2\pi} \cos^{2a_{kj}}(\theta_j) \sin^{2b_{kj}}(\theta_j) \,d\theta_j ,
\end{equation}
where the cross terms between different $k$ have been omitted because their exponents $(a_{kj},b_{kj})$ tend to be more balanced, which lead to exponentially smaller integrals (as will be shown later) and therefore do not affect the leading-order scaling. Moreover, $\gamma_\nu(T)$ collects the prefactors arising from differentiating the trigonometric functions; for the dominant terms of summation, the chain rule yields $\gamma_\nu(T) \propto T^2$.

Defining the basic trigonometric integral
\begin{equation}\label{E39}
    I(a,b) = \frac{1}{2\pi}\int_0^{2\pi} \cos^{2a}\theta \, \sin^{2b}\theta \, d\theta
            = \frac{1}{\pi}\,B\!\left(b+\tfrac12,\,a+\tfrac12\right)
            = \frac{1}{\pi}\,\frac{\Gamma\!\left(a+\tfrac12\right)\Gamma\!\left(b+\tfrac12\right)}{\Gamma(a+b+1)},
\end{equation}
where $B$ denotes the Beta function and $\Gamma$ denotes the Gamma function. Then the variance under parameter sharing can be expressed as
\begin{equation}\label{E39_Var}
    \mathrm{Var}\left(\partial_{\nu}C_{\mathrm{share}}\right) \approx \sum_k \beta_k^2 \,\gamma_\nu(T)\,
       \prod_{j=1}^{\xi} I(a_{kj}, b_{kj}).
\end{equation}

Clearly, the integral in Eq.~(\ref{E39}) is bounded, $I(a,b)\in(0,1)$.  Owing to the prefactor $T^2$, the variance under parameter sharing is at most $\mathcal{O}(T^2)$.  To understand how $I(a,b)$ behaves as a function of $a,b$ (with the worst case: $a+b=T$), we examine the core factor, which is defined as
\begin{equation}\label{E40_new}
    f(a)=\frac{\Gamma\left(a+\tfrac12\right)\Gamma\left(T-a+\tfrac12\right)}{\Gamma(T+1)},\qquad a\in(0,T),
\end{equation}
which is the essential part of Eq.~(\ref{E39}) after using $b=T-a$.  Setting $g(a)=\ln f(a)$ and employing the monotonicity of the digamma function $\psi(z)=\Gamma'(z)/\Gamma(z)$,
\begin{equation}\label{E41}
    g'(a)=\psi\left(a+\frac12\right)-\psi\left(T-a+\frac12\right)=0\;\Longrightarrow\;a=\frac{T}{2}.
\end{equation}
Because $g''(a)=\psi'\left(a+\frac12\right)+\psi'\left(T-a+\frac12\right)>0$ (since $\psi'(z)>0$), $a=T/2$ is the global minimum.  This shows that the closer the values of $a$ and $b$, the smaller the factor $f(a)$ and hence the smaller the expected gradient.  To obtain a larger gradient, one should avoid having $a$ and $b$ too close, i.e., avoid a balanced distribution of the trigonometric powers.

\begin{figure}[ht]
    \centering
    \includegraphics[width=0.45\textwidth]{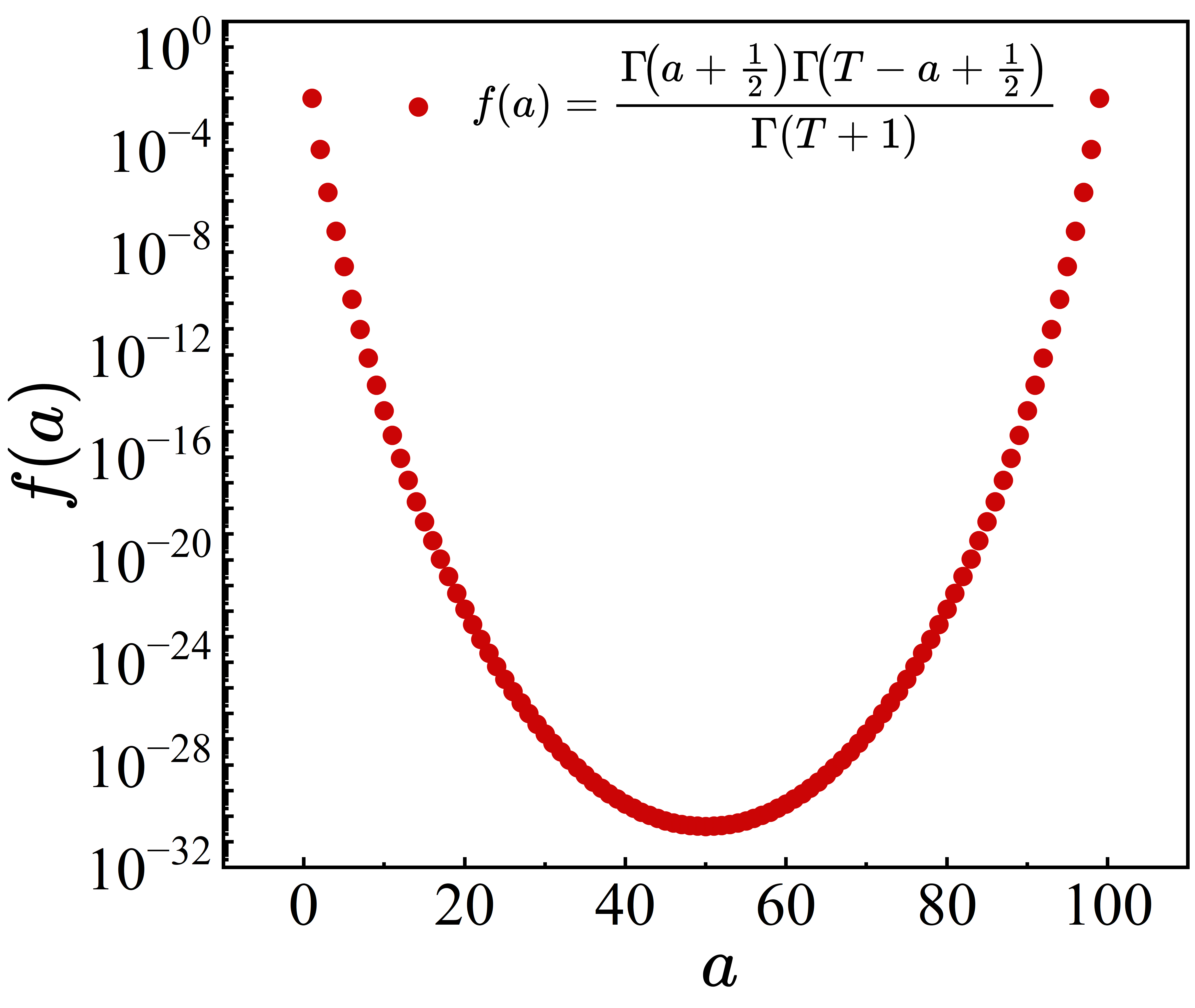}
    \caption{Behavior of $f(a)=\Gamma(a+1/2)\Gamma(T-a+1/2)/\Gamma(T+1)$ for $T=100$ shown on a logarithmic vertical scale. }
    \label{fig:f_a}
\end{figure}

Fig.~\ref{fig:f_a} illustrates the behavior of $f(a)$ for $T=100$, confirming the above analysis. Shown on a logarithmic vertical scale, the figure indicates that $f(a)$ takes relatively large values near the boundaries ($a \approx 0$ or $a \approx T$) and rapidly decays toward the midpoint $a = T/2$. This sharp contrast highlights the dominant contribution of boundary terms to the overall scaling behavior.

Next, we analyze the asymptotic scaling of the function defined in Eq.~(\ref{E40_new}) with respect to $T$. This function can be rewritten as
\begin{equation}\label{E42+}
    f(T)=\frac{\Gamma\left(a+\tfrac12\right)\Gamma\left(T-a+\tfrac12\right)}{\Gamma(T+1)}.
\end{equation}
Two distinct asymptotic regimes must be considered.

\textbf{Case 1}: $a$ fixed (so $b=T-a\sim T$).
Let $a=c$ be a constant. Using Stirling's approximation for the Gamma function,
\begin{equation}\label{E42}
    \Gamma(z)\sim\sqrt{2\pi}\,z^{z-1/2}e^{-z},\qquad z\to\infty,
\end{equation}
we obtain
\begin{equation}\label{E43}
    f(T)\sim\frac{ \sqrt{2\pi}\,(c+1/2)^{c}e^{-c}\;
                  \sqrt{2\pi}\,(T-c+1/2)^{\,T-c}e^{-(T-c)} }
               { \sqrt{2\pi}\,(T+1)^{\,T+1/2}e^{-(T+1)} }.
\end{equation}
Simplifying the exponential and power-law factors, and noting that
$(T-c+1/2)^{\,T-c}/(T+1)^{\,T}\sim e^{-c}T^{-c}$ while the square-root contributions yield a factor of $T^{-1/2}$, we arrive at
\begin{equation}\label{E44}
    f(T) \sim \frac{ \Gamma(c+1/2) }{ T^{c+1/2} }
    \left[1 + \mathcal{O} \left(\frac{1}{T}\right)\right]
    \in \Omega \left(T^{-c-1/2}\right)
    \subset \Omega \left(\frac{1}{\mathrm{poly}(T)}\right).
\end{equation}
Therefore, when $a$ (or equivalently $b$) remains fixed, $f(T)$ decays polynomially as $T^{-c-1/2}$. The same polynomial scaling holds for the integral $I(a,b)$.

\textbf{Case 2}: $a$ and $b$ both scale proportionally with $T$.
Let $a=\alpha T$ and $b=\beta T$, with $\alpha,\beta>0$, $\alpha+\beta=1$, and $\alpha$ bounded away from $0$ and $1$. Applying Stirling's formula again yields
\begin{equation}\label{E45}
    \ln f(T)\sim\alpha T\ln(\alpha T)+\beta T\ln(\beta T)-(T+1)\ln(T+1)-\frac12\ln T+\text{const}.
\end{equation}
Using the condition $\alpha+\beta=1$, this expression simplifies to
\begin{equation}\label{E46}
    \ln f(T)\sim T\left[\alpha\ln\alpha+\beta\ln\beta\right]-\frac12\ln T+\text{const}.
\end{equation}
The function $h(\alpha)=\alpha\ln\alpha+(1-\alpha)\ln(1-\alpha)$ is strictly negative for $\alpha\in(0,1)$, attaining its minimum value $-\ln2$ at $\alpha=1/2$. Consequently,
\begin{equation}\label{E47}
    f(T) \sim e^{T h(\alpha)}\, T^{-1/2}
    \in \mathcal{O} \left(e^{h(\alpha) T}\right), \qquad h(\alpha)<0,
\end{equation}
which implies an exponential decay.

\begin{figure}[ht]
    \centering
    \includegraphics[width=1\textwidth]{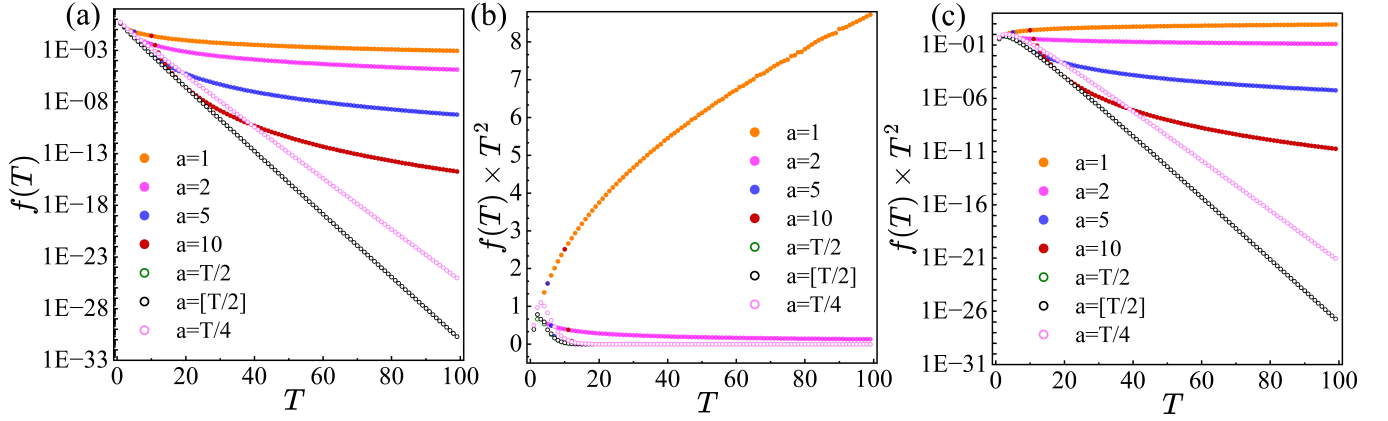}
    \caption{Scaling of $f(T)=\Gamma(a+1/2)\Gamma(T-a+1/2)/\Gamma(T+1)$ for various $a$. (a) $f(T)$ on a logarithmic scale; (b) $f(T)\times T^{2}$ on a linear scale; (c) $f(T)\times T^{2}$ on a logarithmic scale.}
    \label{fig:f_T}
\end{figure}

The asymptotic behavior of $f(T)$ is examined in Fig.~\ref{fig:f_T}. In panel (a), where $f(T)$ is plotted on a logarithmic scale, curves corresponding to fixed $a=1,2,5,10$ exhibit polynomial decay, appearing as concave curves whose slopes become shallower as $T$ increases. In contrast, the curves for $a=\lfloor T/2\rfloor$, $a=T/2$ and $a=T/4$ fall along nearly straight lines, reflecting the exponential decay. To further highlight the distinction between polynomial and exponential scaling, we multiply $f(T)$ by $T^{2}$ and display the results on both linear (Fig.~\ref{fig:f_T}(b)) and logarithmic (Fig.~\ref{fig:f_T}(c)) scales. While the fixed-$a$ curves are lifted to functions with much slower decay, becoming nearly constant or even mildly increasing, the curves with $a=\mathcal{O}(T)$ exhibit little qualitative change and remain exponentially small, confirming the fundamental difference between the two regimes.

Based on theoretical derivation and numerical experiments, We can obtain two key scaling laws:

\textbf{Polynomial decay of Case 1}: when the trigonometric powers are asymmetric, i.e., one of $a$ or $b$ is fixed, the integral $I(a,b)$ decays polynomially with $T$; for example, with $a=1$, $b=T-1$ one has $I\sim T^{-3/2}$.

\textbf{Exponential decay of Case 2}: when the powers are balanced, i.e., $a$ and $b$ both scale proportionally with $T$ and are close to each other, $I(a,b)$ decays exponentially; e.g., for $a=b=T/2$, $I\sim 2^{-T}/\sqrt{T}$.

Since our primary interest lies in the scaling behavior of the gradient variance, it suffices to focus on the terms in Eq.~(\ref{E39_Var}) that dominate the asymptotic scaling. In particular, if there exists an index $k$ such that for all $i$, $a_{k i} = \mathcal{O}(1)$ (or equivalently $a_{k i} \ll T$), then all integrals associated with this term decay at most polynomially with $T$. Taking into account the external factor $\gamma_\nu(T)\propto T^2$ arising from the chain rule, as well as the product over the $\xi$ shared parameters, the gradient variance in the parameter-sharing case satisfies
\begin{equation}\label{E51}
    \mathrm{Var}\left(\partial_{\nu}C_{\mathrm{share}}\right)
    \approx \gamma_\nu(T) \prod_{j=1}^{\xi} I(a_{k j}, b_{k j})
    \in \Omega\left(\frac{T^2}{\mathrm{poly}(T)}\right).
\end{equation}
The above condition on $a_{k i}$ is not restrictive and can be naturally satisfied in quantum circuits. As illustrated by the general ansatz expansion in Eq.~(\ref{E36-1}), whenever shared parameterized gates jointly influence the loss function through multiplicative interactions, the trigonometric expansion inevitably contains asymmetric terms in which the powers $a_{k i}$(or $b_{k i}$) remain independent of $T$. This situation is particularly prevalent in QRNN architectures, where circuit symmetry and the presence of entanglement allow shared parameters at different time steps to couple together and collectively affect the loss function through time. Such asymmetric terms dominate the overall scaling behavior of the variance, ensuring polynomial scaling and avoiding exponential decay. Note that We retain the $T^2$ factor explicitly to indicate that when the polynomial order in the denominator is sufficiently small, the gradient variance may even exhibit a growing behavior with $T$. This completes the proof of Theorem~2 in the main text. 

\hfill$\square$

This insight provides a useful guideline for circuit design: to achieve polynomial gradient complexity, one should aim to make the exponent pairs $(a,b)$ in each trigonometric term of the loss function \emph{asymmetric}, i.e., keep one of $a$ or $b$ as a fixed small number. In the specific QRNN architecture studied here, this corresponds to a loss function containing only cosine terms, $C_T=(1+\cos^T\theta)/2$ (Eq.~\ref{E31}). Its derivative produces a factor $\cos^{2(T-1)}\theta\sin^2\theta$, which corresponds to $a=T-1$ and $b=1$. The associated integral then decays as $\Omega(T^{-3/2})$, and after multiplication by the prefactor $T^2$ yields an overall complexity of $\mathcal{O}(\sqrt{T})$.

For more general PQC-based QRNN architectures, such as the highly-entangled ansatz employed in Section~F, the variance exhibits constant complexity. In this case, the combination of parameter sharing, entanglement, and symmetric circuit structures naturally tends to generate asymmetric distributions of the exponents in the trigonometric expansion. Although many summation terms may be present, the asymptotic scaling is governed by the term that decays most slowly, namely the one with the most asymmetric exponent orientation. As long as at least one term exhibits a sufficiently asymmetric, the overall gradient variance remains polynomial. The constant complexity observed in numerical experiments (variance $\mathcal{O}(1)$) can thus be understood as the statistical consequence of such asymmetric exponent configurations.

In summary, parameter sharing suppresses exponential gradient dilution by fixing the number of trainable parameters, while symmetric and entangled circuit structures drive the exponent pairs $(a,b)$ in the trigonometric expansion toward asymmetric values, ensuring that the associated integrals decay only polynomially. Together, these effects transform the gradient-variance scaling from exponential decay to polynomial behavior. The central design principle is therefore clear: under parameter sharing, one should exploit circuit symmetry to induce asymmetric exponent orientations, thereby guaranteeing polynomial gradient scaling.

\textit{Qualitative discussion based on the Haar measure}---The following discussion is intended only to provide qualitative intuition rather than a rigorous proof. As mentioned in the main text, the sum of Weingarten functions over different permutation combinations from the symmetric group $S_n$ \cite{puchala2017symbolic},
\begin{equation}\label{E40}
    \begin{split}
        \int_{U(d)}
        d\mu(W) &\prod_{k=1}^n w_{i_k j_k^{\prime } } w_{i_k j_k^{\prime } }^* = \sum_{\sigma, \tau \in S_n} \prod_{k=1}^n\delta_{i_k i_{\sigma(k)}^{\prime } } \delta_{j_k j_{\tau(k)}^{\prime } } \text{Wg}(\tau\sigma^{-1}, d).
    \end{split}
\end{equation}
In the parameter non-sharing case, the variational circuit consists of $N$ independent Haar-random unitary matrices $W_1, W_2, \dots, W_N$, where $N \propto T$ scales with the system size. Computing the gradient variance involves performing Haar integration over each $W_i$ independently. The integral for each $W_i$ is a second-order moment. According to Eq. (\ref{E40}), when $n=2$, the dominant contribution comes from the identity permutation, yielding 
\begin{equation}\label{E41}
    \int_{\mathcal{U}(d)} d \mu(W) w_{\boldsymbol{i}_1, \boldsymbol{j}_1} w_{\boldsymbol{i}_2, \boldsymbol{j}_2} w_{\boldsymbol{i}_1^{\prime}, \boldsymbol{j}_1^{\prime}}^* w_{\boldsymbol{i}_2^{\prime}, \boldsymbol{j}_2^{\prime}}^*
        \approx  \frac{1}{d^2}\left(\delta_{\boldsymbol{i}_1, \boldsymbol{i}_1^{\prime}} \delta_{\boldsymbol{i}_2, \boldsymbol{i}_2^{\prime}} \delta_{\boldsymbol{j}_1, \boldsymbol{j}_1^{\prime}} \delta_{\boldsymbol{j}_2, \boldsymbol{j}_2^{\prime}}+\delta_{\boldsymbol{i}_1, \boldsymbol{i}_2^{\prime}} \delta_{\boldsymbol{i}_2, \boldsymbol{i}_1^{\prime}} \delta_{\boldsymbol{j}_1, \boldsymbol{j}_2^{\prime}} \delta_{\boldsymbol{j}_2, \boldsymbol{j}_1^{\prime}}\right),
\end{equation}
which is the approximate version of Eq. (\ref{A7}). That is, each independent module contributes a factor of $1/d^2$. Since there are $N \propto T$ independent modules, these factors multiply to give 
\begin{equation}\label{E42}
    \text{Var}(\partial_{\theta} C_{\text{non}}) \sim \left( \frac{1}{d^2} \right)^T \sim d^{-2T},
\end{equation}
which explains the origin of the exponential decay.

In the parameter sharing case, a single Haar-random unitary $W$ is repeated across all $T$ time steps. Here, computing the gradient variance involves integrating a $2T$-th moment of the same $W$. By the product rule, the gradient is a sum of $T$ terms $\partial_\theta C = \sum_{t=1}^T g_t$, where $g_t$ corresponds to the derivative term at time step $t$. The variance is 
\begin{equation}\label{E43}
    \text{Var}(\partial_{\theta} C_{\text{share}}) = \mathbb{E}\left[ \left( \sum_{t=1}^T g_t \right)^2 \right] = \sum_{t=1}^T \sum_{t'=1}^T \mathbb{E}[g_t g_{t'}],
\end{equation}
which contains $T^2$ cross-terms. However, due to the symmetry of the circuit structure, the actual number of non-zero terms is often less than $T^2$. 

To evaluate $\mathbb{E}[g_t g_{t'}]$, we can expand the trace and obtain
\begin{equation}\label{E44}
\mathbb{E}[g_t g_{t'}] = \sum_{\alpha,\beta} c_{\alpha\beta}(t,t') \int_{U(d)} dW \, \prod_{k=1}^{2T} w_{i_k^{(\alpha)} j_k^{(\alpha)}} {w}_{i_k^{'(\beta)} j_k^{'(\beta)}}^*,
\end{equation}
where the coefficients $c_{\alpha\beta}(t,t')$ arise from the traces involving the operators $O$, $\rho_{\text{in}}$, and other quantum gates unrelated to $W$, and the multi-indices $\{i_k^{(\alpha)}, j_k^{(\alpha)}, i_k^{'(\beta)}, j_k^{'(\beta)}\}$ encode the specific contraction pattern for each term in the expansion. The product contains exactly $2T$ factors of $W$ or ${W}^*$, confirming that the integral is a $2T$-th moment of the Haar measure, which requires $W$ to form $2T$-design. Applying the Weingarten formula (Eq. (\ref{E40})) to each such moment gives
\begin{equation}\label{E45}
    \mathbb{E}[g_t g_{t'}] = \sum_{\alpha,\beta} c_{\alpha\beta}(t,t') \sum_{\sigma, \tau \in S_{2T}} \left( \prod_{k=1}^{2T} \delta_{i_k^{(\alpha)}, i_{\sigma(k)}^{'(\beta)}} \delta_{j_k^{(\alpha)}, j_{\tau(k)}^{'(\beta)}} \right) \text{Wg}(\tau\sigma^{-1}, d).
\end{equation}
Here, the Kronecker deltas enforce the index contractions determined by the circuit and measurement structure, and the Weingarten function $\text{Wg}(\sigma, d)$ decays predominantly as $d^{-|\sigma|}$ \cite{puchala2017symbolic}, with other terms being higher-order corrections in the large-$d$ limit. The key to avoiding BPs lies in how parameter sharing alters the structural properties of the integral:

(1) Number of effective permutation patterns is $\mathcal{O}(\text{poly}(T))$: Among the $(2T)!^2$ permutation pairs $(\sigma,\tau)$, the dominant contributions to the decay factor $d^{-|\sigma|}$ come only from the identity permutation $\sigma = \tau = e$ and certain simple cyclic permutations (e.g., adjacent swaps). Consequently, the number of effective permutation patterns is only $\mathcal{O}(\text{poly}(T))$.

(2) Loop cancellation mechanism from $\delta$-functions: For each permutation pattern, the Weingarten function provides a suppression factor scaling as $d^{-2T}$. However, in the parameter sharing architecture, the repeated appearance of the same unitary $W$ across time steps generates exactly $2T$ closed index loops in the tensor-network diagram due to the $\delta$-function constraints. Each loop corresponds to a trace over a subsystem. When the traced operator is the identity, each loop contributes a factor of $d$. Since most loops arise from tracing identity operators, the total loop enhancement factor is approximately $d^{2T}$, which nearly cancels the Weingarten suppression $d^{-2T}$. The cancellation leaves a residual dimensional dependence that is at most polynomial in $d$ (e.g., $d^{-2}$ due to traces of the observable $O$ and the input state $\rho_{\mathrm{in}}$ replacing two of the identity traces). Crucially, since any polynomial factor in $d$ is constant with respect to $T$, this residual factor is $\mathcal{O}(1)$ in terms of $T$-scaling. Hence, each effective permutation pattern contributes $\mathcal{O}(1)$ to the variance as a function of $T$.

(3) Number of effective cross-terms is $\mathcal{O}(\text{poly}(T))$: Although the variance expression contains $T^2$ cross-terms, many of them vanish or become negligible under Haar averaging due to the temporal symmetry of the circuit and the locality of the observable. The actual number of effective cross-terms is only $\mathcal{O}(\text{poly}(T))$.

Combining the above three points, the total variance can be estimated as
\begin{equation}\label{E45}
    \text{Var}(\partial_\theta C_{\text{share}}) \sim \mathcal{O}(\text{poly}(T))/d^{2},
\end{equation}
where the factor $\mathcal{O}(\text{poly}(T))$ comes from points (1) and (3), and the factor $d^{-2}$ comes from point (2).

From this perspective, we find that the core mechanism by which parameter sharing avoids BPs is that it transforms the product of multiple independent Haar integrals (which leads to factorized exponential decay $d^{-2T}$) in the non-sharing case into a high-order moment integral of a single unitary matrix. In the Weingarten expansion of this high-order moment, (1) the number of index loops grows linearly with $T$, almost canceling the exponential decay from the Weingarten function; (2) the number of effective permutation patterns is only $\mathcal{O}(\text{poly}(T))$, rather than super-exponential; (3) the number of effective cross-terms is only $\mathcal{O}({\text{poly}(T)})$. The combined effect of these three factors ultimately renders the gradient variance polynomially complex, rather than exponentially decaying.

\textit{Further discussion.}---Whether the number of parameters scales with the system size---such as the sequence length $T$ or the number of qubits $m$---is the key factor determining the scaling behavior of the gradient variance. In the parameter non-sharing setting, the number of trainable parameters $N$ grows linearly with the system size, $N \propto T$. Consequently, the bounded expectation value of the observable is spread over an increasing number of degrees of freedom, so that each parameter receives an exponentially shrinking share of the total sensitivity. This effect persists even in highly randomized circuits that form a 2‑design. That is, the Haar average over each independent parameter introduces a suppression factor of order $1/d$, and the product of these factors over multiple modules leads to a gradient variance that decays as $\sim (1/d)^T$, i.e., a BP phenomenon. In contrast, parameter sharing keeps the number of effective trainable parameters fixed, $N =\mathcal{O}(1)$, independent of the system size. The bounded expectation value is therefore not diluted, thereby avoiding the BPs. Moreover, because a shared parameter influences the output repeatedly through $T$ time steps, its contributions can coherently interfere for suitably chosen circuit structures. In such cases, the influence of a single parameter can even be enhanced, leading to a complexity that grows polynomially with the system size.

Although parameter sharing lifts the gradient scaling from exponential to polynomial, the precise polynomial degree (e.g., $\mathcal{O}(1)$, $\mathcal{O}(\sqrt{T})$, or $\mathcal{O}(T^2)$) is determined by the geometric structure of the circuit. The difference originates from the degree of coherence among the terms $g_t$ in the gradient expansion $\partial_\theta C = \sum_{t=1}^T g_t$. If the circuit consists of a chain of identical single-qubit rotations (e.g., $R_y(\theta)$) along the depth direction, their coherent superposition produces a $T^2$ amplification of the variance. Introducing transverse (qubit count) extensions or multi-qubit entanglement breaks this perfect synchrony, weakens the correlations among the $g_t$, and reduces the amplification, leading to a scaling of $\mathcal{O}(\sqrt{T})$ or $\mathcal{O}(1)$. The same mechanism explains why even in certain highly symmetric non-sharing circuits, such as the circuit depth scaling model, an “effective merging” of parameters across layers can also avoid BPs.

Therefore, the design of trainable quantum sequential models should adhere to the following principles. First, one must strictly control the number of effective parameters, for instance through parameter sharing or other structural designs such as translation invariance, to prevent the exponential dilution of gradients. Second, the temporal coherence of the circuit should be consciously exploited to obtain the desired gradient amplification. Finally, a balance between expressivity and trainability must be struck: overly simple repeating units may facilitate training but limit the model’s capacity, while introducing excessive randomization or independent parameters reintroduces the risk of BPs.

\section{\label{F}Section F. Numerical Experiments}
This section presents additional experimental results to validate the theoretical predictions. We perform two sets of experiments, as described in the main text. The first set examines the scaling of gradient variance in the three examples discussed in Section E.1 and Section E.2. The second set focuses on a general QRNN model.

\subsection{\label{F1}1. Results for the Gradient Variance in the Three Examples}
\begin{figure}[ht]
    \centering
    \includegraphics[width=5.5in]{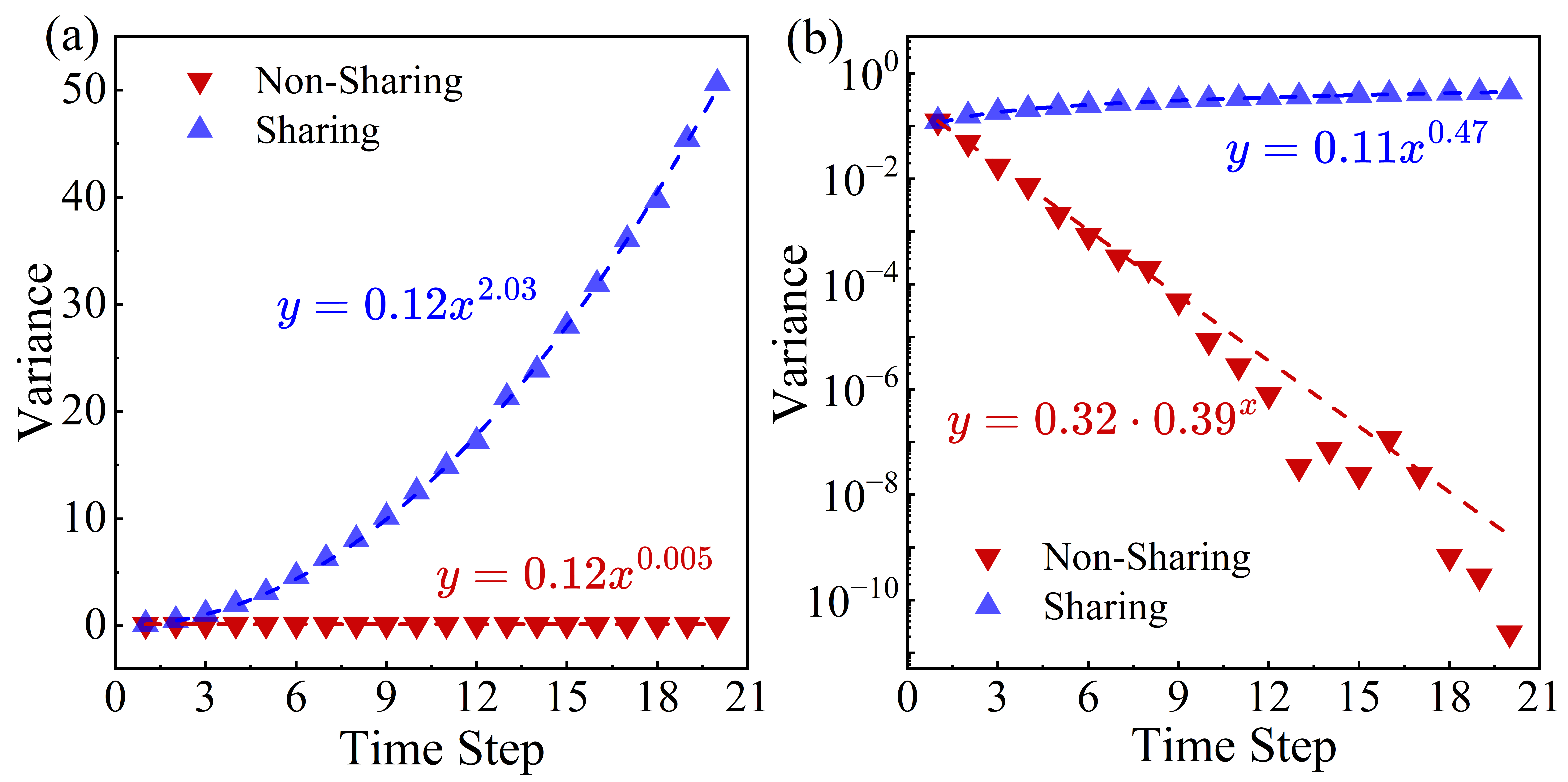}
    \caption{Gradient variance of the loss function as a function of the number of time steps ($T$) for (a) the circuits in Fig. \ref{FE1} and (b) the circuits in Fig. \ref{FE2}. Fitted curves (with a coefficient of determination $R^2 > 0.999$) are included to illustrate the scaling trend.}
    \label{FF11}
\end{figure}

Fig. \ref{FF11} shows the gradient variance for the ansatz circuit in Section E.1. The numerical results for the loss function gradient variance of the circuits in Fig. \ref{FE1} and Fig. \ref{FE2} are presented in Fig. \ref{FF11}. Partial derivatives of the loss function are estimated using a differential approximation method, and the variances are computed over 3000 uniformly sampled parameter values. Fitted curves are included to illustrate scaling trends. The results confirm that in the circuit depth dimension, the gradient variance scales as $\mathcal{O}(T^2)$ under parameter sharing and $\mathcal{O}(1)$ without sharing; in the qubit count dimension, the variance scales as $\mathcal{O}(\sqrt{T})$ with parameter sharing and $\mathcal{O}((3/8)^T)$ without sharing. These findings are consistent with our theoretical conclusion that parameter sharing amplifies the gradient variance in both the circuit depth and qubit count dimensions.

\begin{figure}[ht]
    \centering
    \includegraphics[width=5.5in]{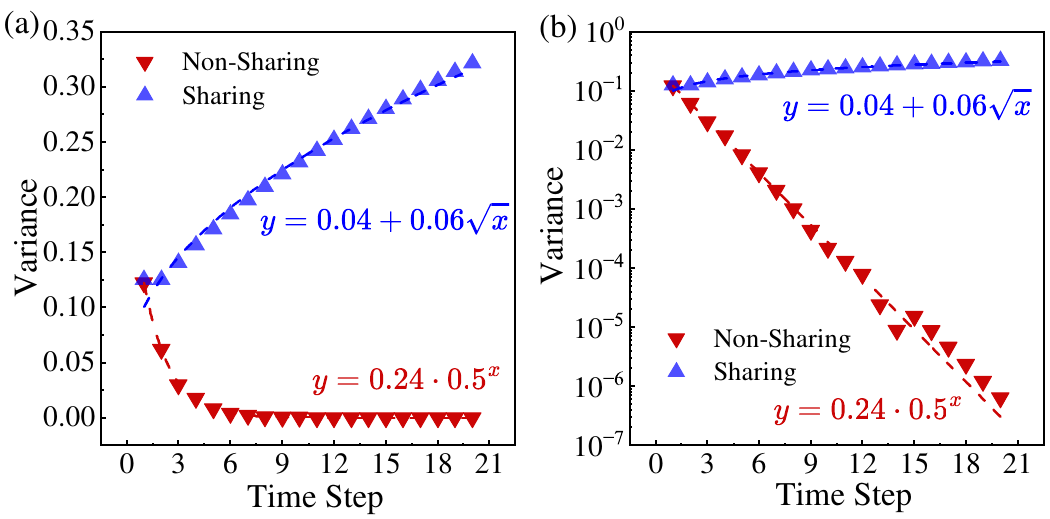}
    \caption{Gradient variance of the loss function as a function of the number of time steps ($T$) for the QRNN architecture shown in Fig. \ref{FE3}, presented in (a) linear and (b) logarithmic axes. Fitted curves (with a coefficient of determination $R^2 \approx 0.99$) are included to illustrate the scaling trend.}
    \label{FF12}
\end{figure}

Fig. \ref{FF12} presents the numerical results for the gradient variance of the specific QRNN architecture discussed in Section E.2 (Fig. \ref{FE3}). As shown, with parameter sharing, the variance exhibits a scaling trend of $\mathcal{O}(\sqrt{T})$, while without parameter sharing, it follows $\mathcal{O}((1/2)^T)$. These results indicate that the recurrent architecture alone is insufficient to mitigate BPs in QRNNs, but parameter sharing can effectively alter the gradient variance scaling, helping to mitigate BPs.
  
\subsection{\label{F2}2. Results for the Gradient Variance of General QRNN Models}
\begin{figure}[ht]
    \centering
    \includegraphics[width=5.2in]{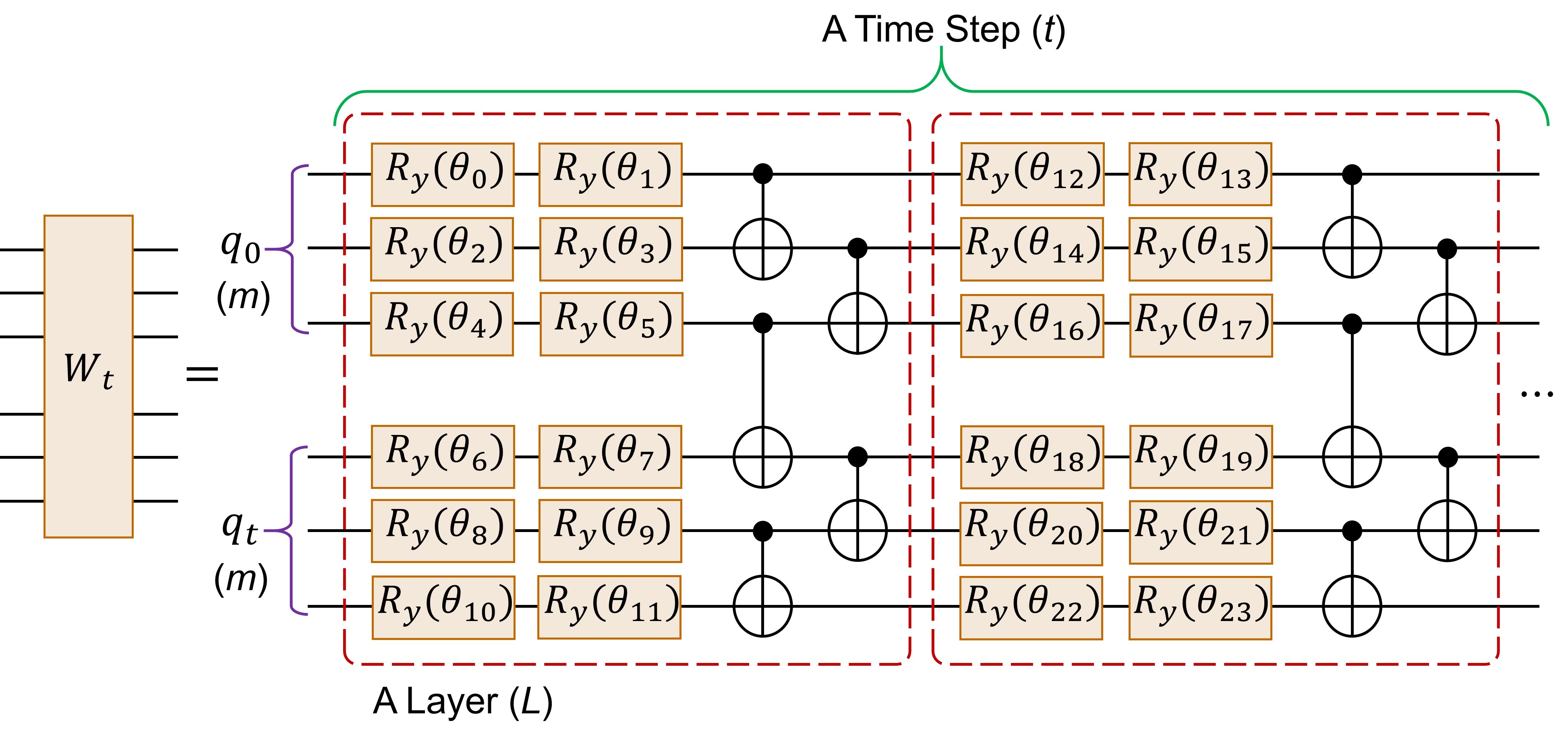}
    \caption{A strongly entangled, multi-layer ($L$) ansatz circuit used as the module $W_t$ to build a general QRNN model. It acts on the $q_0$ and $q_t$ registers, each with $m$ qubits, and implements one time step in the QRNN architecture.}
    \label{FF13}
\end{figure}

We generalize the QRNN model in Fig. \ref{FE3} by replacing each PQC module $W_t$ with a hardware-efficient, strongly entangled ansatz circuit, as shown in Fig. \ref{FF13}. Each such circuit acts as a recurrent block (i.e., a time step) in the QRNN. By increasing the number of layers ($L$), the expressive power of the circuit is enhanced, and its output state approaches the Haar distribution; hence the QRNN model constructed in this manner exhibits general learning capability. The resulting QRNN architecture is characterized by three variables: the number of time steps $T$, the number of layers $L$, and the number of qubits $m$ in the Storer and Loader registers, as illustrated in Fig. \ref{FF13}.

\begin{figure}[ht]
    \centering
    \includegraphics[width=1.0\linewidth]{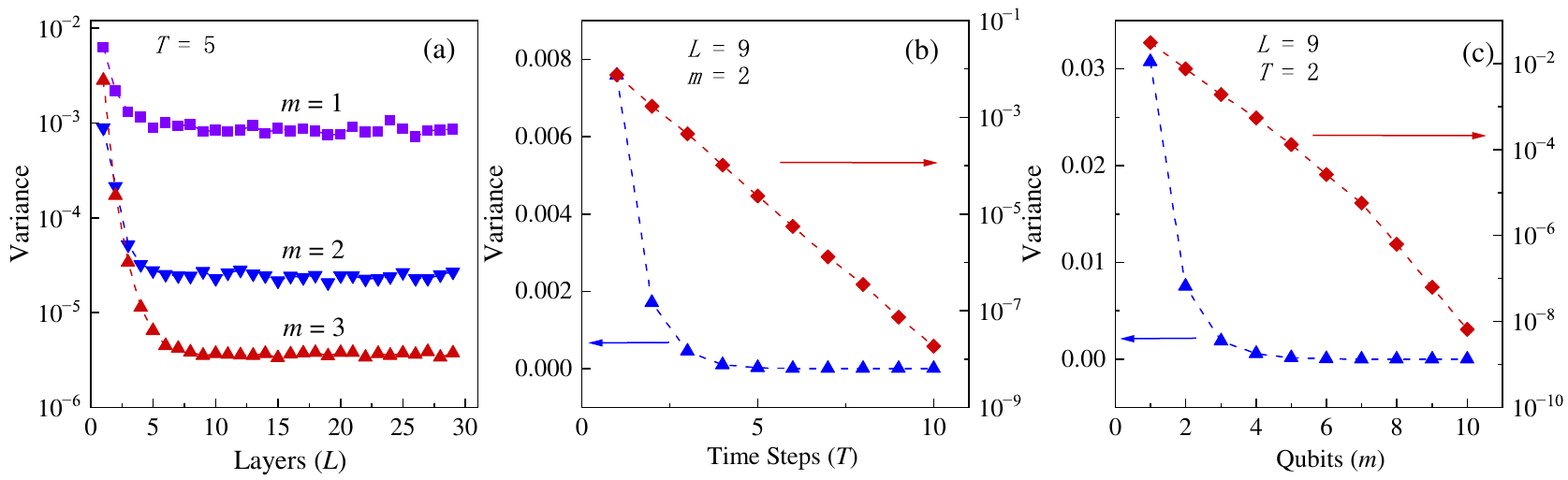}
    \caption{Gradient variance of the loss function for a general QRNN architecture under parameter non-sharing. (a) Variance versus the number of layers ($L$), with time step count fixed at $T = 5$ and register sizes $m=1,2,3$ (see Fig.~\ref{FF13}). The variance decays exponentially with increasing $L$, and saturates as the ansatz circuit approaches the Haar distribution. Under the $2$-design condition (i.e., $L=9$), the variance also decays exponentially with (b) the number of time steps $T$ and (c) the register size $m$ (both linear and log axes are shown). These scaling trends agree with Theorem 1, confirming the presence of a BP in the non-sharing setting.}
    \label{FF14}
\end{figure}

In the parameter non-sharing case, all rotation angles of the single-qubit gates are distinct, resulting in a total of $N_{\text{non}} = 4mLT$ parameters in the QRNN. The angle values $\theta_i$ are randomly sampled from the interval $[0, 2\pi]$, and the partial derivatives of the loss function are computed using the parameter-shift method \cite{schuld2019evaluating}. The gradient variance is then estimated from 3000 random parameter sets. The numerical results for the gradient variance are presented in Fig. \ref{FF14}.

Fig. \ref{FF14}(a) shows that when the number of layers is $L=9$, the ansatz circuit approximates the 2-design condition. Under this setting, Fig. \ref{FF14}(b) and Fig. \ref{FF14}(c) show that the gradient variance decays exponentially with increasing time steps $T$ and the number of qubits $m$. These results are consistent with the predictions of Theorem 1, as described in Eq. (\ref{D25}) and Eq. (\ref{D26}), demonstrating that in the case of parameter independence, the QRNN architecture exhibits the BP phenomenon.

\begin{figure}[ht]
    \centering
    \includegraphics[width=1.0\linewidth]{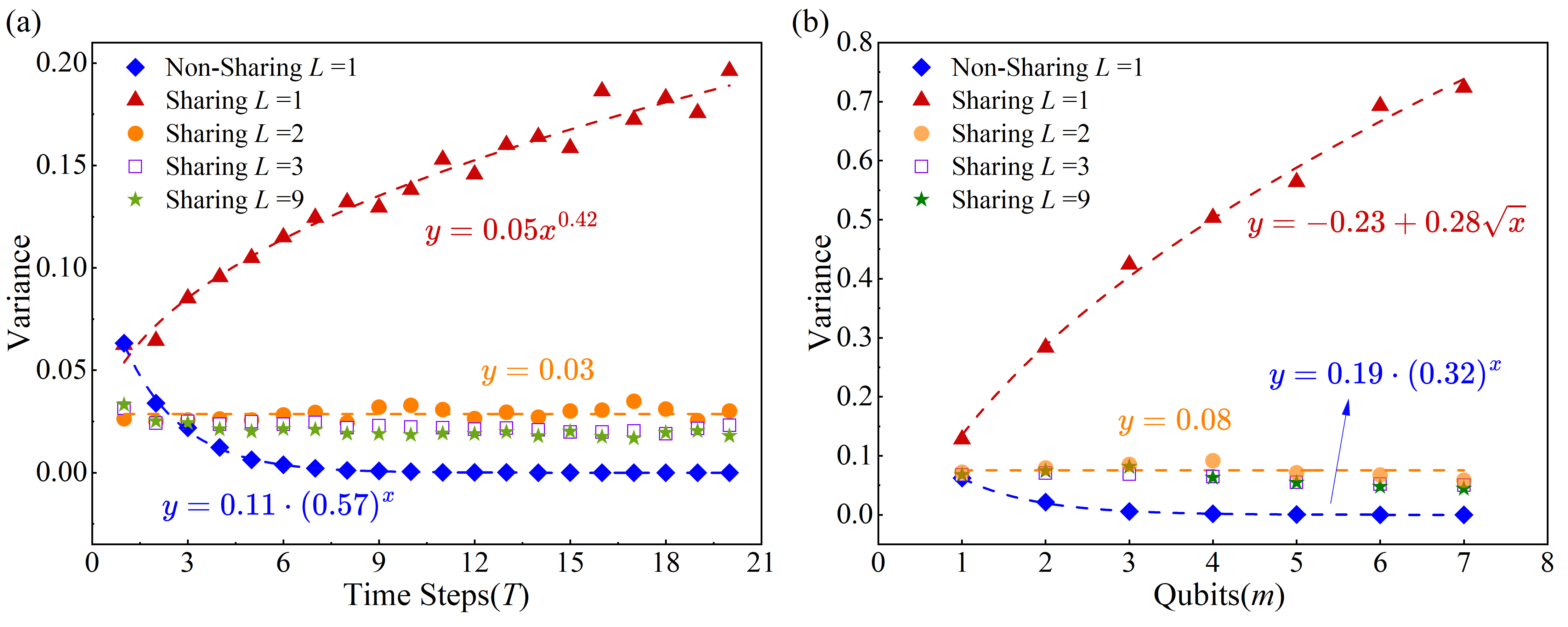}
    \caption{Gradient variance of the loss function for a general QRNN architecture under parameter sharing. (a) Variance versus time steps ($T$), with register size fixed at $m = 1$ and layers $L=1,2,3,9$; (b) variance versus register size ($m$), with time step fixed at $T = 2$ and the same set of $L$ values. For comparison, the corresponding non‑sharing results ($L=1$) are also plotted. Fitted curves illustrate the scaling trends, confirming the absence of a BP under parameter sharing.}
    \label{FF15}
\end{figure}

In the parameter sharing case, the rotation angles of the single-qubit gates are identical across different recurrent blocks, leading to a total of $N_{\text{share}} = 4mL$ parameters in the QRNN. Here, the partial derivatives of the loss function are calculated using the differential approximation method, and the variance is estimated over 3000 samples. The numerical results for the gradient variance are presented in Fig.~\ref{FF15}.

As shown in Fig.~\ref{FF15}, even when the number of layers is $L =1$ (below the 2-design threshold), the non-sharing variance already decays exponentially with respect to the time step $T$ and register size $m$. In contrast, the variance in the parameter-sharing case exhibits an increasing trend. As $L$ increases from 1 to 9 (as the circuit approaches the Haar distribution), the magnitude of the parameter-sharing variance decreases; however, it stabilizes within a certain range and does not decay exponentially. Regardless of the number of layers, the variance remains at least of constant complexity, $\mathcal{O}(1)$, preventing the loss function from experiencing BP. These results support Theorem 2 in the main text, confirming that parameter sharing significantly alters the gradient scaling behavior in QRNNs, effectively preventing gradient vanishing during training.

